\documentclass{aa}  
\usepackage{CJKutf8}
\usepackage{graphicx}
\usepackage[breaklinks,colorlinks,citecolor=blue,linkcolor=blue]{hyperref}
\usepackage{txfonts}
\usepackage{lscape}
\usepackage{epstopdf}
\usepackage{CJK}
\usepackage{float}
\usepackage[absolute]{textpos}
\usepackage{amsmath}
\usepackage{mathrsfs}

\usepackage{arydshln}

\begin{document} 
   \title{PMO Polaris CO survey. I. A 100 deg$^2$ view of the Polaris Flare}
   \author{Xunchuan Liu \begin{CJK}{UTF8}{gbsn}(刘训川)\end{CJK}
          \inst{1,3}\thanks{liuxunchuan001@gmail.com}
          \and
          Bing-Gang Ju
          \inst{2}\thanks{bgju@pmo.ac.cn}
          \and 
          Fujun Du\inst{2}
          \and
          Tianwei Zhang\inst{4}
          \and 
          Lixia Yuan\inst{2}
          \and 
          Paul F. Goldsmith\inst{5}
          \and 
          Lianghao Lin\inst{6}
          \and
          Zhihong He\inst{7}
          \and 
          Chao Zhang\inst{8}
          \and 
          Ping Yan\inst{2}
          \and 
          Shengyu Jin\inst{2}
          \and 
          Yongxing Zhang\inst{2}
          \and 
          Dengrong Lu\inst{2}
          } 
   \institute{
Leiden Observatory, Leiden University, P.O. Box 9513, 2300RA Leiden, The Netherlands 
\and
Purple Mountain Observatory, Chinese Academy of Sciences, Nanjing 210023
\and
Shanghai Astronomical Observatory, Chinese Academy of Sciences, Shanghai 200030, PR China
\and 
Research Center for Computational Earth and Space Science, Zhejiang Laboratory, Hangzhou 311100, China
\and
Jet Propulsion Laboratory, California Institute of Technology, 4800 Oak Grove Drive, Pasadena, CA 91109, USA
\and
Max-Planck-Institut f\"ur Radioastronomie, Auf dem Hügel 69, 53121 Bonn, Germany
\and 
School of Physics and Astronomy, China West Normal University, No. 1 Shida Road, Nanchong 637002, China
\and 
Institute of Astronomy and Astrophysics, School of Mathematics and Physics, Anqing Normal University, Anqing, China
              }
              
\date{Received xxx xx}
\abstract{
Large-area CO surveys are essential for studying molecular cloud dynamics and evolution; however, most have focused on the Galactic plane, leaving high-latitude clouds less explored. We present the PMO Polaris CO Survey (PPCOS), which maps a 100~deg$^2$ region of the Polaris Flare in the $J=1-0$ transitions of $^{12}$CO, $^{13}$CO, and C$^{18}$O using the Delingha 13.7~m telescope. As the first large-area CO survey at high Galactic latitude ($|b| > 20\degr$) with sub-arcminute resolution, PPCOS achieves sensitivities of $\sim$0.46~K for $^{12}$CO and $\sim$0.23~K for $^{13}$CO and C$^{18}$O at a spectral resolution of 0.16~km~s$^{-1}$ and an angular resolution of 50\arcsec.
The $^{12}$CO emission reveals seven distinct complexes, where only $\sim$10\% of pixels display multiple velocity components, alongside a global velocity gradient of 0.18~km~s$^{-1}$~pc$^{-1}$. Typical line widths are $1.2 \pm 0.6$~\mbox{km~s$^{-1}$} for $^{12}$CO, while $^{13}$CO components are systematically narrower ($\lesssim 0.7\,\Delta V_{\rm ^{12}CO}$). The $^{12}$CO/$^{13}$CO intensity ratios (5--25) indicate widespread $^{12}$CO optical thickness, resembling conditions found in giant molecular clouds (GMCs).
Globally, the CO emission divides into two groups: a major group aligned with the velocity gradient and a secondary group elongated perpendicular to it, possibly regulated by large-scale coherent dynamics. We propose a three-layer hierarchy: a dynamically assembling and dispersing periphery traced by $^{12}$CO, a more stable intermediate kernel traced by $^{13}$CO, and gravitationally bound compact cores traced by C$^{18}$O. No young stellar objects are firmly associated with the molecular gas. PPCOS provides an ideal laboratory for studying turbulence, hierarchical structure, and early cloud evolution in a nearby, relatively simple molecular cloud.
}


\keywords{ISM: kinematics and dynamics --- ISM: clouds --- ISM: evolution }

 \maketitle
 

\section{Introduction} \label{sec_intro}

Molecular gas provides the fundamental fuel for star formation and plays a critical role in galaxy evolution. Understanding where and how atomic hydrogen (\ion{H}{I}) converts into molecular hydrogen (H$_2$), the most abundant molecule in the interstellar medium (ISM), is essential for advancing our knowledge of cloud physics and Galactic structure \citep[e.g.,][]{Spitzer1978,1979ApJ...232L..89S,1998ApJ...498..541K,2009ApJ...693..216K}. The 21~cm line of \ion{H}{I} allows for efficient all-sky and wide-area surveys \citep{1954BAN....12..117V,1993AIPC..278..279H,2016A&A...594A.116H}, providing comprehensive coverage of the atomic gas in the Milky Way \citep{2003ApJ...586.1067H,2018MNRAS.474..289W,2020A&A...639A..26K,2025NatAs.tmp..148L}. In contrast, the symmetric H$_2$ molecule lacks a permanent dipole moment, making direct observations of cold molecular gas extremely challenging \citep[e.g.,][]{1977ApJ...216..291S,2017ApJ...841...25G}.
Carbon monoxide (CO) is the second most abundant interstellar molecule, and its $J=1-0$ ground-state transition is easily excited at low temperatures. Therefore, CO is widely used as the primary tracer for cold molecular gas mass \citep{2001ApJ...547..792D,2015ARA&A..53..583H}. However, large-area CO mapping projects remain highly time-consuming, even when using single-dish telescopes with relatively large beams. Consequently, most systematic surveys have been restricted to the dense Galactic plane, leaving high-Galactic-latitude regions relatively poorly explored (see Table~\ref{tab:cloud_surveys}).
Major Galactic plane surveys include the Galactic Ring Survey \citep[GRS;][]{2006ApJS..163..145J}, the Three-mm Ultimate Mopra Milky Way Survey \citep[ThrUMMS;][]{2015ApJ...812....6B}, the FOREST Unbiased Galactic plane Imaging survey \citep[FUGIN;][]{2017PASJ...69...78U}, and the Milky Way Imaging Scroll Painting \citep[MWISP;][]{2019ApJS..240....9S}. Mapping clouds outside of the busy Galactic plane is crucial for studying the early stages of molecular cloud formation, before active star formation begins to disrupt the natal gas \citep[e.g.,][]{1985ApJ...295..402M,2008ApJ...680..428G}.

\begin{table*}[htbp]
\centering
\caption{Summary of large-area ($\gtrsim 10$ deg$^2$) molecular cloud CO (1–0) surveys$^{(1)}$}
\label{tab:cloud_surveys}
\begin{tabular}{l c l c c c l}
\hline\hline
Survey \& Region & Map area & Telescope & $\Theta^{(2)}$ & $\delta V$$^{(3)}$ & Noise$^{(4)}$ & Ref. \\
       & (deg$^2$) &          & (\arcmin) & (km\,s$^{-1}$) & (K) &  \\
\hline
PPCOS (Polaris Flare) & $\sim$100 & Delingha 13.7m & 50 & 0.16 & 0.46$^{(9)}$ & This work \\
FCRAO (Taurus)$^{(5)}$ & $\sim$100 & FCRAO 14m & 45 & 0.076 & 0.58 & \citealt{2008ApJS..177..341N} \\
COMPLETE (Perseus A)$^{(6)}$ & $\sim$10 & FCRAO 14m & 45 & 0.064 & 0.35 & \citealt{2006AJ....131.2921R} \\
COMPLETE (Ophiuchus A) & $\sim$10 & FCRAO 14m & 45 & 0.064 & 0.98 & \citealt{2006AJ....131.2921R} \\
ThrUMMS (Galactic Plane)$^{(7)}$ & 120 & Mopra 22m & 33 & 0.33 & 1.3 & \citealt{2015ApJ...812....6B} \\
MWISP (Galactic Plane) & $\sim$2500$^{(8)}$ & Delingha 13.7m & 50 & 0.16 & 0.45 & \citealt{2018AcASn..59....3S} \\
FUGIN (Galactic Plane) & $\sim$150 & Nobeyama 45m & 14 & 1.3 & 1.47 & \citealt{2017PASJ...69...78U} \\
DHT16 (Polaris Flare)$^{(10)}$ & $\sim$50 & CfA 1.2m & 8.5 & 0.65 & 0.25 & \citealt{1990ApJ...353L..49H} \\
\hline\hline
\end{tabular}\\
\vspace{0.5em}
\footnotesize{
$^{(1)}$ Only surveys that mapped $^{12}$CO (1--0) and were conducted with telescopes of aperture larger than 10\,m 
(i.e., with a native spatial resolution better than $\sim$1\arcmin) are listed, with DHT16 being the exception.
$^{(2)}$ Full width at half-power (FWHP) of the telescope's main beam.
$^{(3)}$ $\delta V$ indicates the velocity resolution (usually the velocity channel width) of the survey.  
$^{(4)}$ The $1-\sigma$ noise level of $^{12}$CO (1--0) at a channel width of $\delta V$ shown in the 5$_{\rm th}$ column. The noise level of $^{13}$CO is typically 50\% lower than that of $^{12}$CO due to higher atmospheric transparency at its frequency.  
$^{(5)}$ The FCRAO map of Taurus is referred to as the FCRAO survey for simplicity.  
$^{(6)}$ The label ``A'' in Perseus~A and Ophiuchus~A denotes the main part of each cloud.
$^{(7)}$ The intrinsic velocity resolution is 0.085~km\,s$^{-1}$, but the data were smoothed to 0.33~km\,s$^{-1}$ to improve sensitivity; the listed noise corresponds to the smoothed data.  
In addition to ThrUMMS, another Mopra survey, the Southern Galactic Plane CO Survey \citep{2015PASA...32...20B}, covered a relatively larger area of $\sim$200~deg$^2$; however, it was not Nyquist sampled and is therefore not included here.  
$^{(8)}$ MWISP is gradually extending to $|b|\lesssim 10^\circ$, beyond the original $|b|\lesssim 5^\circ$ coverage.  
$^{(9)}$ For the PMO Polaris CO survey (this work), the quoted noise corresponds to the native spatial and spectral resolution (see Sect.~\ref{sec_obs} and Figure~\ref{fig:cellnoise}).  
$^{(10)}$ The survey by the CfA 1.2~m telescope, HT90 \citep{1990ApJ...353L..49H,1993A&A...268..265H}, and its extended version DHT16 \citep[][see also Sect.~\ref{sec_intro}]{2001ApJ...547..792D} are limited in spatial resolution and only approximately half-Nyquist sampled. They are included here because HT90/DHT16 represents the first large-area CO survey of the Polaris Flare.
}
\end{table*}

The Polaris Flare is an ideal target for wide-field CO observations among high-Galactic-latitude molecular clouds \citep{1990ApJ...353L..49H}. It is a classic high-latitude cirrus cloud located at a distance of approximately 150~pc \citep{1998A&A...331..669F}. Although historical distance estimates range from 105--125~pc \citep{1999A&A...352..645Z} up to 240~pc \citep{1990ApJ...353L..49H}, the value of 150~pc is widely adopted in modern studies. This close proximity allows sub-arcminute observations to resolve physical structures down to linear spatial scales of $\sim$0.04~pc. Consequently, researchers can perform detailed investigations of both the diffuse envelope and the dense embedded components of the cloud.
The cloud complex is characterized by cold dust temperatures \citep{1998A&A...333..709L,1999A&A...347..640B}, substantial CO line emission \citep{1990ApJ...353L..49H,1996A&A...313..929M}, and a total lack of active star formation. Deep \textit{Herschel}/SPIRE surveys have identified only pre-stellar cores within the region \citep{2010A&A...518L.102A}. This absence of internal stellar feedback makes the Polaris Flare a pristine observational laboratory for exploring the earliest stages of molecular cloud formation \citep{2010A&A...518L.104M}. Furthermore, it offers a remarkably unconfused background. This clear background facilitates rigorous investigations into the diffuse interstellar medium, including the atomic-to-molecular phase transition and its associated hydrodynamical processes \citep[e.g.,][]{1998A&A...336..697S,2002A&A...390..307O,2010MNRAS.406.2713B}.
In this context, a systematic survey mapping the lowest rotational $J=1-0$ transition of $^{12}$CO and its primary isotopologues is ideal. While $^{12}$CO traces the widespread diffuse periphery, the rarer isotopologues selectively probe the embedded denser sub-structures. Together, these tracers can yield a comprehensive and continuous view of the cloud's hierarchical architecture.

\begin{figure*}[!t]
    \centering
    \hspace{0.5cm}\includegraphics[width=0.8\linewidth]{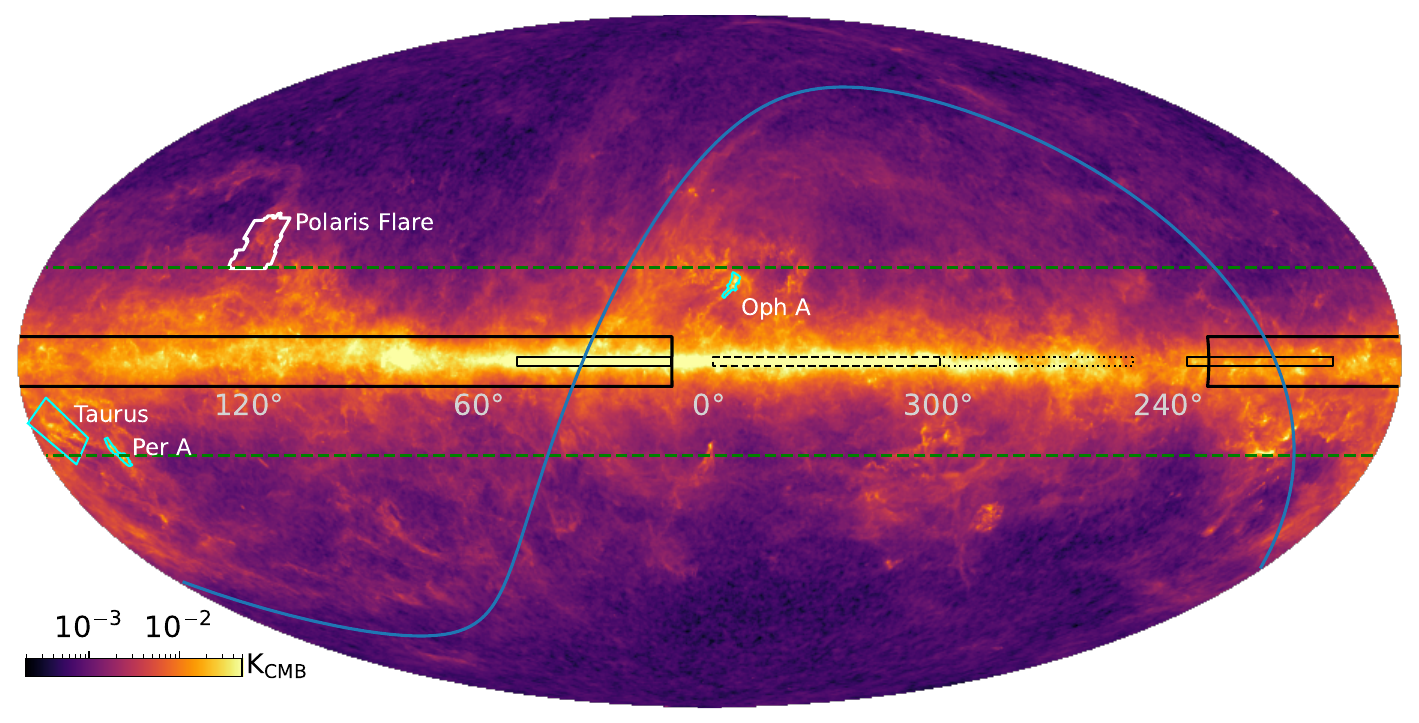}\\
    \includegraphics[width=0.49\linewidth]{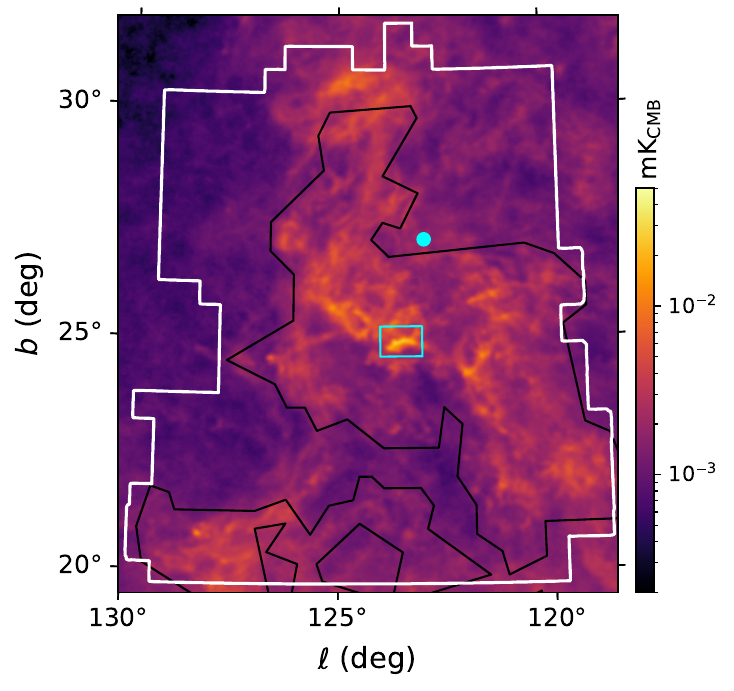}
    \includegraphics[width=0.45\linewidth]{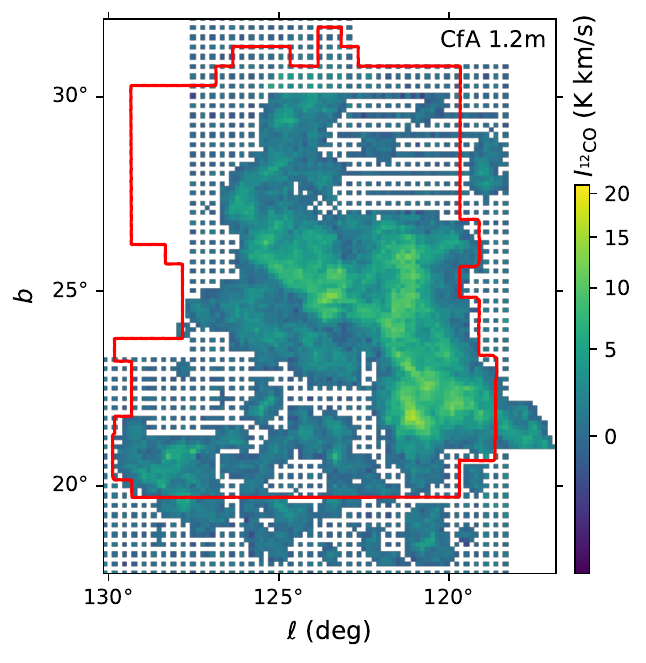}
\caption{Top: All-sky \textit{Planck} 353~GHz continuum image \citep{2014A&A...571A..11P} in a Galactic Mollweide projection. The white boundary encloses the target footprint of the PPCOS (this work). Thick solid, thin solid, and dashed black lines indicate the sky coverages of MWISP, FUGIN, and ThrUMMS, respectively (see Table~\ref{tab:cloud_surveys}). The dotted black line delineates the extended coverage of the Mopra Southern Galactic Plane CO Survey, which is not Nyquist sampled, in comparison with ThrUMMS (see the footnote of Table~\ref{tab:cloud_surveys}). Surveys of nearby clouds (Taurus, Perseus~A, and Ophiuchus~A) conducted by FCRAO (Table~\ref{tab:cloud_surveys}) are shown in cyan. The blue line marks the celestial equator, while the green lines indicate Galactic latitudes of $b = \pm20\degr$.
Bottom left: Zoomed-in view of the 353~GHz continuum across the Polaris Flare. The black contour outlines the approximately half-Nyquist sampled region of the DHT16 survey (detailed in the bottom-right panel), and the white boundary corresponds to that in the top panel. The cyan dot marks the North Celestial Pole. The cyan rectangle, covering roughly 0.6~deg$^2$, highlights the brightest continuum region of the complex, designated as the Polaris cloud \citep{2001A&A...366..636B}.
Bottom right: $^{12}$CO integrated intensity (Moment 0) map from the DHT16 survey \citep{1990ApJ...353L..49H,1993A&A...268..265H}. The contiguous central region features half-Nyquist sampling (with a pixel size comparable to the $\sim$8.5\arcmin\ beam), while the surrounding pixels exhibit sparser coverage. The red boundary encloses the target footprint of PPCOS. Note that the bottom-left panel utilizes a gnomonic (TAN) projection, whereas the bottom-right panel utilizes a plate carrée (CAR) projection.}
    \label{fig:planck}
\end{figure*}

To date, CO observations of the Polaris Flare remain scarce. Existing data are heavily fragmented between low-resolution wide-field maps and highly localized small-scale studies. The most comprehensive wide-field effort was conducted by \citet{1990ApJ...353L..49H} (hereafter HT90\footnote{The HT90 survey was later integrated into the composite Milky Way CO survey \citep{2001ApJ...547..792D} as regional survey No.~16, commonly referred to as DHT16. These data were made publicly available via the CfA Molecular Line Survey website at \url{https://harvard.edu}.}) using the CfA 1.2~m telescope. They mapped the region in $^{12}$CO ($1-0$) with a half-power beam width (HPBW) of 8.7\arcmin.
This survey was executed via a balanced on--off position-switching mode on a regular 7.5\arcmin\ grid in $l$ and $b$. Because this sampling interval is larger than half the HPBW, the dataset is only approximately half-Nyquist sampled, or even coarser. This undersampled region of DHT16 spans an aggregate area of approximately 65~deg$^2$ (Figure~\ref{fig:planck}). Furthermore, its limited spectral resolution of 0.65~km~s$^{-1}$ is insufficient to fully resolve the remarkably narrow CO line profiles characteristic of the Polaris Flare \citep[$\lesssim$0.5~km~s$^{-1}$;][]{2006JKAS...39....9C,2010A&A...518L..92W} (see Figure~\ref{fig:planck}).

Subsequent CO investigations across the Polaris Flare have almost exclusively targeted small, localized regions. These studies focus tightly on the densest pockets of the cloud. For instance, the brightest dust continuum feature is designated as the Polaris cloud and spans a diameter of roughly one degree (Figure~\ref{fig:planck}). This structure was mapped in $^{13}$CO ($1-0$) with the FCRAO 14~m telescope over a spatial footprint of just $\sim$0.6~deg$^2$ \citep{2001A&A...366..636B,2003ApJ...591.1013B}. The densest node within this structure, MCLD~123.5+24.9, harbors several pre-stellar cores \citep{2001A&A...366..636B}. Using the IRAM 30~m telescope, \citet{1998A&A...331..669F} targeted a narrow $5\arcmin \times 7\arcmin$ field within MCLD~123.5+24.9 to map three distinct CO lines. This work was part of the IRAM key project ``Small-scale structure of pre-star forming regions''. More recently, \citet{2025ApJ...981..158S} mapped a 0.5~deg$^2$ portion of the Polaris cloud in $^{12}$CO ($1-0$) using the Nobeyama 45~m telescope.
Consequently, a major observational gap remains. No contiguous, large-area ($>100$~deg$^2$) CO survey of the Polaris Flare has yet been executed at sub-arcminute angular resolution. In fact, this is true for any high-Galactic-latitude cloud complex ($|b| > 20\degr$; see Table~\ref{tab:cloud_surveys}).

\begin{figure*}
    \centering
    \includegraphics[width=0.99\linewidth]{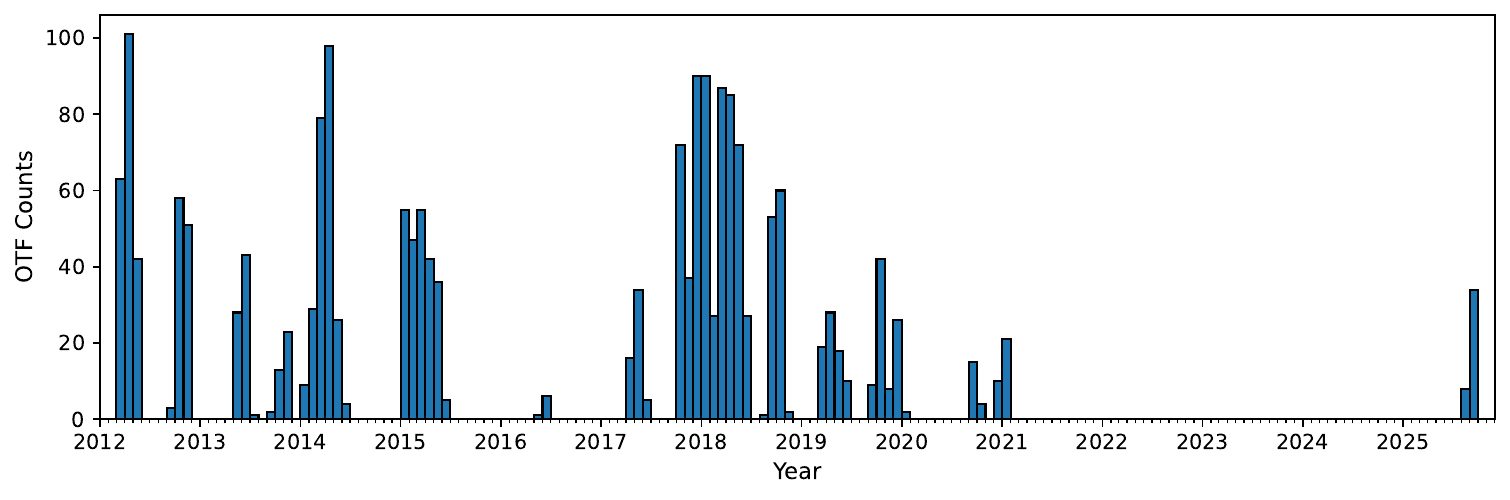}
    \caption{Monthly distribution of the OTF scans for the 424 cells of the PMO Polaris CO survey (Sect.~\ref{sec_obs}). Most observations were conducted between March 2012 and January 2021, with ten additional cells observed in 2025. 
    \label{fig:obsdate}}
\end{figure*}

Here, we present the PMO Polaris CO Survey (hereafter PPCOS), a large-area mapping project of the Polaris Flare in the $J=1-0$ transitions of $^{12}$CO, $^{13}$CO, and C$^{18}$O. These observations were obtained using the Delingha 13.7~m telescope, operated by the Purple Mountain Observatory (PMO). The survey covers approximately 100~deg$^2$ with sub-arcminute angular resolution and simultaneous multi-line coverage. Adopting a distance of $d=150$~pc, where one degree corresponds to 2.62~pc, the mapped footprint spans a total physical area of $\sim$686~pc$^2$.
This unique combination of a wide field of view, high Galactic latitude ($|b|>20\degr$), fine angular resolution, and simultaneous multi-isotopologue data distinguishes PPCOS from previous datasets. In particular, it constitutes the first CO survey of a high-latitude molecular cloud extending over more than 100~deg$^2$ at sub-arcminute resolution (see Table~\ref{tab:cloud_surveys}). These characteristics enable studies that were previously impossible. For example, it enable us to resolve the small-scale turbulent and filamentary structures of the Polaris Flare while simultaneously capturing the global environment of the cloud complex.

This paper is organized as follows. In Sect.~\ref{sec_obs_data}, we describe the survey observations, calibration strategy, and data reduction procedures. In Sect.~\ref{sec_glimpse}, we present a global overview of the CO line emission across the Polaris Flare, focusing on large-scale spatial distributions (Sects.~\ref{sec_codistri} and \ref{sec_C18O}), dynamic patterns (Sect.~\ref{sec_highmoms}), and position--velocity projections (Sect.~\ref{sec_pvandoverall}) revealed by the data cubes.
In the discussion section (Sect.~\ref{sec_discuss}), we highlight an intriguing localized nested structure as a case study (Sect.~\ref{sec_complexD}), examine the globally starless nature of the Polaris Flare (Sect.~\ref{sec_starless}), and outline the broader scientific topics enabled by this wide-field dataset (Sect.~\ref{sec_topics}). Finally, we summarize our primary conclusions in Sect.~\ref{sec_summary}.

\begin{figure}
    \centering
    \includegraphics[width=0.85\linewidth]{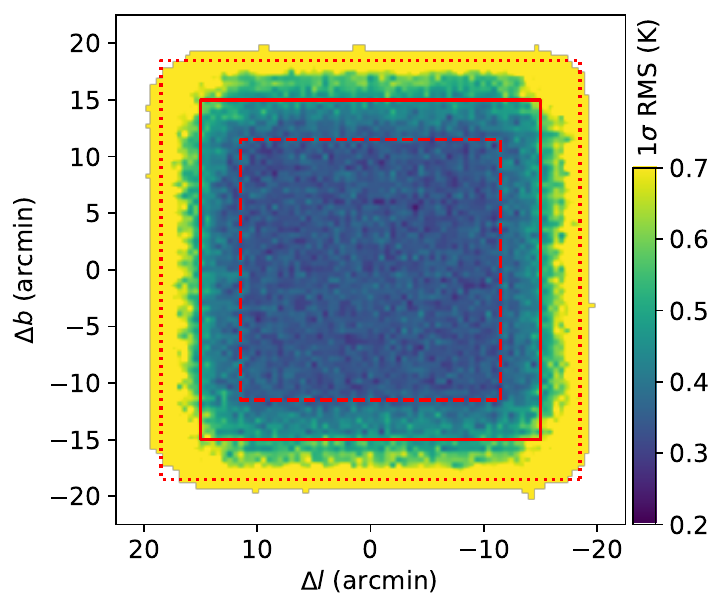}\\
    \includegraphics[width=0.75\linewidth]{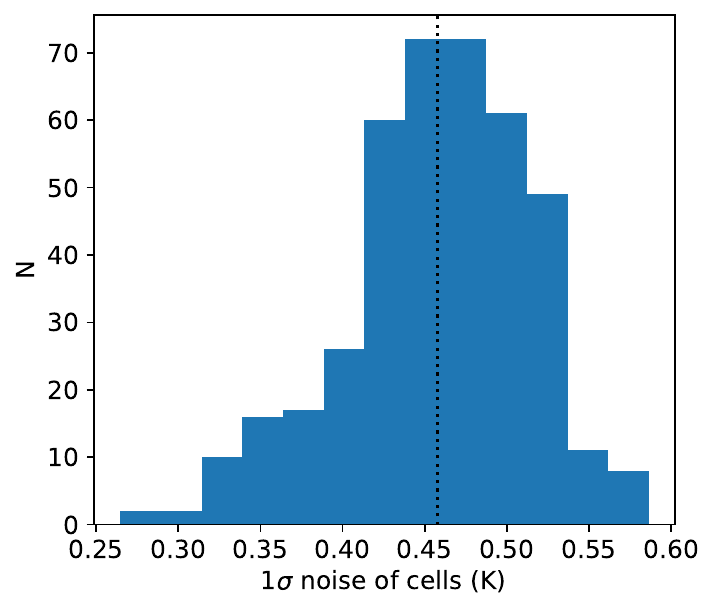}
    \caption{Upper: Example $1\sigma$ noise map of a single cell (centered at $l = 122.5\degr$, $b = 25.0\degr$) for $^{12}$CO (1--0). No smoothing was applied here, in contrast to Figure~\ref{fig:largecubenoise} (see Sect.~\ref{sec_dr}). The red dashed, solid, and dotted boxes mark the innermost $23\arcmin \times 23\arcmin$ region, the nominal $30\arcmin \times 30\arcmin$ cell, and the $37\arcmin \times 37\arcmin$ high-noise margin, respectively. Lower: Distribution of the mean noise across all cells, where the mean for each cell is calculated within its central region.}
    \label{fig:cellnoise}
\end{figure}



\begin{figure*}
    \centering
    \includegraphics[width=0.99\linewidth]{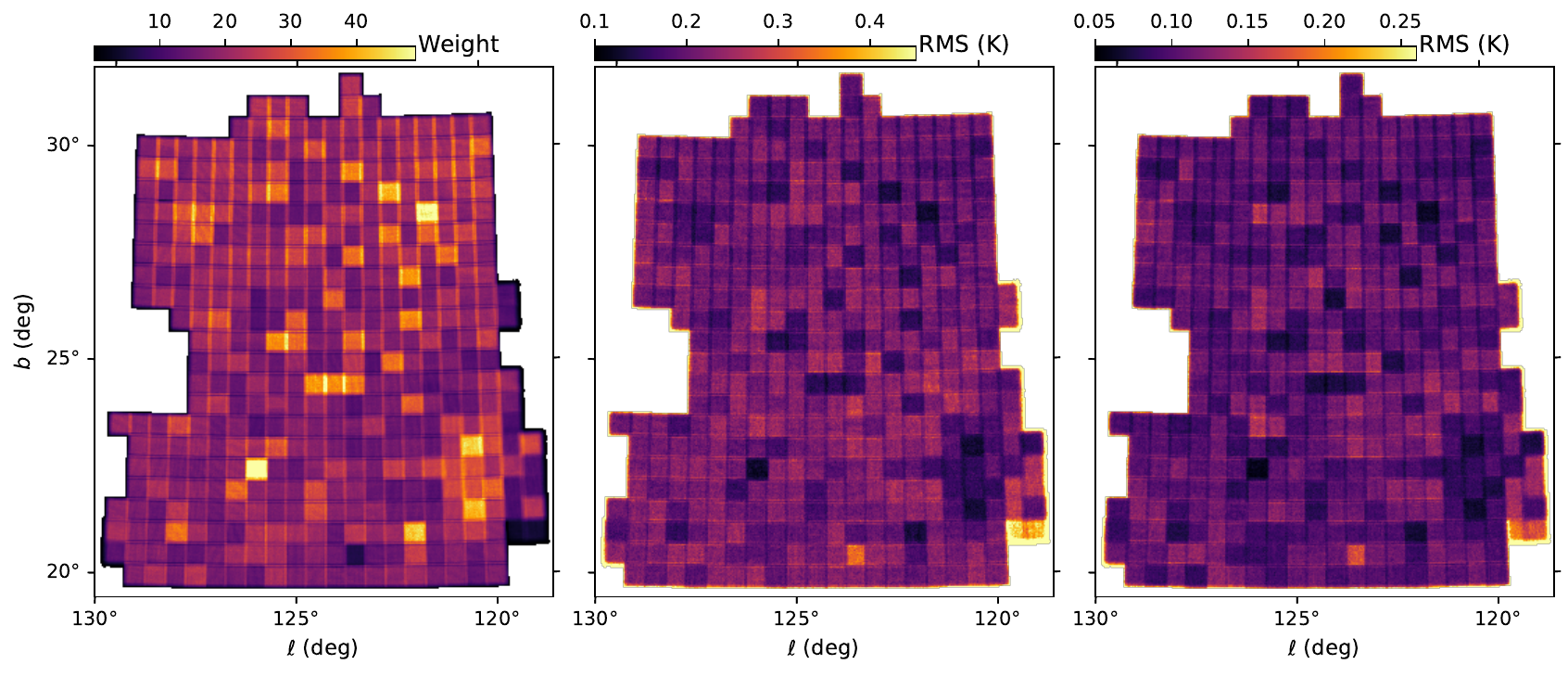}
    \caption{Left: Weight map of $^{12}$CO merged cube. Middle and right: RMS (1$\sigma$) maps of the $^{12}$CO and $^{13}$CO merged cubes, respectively. A $\sigma_{\rm L}$ of $0.4\arcmin$ was adopted to create the merged cubes shown here (Sect. \ref{sec_dr}).
    \label{fig:largecubenoise}
}
\end{figure*}
 
\section{Observations and data reduction}\label{sec_obs_data}
\subsection{Delingha 13.7 m}

The Delingha 13.7\,m telescope, an alt-azimuth instrument, is located near Delingha, the third-largest city in Qinghai Province, western China, at an elevation of approximately 3200\,m (latitude $37^{\circ}22.4',\mathrm{N}$, longitude $97^{\circ}33.6',\mathrm{E}$). The site features dry and stable atmospheric conditions, providing excellent observing conditions at millimeter wavelengths \citep{2016PASP..128j5003T}. Its alt-azimuth design enables efficient On-The-Fly (OTF) mapping of sources near the north celestial pole, such as the Polaris Flare\footnote{For an equatorial mount, the instrument rotation axis is aligned with the pole, making continuous scanning mechanically challenging in this region.}.

The telescope's main reflector has a routinely measured surface accuracy of $\sim 73~\mu$m \citep{2012SPIE.8444E..4BY,2019ApJS..240....9S}. 
Its tracking performance is typically $1''$--$3''$ in both azimuth and elevation, while the overall pointing accuracy across the sky is better than $5''$. 
This corresponds to roughly one-tenth of the HPBW ($\theta_{\rm natal}$) at 115~GHz, which is $\sim 50''$ \citep{2004ChJAA...4..390Z}.

The Delingha 13.7 m telescope is equipped with a $3 \times 3$ multibeam Superconducting Spectroscopic Array Receiver (SSAR), a sideband-separating superconductor–insulator–superconductor (SIS) system installed in 2010 \citep{2012ITTST...2..593S}. Each beam measures only the total intensity (Stokes I); the two orthogonal polarization components (e.g., H/V or L/R) are not separated. The nine beams are arranged with a regular centroid spacing of $\sim172\arcsec$, which is routinely verified before each observing season, enabling efficient fully sampled mapping. Each dual-sideband (2SB) SIS mixer provides a typical receiver noise temperature of $\sim60$ K, with an image-rejection ratio exceeding 10 dB across the \mbox{85–115 GHz} band \citep{2012ITTST...2..593S}. The image rejection ratio for this survey, estimated from the $^{12}$CO line leaked into the lower sideband (image of the upper-sideband line), is $\sim$18 dB. The intermediate-frequency (IF) band, spanning 2.6–4.0 GHz, is processed through an independent IF module and digital control system.

The backend system consists of 18 high-resolution fast Fourier transform spectrometers (FFTSs), enabling simultaneous data recording from 9 beams and 2 sidebands (LSB and USB). Each FFTS provides 16,384 spectral channels and can operate with either a 1 GHz or 200 MHz bandwidth \citep{ZHENQIANG2021559}. For this survey, the 1 GHz mode was adopted with the local oscillator (LO) set at 112.6 GHz. This configuration simultaneously covers the $^{12}$CO, $^{13}$CO, and C$^{18}$O $J=1$–0 lines at 115.271, 110.201, and 109.782 GHz, respectively. In this mode, the channel spacing is approximately 61 kHz, corresponding to a velocity resolution of 0.16–0.17 km s$^{-1}$. 


\subsection{Survey coverage} \label{sec_mapregions}
PPCOS spans approximately $l = 120\degr$ to $130\degr$ and $b = 20\degr$ to $30\degr$, mapping a single contiguous region of 100~deg$^2$ toward the Polaris Flare. Due to its high declination, the Polaris Flare remains observable by the Delingha 13.7~m telescope almost continuously throughout the day. This favorable visibility enabled PPCOS to gather data over a ten-year period from 2012 to 2021 (see Figure~\ref{fig:obsdate}) during the filler time of the high-priority MWISP survey. Specifically, observations were taken when the Galactic plane was below the horizon or otherwise unobservable. The observational strategy and setup were closely matched to those of the MWISP survey \citep{2019ApJS..240....9S}.

PPCOS partitions the target region into 424 individual cells, each with a nominal size of $30\arcmin \times 30\arcmin$ (Figure~\ref{fig:cellnoise}), centered on integer or half-integer values of $l$ and $b$. The primary observations were conducted between 2012 and 2021, during which 414 cells were completed. To minimize gaps and ensure a highly contiguous final map, ten additional cells were observed along the survey boundary in 2025. Because these supplementary cells primarily target edge gaps, the resulting aggregate footprint is not a perfect rectangle; however, it fully encompasses the main body of the legacy DHT16 survey (Figure~\ref{fig:planck}). While these 424 cells represent a total nominal coverage of approximately 106~deg$^2$, spatial overlap between neighboring cells yields a final net mapped area of $\sim$100~deg$^2$ (Figure~\ref{fig:planck}).

\subsection{OTF Observations} \label{sec_obs}
Observations of each cell were conducted using the position-switching on-the-fly (OTF; see \citealt{2018AcASn..59....3S}) scanning mode. The telescope scanned at a rate of 50\arcsec~s$^{-1}$ with a data dump time of 0.3~s, resulting in a sampling interval of 15\arcsec\ along the scan direction. Individual scan rows were separated by 10\arcsec\ in the perpendicular direction, providing excellent oversampling of the 50\arcsec\ telescope beam and ensuring uniform sensitivity across the mapped area. This configuration achieves full Nyquist sampling, in direct contrast to the low-resolution, undersampled strategy of the legacy HT90 survey \citep{1990ApJ...353L..49H}.

At the cell borders, redundant sampling was performed. For example, when scanning a cell, the left edge of the nominal region is also covered by the rightmost beam plus an additional half-beam width, producing an extra $\sim$3.5\arcmin\ noisy margin beyond the nominal boundary (Figure~\ref{fig:cellnoise}). Each $30\arcmin \times 30\arcmin$ cell can be fully mapped within half an hour along either $l$ or $b$. To further reduce noise fluctuations, each cell was scanned along both $l$ and $b$ at least twice. The only exception is several cells at the western margin of the mapped region, which were observed only once and therefore exhibit higher noise (Figure~\ref{fig:largecubenoise}).

Standard calibration sources were observed roughly every two hours to monitor spectral profiles and line intensities, ensuring long-term observational stability. Additional mapping scans were performed when the primary dual-direction scans did not achieve our target rms noise threshold (0.5~K for the upper sideband and 0.3~K for the lower sideband). In total, approximately two thousand individual OTF scans were executed (Figure~\ref{fig:obsdate}), consuming over one thousand hours of telescope time including all system overheads.

Observations were calibrated using the standard ambient temperature chopping wheel method, which alternates measurements between the cold sky and a warm load. The typical system temperatures ($T_{\rm sys}$), including contributions from the receiver, antenna optics, dome membrane, and atmosphere, average $310 \pm 30$~K for $^{12}$CO at the upper sideband (USB) and $165 \pm 15$~K for $^{13}$CO and C$^{18}$O at the lower sideband (LSB). These values are slightly higher than those reported for the main MWISP survey \citep{2019ApJS..240....9S}. This difference is primarily due to the lower elevation angles of the PPCOS observations, which result from the high declination of the Polaris Flare and the lower relative scheduling priority of this filler-time survey.

\begin{figure}[!t]
    \centering
    \includegraphics[width=0.9\linewidth]{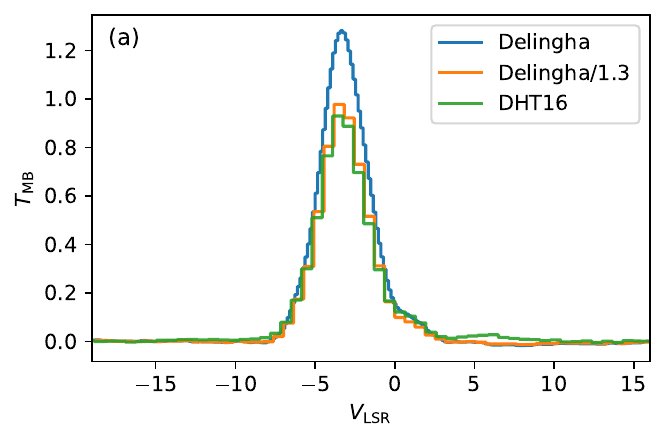}
    \includegraphics[width=0.9\linewidth]{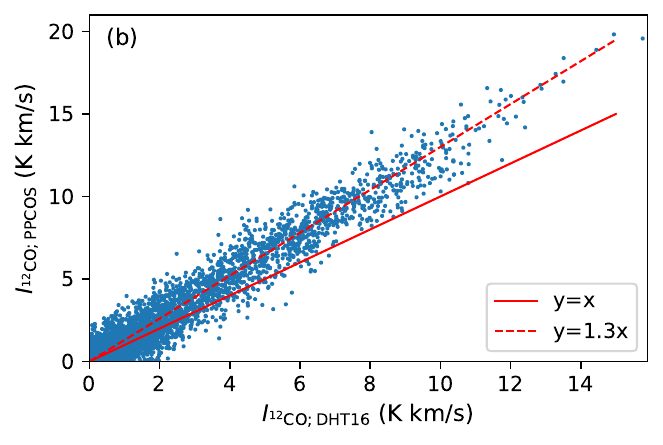}
    \includegraphics[width=0.9\linewidth]{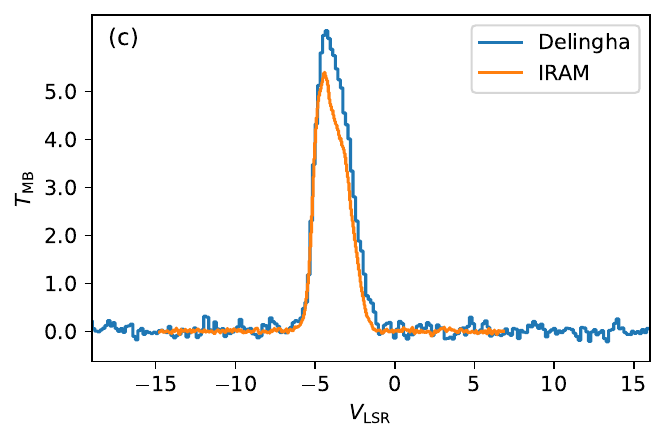}
    \caption{Panel (a): Mean spectra of $^{12}$CO averaged over the overlapping region between PPCOS and DHT16 (see right panel of Figure~\ref{fig:planck}). 
The blue and green lines correspond to this survey and DHT16, respectively. 
The orange line shows the mean spectrum of PPCOS as well, scaled down by a factor of 1.3 and smoothed to the spectral resolution of DHT16.  
Panel (b): Pixel-by-pixel relation between the $^{12}$CO integrated intensities of PPCOS and DHT16. 
PPCOS has been spatially smoothed to match the resolution of DHT16. 
The red solid and dashed lines represent $y=x$ and $y=1.3x$, respectively.  
Panel (c): Mean spectra of $^{12}$CO averaged over the mapped region of the IRAM 30\,m observation \citep{1998A&A...331..669F}, 
covering a region of only about $5\arcmin\times 7\arcmin$.
    }
    \label{fig:avespec}
\end{figure}

\begin{figure}
    \centering
    \includegraphics[width=0.95\linewidth]{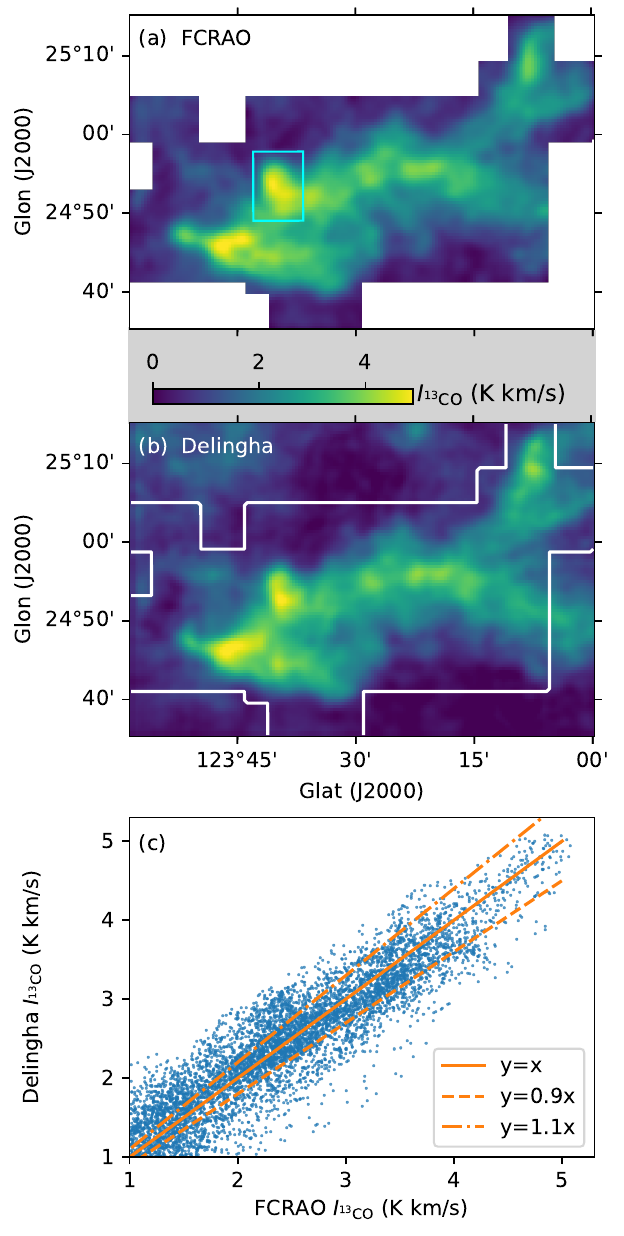}
    \caption{The $^{13}$CO (1–0) integrated intensity maps of Polaris cloud, measured by the FCRAO 14 m telescope \citep[panel a;][]{2001A&A...366..636B} and the Delingha 13.7 m telescope (panel b; this work). Panels (a) and (b) share the same color scale, as indicated by the colorbar between them. 
    Note that MCLD 123.5+24.9 \citep[e.g.,][]{2002A&A...383..591H,2012ApJ...745..195S} is denoted by a cyan box in Panel (a), and the Polaris clouds serve as its host clump. 
    In panel (c), the intensities of the two maps are compared, showing a consistent intensity ratio of approximately unity, as indicated by the orange lines.}
    \label{fig:compare_fcrao}
\end{figure}

\subsection{Data reduction}\label{sec_dr}
\subsubsection{Individual Cells} \label{sec:dr_cells}

Raw data reduction for each individual cell was performed using the GILDAS software package \citep{2005sf2a.conf..721P}\footnote{\url{http://www.iram.fr/IRAMFR/GILDAS}}, with the outputs gridded into data cubes spanning $45.5\arcmin \times 45.5\arcmin$.
Each individual cube comprises three distinct concentric zones (Figure~\ref{fig:cellnoise}): a central, uniformly low-noise core ($23\arcmin \times 23\arcmin$) fully sampled by all nine receiver beams during a single scan; the nominal cell area ($30\arcmin \times 30\arcmin$), where pixel integration times exceed half of the core value; and a peripheral high-noise margin ($37\arcmin \times 37\arcmin$) designed to ensure spatial uniformity across the final merged mosaic (Sect.~\ref{sec:dr_merge}).
The antenna temperature, $T_A^*$, was converted to the main-beam brightness temperature scale via $T_\mathrm{MB} = T_A^*/\eta_\mathrm{MB}$, assuming a beam-filling factor of unity for the extended CO emission.
The main-beam efficiency, $\eta_\mathrm{MB}$, ranged between 40\% and 50\% across the different observing epochs \citep{2019ApJS..240....9S}.

A first-order polynomial baseline was fitted to and subtracted from the spectrum of each pixel after masking the velocity range containing real emission (from $-10$ to 10~km~s$^{-1}$; Sect.~\ref{sec_12co_dist}).
Channels affected by instrumental artifacts or bad performance were flagged and removed, and the root-mean-square (rms) noise was estimated from the remaining emission-free velocity coverage.
The noise properties of each cell were characterized within its central $23\arcmin \times 23\arcmin$ zone.
For $^{12}$CO, the native rms noise ranges from 0.25 to 0.65~K, with a survey-wide mean of 0.46~K.
This spatial variation reflects differences in integration time, atmospheric conditions, target elevation, and system performance during the on-the-fly (OTF) scans.
The native noise levels for $^{13}$CO and C$^{18}$O (recorded simultaneously in the same sideband) are systematically lower, typically averaging $\sim$0.23~K; this is roughly half the $^{12}$CO noise due to a lower system temperature in that sideband.
These baseline sensitivities are comparable to those achieved by the FCRAO survey of Taurus \citep[][see also Table~\ref{tab:cloud_surveys}]{2008ApJS..177..341N}, guaranteeing sufficient depth to map weak and diffuse molecular structures.

\begin{figure*}[!t]
\centering
\includegraphics[width=0.95\linewidth]{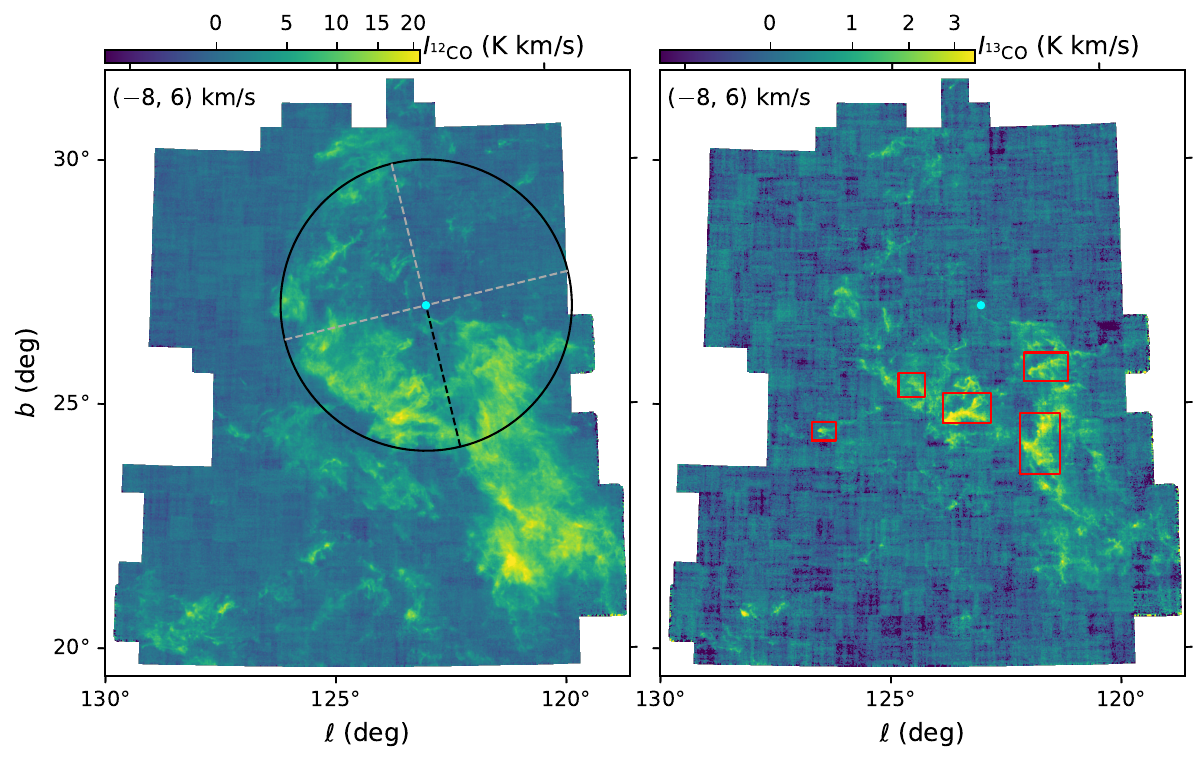}
\caption{Moment-0 maps of $^{12}$CO (left) and $^{13}$CO (right) from the PMO Polaris CO survey, integrated over the velocity range $-8$ to $6~\mathrm{km\,s^{-1}}$.  
Note that these overall moment-0 maps cannot fully display the detailed CO distributions due to image compression from limited figure size, the broad velocity range used for integration, and color maps that are not optimized for individual complexes (see Figure~\ref{fig:Dmom0} for an example).  
The cyan dot in each panel marks the north celestial pole. The black circle indicates the declination $DEC = 87\degr$. The dashed black line denotes $RA = 0\degr$, while the three dashed gray lines denote $RA = 90\degr$, $180\degr$, and $270\degr$; note that $RA$ increases clockwise.  
The red boxes in the right panel indicate the detected emission regions of C$^{18}$O, which are highlighted by the orange boxes in Figure~\ref{fig:mom0_18}.
\label{fig:mom0_12_13}}
\end{figure*}

\subsubsection{Data merging and mosaicking} \label{sec:dr_merge}

Due to the spatial overlaps between adjacent cells, the cumulative integration time in these overlapping regions increases, bringing their noise characteristics closer to those of the cell cores.
To generate a spatially continuous and uniform map, the individual cubes for each of the three CO isotopologues were co-added and reprojected onto a single wide-field mosaic (Figure~\ref{fig:largecubenoise}).

The final grid was generated with a pixel step size of $L_{\rm pix} = 0.5\arcmin$ in both Galactic longitude and latitude, where each input spectrum was assigned a weight defined as
\begin{equation}
W = W_{\rm noise}\, W_{\rm dist},
\end{equation}
where
\begin{equation}
W_{\rm noise} = \frac{1}{\sigma_{\rm noise}^2}, \qquad
W_{\rm dist} = \exp\left(-\frac{\Delta L^2}{2\sigma_{L}^2}\right).
\end{equation}
Here, $\sigma_{\rm noise}$ is the rms noise of the input spectrum, $\Delta L$ represents the angular distance between the original coordinate of the input spectrum and the target pixel in the merged grid, and $\sigma_{L}$ is the standard deviation of the Gaussian kernel used for the resampling.
Specifically, the final mosaic cube $\mathcal{M}$ is interpolated from the individual cell cubes (denoted as $\mathcal{C}$) according to
\begin{equation}
    \mathcal{M}_i = \frac{\sum_{j,k} W_{i,j,k} C_{j,k}}{\sum_{j,k} W_{i,j,k}},
\end{equation}
where $i$ is the pixel index of the merged cube, $j$ is the cell index, and $k$ is the pixel index within that specific cell.

This image-mosaicking process was implemented using the \texttt{Astropy} Python package \citep{2022ApJ...935..167A}.
The resulting mosaic, $\mathcal{M}$, is projected using a gnomonic (TAN) geometry centered at $l = 124.75\degr$, $b = 25.75\degr$.
In this work, we adopt a kernel width of $\sigma_{L} = 0.4\arcmin$, which yields an effective angular resolution for the final dataset of
\begin{equation}
    \Theta_{\rm merge} = \sqrt{\Theta_{\rm native}^2 + 8\ln(2)\,\sigma_L^2 +  L_{\rm pix}^2} \sim 1.5\arcmin.
\end{equation}
The rms noise levels of the final mosaic, evaluated at the native spectral resolution following the method in Sect.~\ref{sec:dr_cells}, are $0.21 \pm 0.04$~K for $^{12}$CO and $0.11 \pm 0.03$~K for $^{13}$CO and C$^{18}$O (see Table~\ref{tab:cloud_surveys}).

\begin{figure*}
    \centering
    \includegraphics[width=0.52\linewidth]{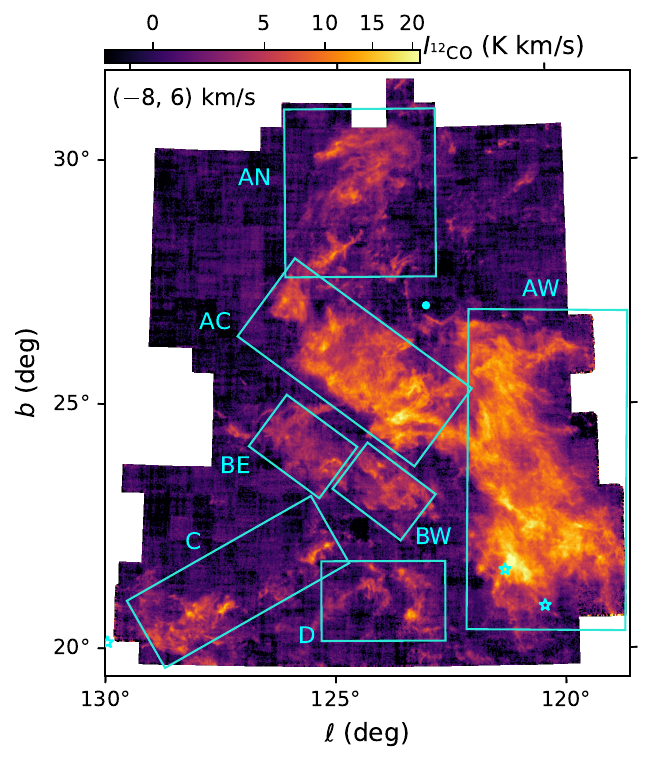}
    \includegraphics[width=0.37\linewidth]{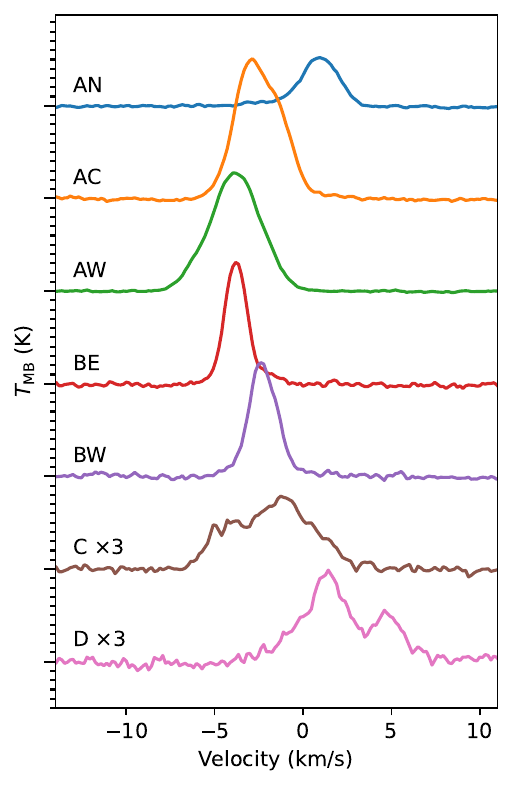}
    \caption{Left: Subregions (also referred to as complexes) of the Polaris Flare, enclosed by cyan boxes. 
    The corresponding labels are indicated next to each box. The background shows the moment~0 map of $^{12}$CO, identical to the left panel of Figure~\ref{fig:mom0_12_13}, but displayed with a different colormap. 
Two WISE Class I/II young stellar object candidates \citep{2016MNRAS.458.3479M} are indicated by cyan pentagrams, with the northeastern and southwestern sources denoted as YSO1 and YSO2, respectively (Sect. \ref{sec_starless}).
Right: Mean $^{12}$CO spectra of the different complexes of the Polaris Flare. The major and minor ticks on the y-axis are separated by 1~K and 0.1~K, respectively. The spectra for complexes C and D have been multiplied by a factor of three.}
    \label{fig:subpart}
\end{figure*}

\subsection{Flux check} \label{sec:flux_check}

To evaluate the reliability of our calibration, we compared our data with previous surveys. These older datasets were obtained using different telescopes, apertures, calibration strategies, and observing epochs (Table~\ref{tab:cloud_surveys}). Panel (a) of Figure~\ref{fig:avespec} presents the mean $^{12}$CO spectra from PPCOS and DHT16, averaged over their spatial intersection. This intersection corresponds to the overlapping area enclosed by the black and white contours in the lower-left panel of Figure~\ref{fig:planck}. The global profile shapes exhibit excellent agreement. A Gaussian fit to our averaged profile yields a systemic velocity of $-3.2$~km~s$^{-1}$ and a global line width (FWHM) of 3.5~km~s$^{-1}$, which is fully consistent with the DHT16 values.

However, the line intensities from our survey are systematically higher than those of DHT16 by approximately 30\%. This systematic discrepancy is also evident in a pixel-by-pixel comparison of the integrated intensities (panel b of Figure~\ref{fig:avespec}). For this comparison, the PPCOS data were smoothed to an 8.5\arcmin\ beam to match the angular resolution of DHT16. Differences in sensitivity likely explain this variation. For instance, \citet{2021ApJS..256...32S} demonstrated that the $^{12}$CO flux recovered by the MWISP survey in the Galactic plane can exceed the CfA 1.2~m observations by a factor of up to 1.6. This trend is driven by the ability of deeper surveys to recover faint, extended emission.


\begin{table}
\centering
\caption{Basic parameters of the subregions of Polaris Flare$^{(1)}$.}
\begin{tabular}{c c c c c c}
\hline
Subregion & $l$ & $b$ & Size & PA & Velocity$^{(2)}$ \\
          & $\degr$ & $\degr$ & $\degr$ & $\degr$ & km/s \\
\hline
AN & 124.5 & 29.4 & $3.1 \times 3.5$ & 0   & 0.95 \\
AC & 124.6 & 26.0 & $4.5 \times 2.0$ & 145 & –2.54 \\
AW & 120.3 & 23.7 & $3.3 \times 6.6$ & 0 & –3.83 \\
BE & 125.8 & 24.2 & $1.8 \times 1.3$ & 145 & –3.79 \\
BW & 123.9 & 23.3 & $1.2 \times 1.8$ & 50  & –2.27 \\
C  & 127.2 & 21.4 & $4.4 \times 1.6$ & 30  & –1.7 \\
D  & 124.0 & 21.0 & $2.6 \times 1.6$ & 0   & 1.3; 4.9 \\
\hline
\end{tabular}
\label{tab:subregions}
\vspace{0.25em}
\footnotesize{
$^{(1)}$ The central locations ($l$, $b$), sizes, and position angles ($PA$) correspond to the cyan boxes shown in Figure~\ref{fig:subpart}. The $PA$ is measured anticlockwise from west.
$^{(2)}$ Line centers are derived from Gaussian fitting. For complex D, a double-peak Gaussian fit is applied (Figure~\ref{fig:subpart}). The number of decimal places varies to indicate the precision of each Gaussian fit.}
\end{table}

Panel (c) of Figure~\ref{fig:avespec} compares the mean $^{12}$CO spectra obtained with the Delingha 13.7~m telescope (this work) and the IRAM 30~m telescope toward MCLD~123.5+24.9 \citep{1998A&A...331..669F}, averaged over a $5\arcmin \times 7\arcmin$ field.
The line profiles exhibit close agreement, with our measured intensities exceeding those of IRAM by only $\sim$10\%.
This minor discrepancy is well within the expected variations arising from telescope-specific calibration schemes and is likely explained by the error-beam correction applied to the legacy IRAM dataset \citep{2001A&A...365..285B}, which accounts for forward-hemisphere sidelobe flux.
In contrast, other single-dish datasets—including those from the Delingha 13.7~m, FCRAO 14~m, and CfA 1.2~m telescopes—did not undergo equivalent detailed error-beam decompositions.

In Figure~\ref{fig:compare_fcrao}, we compare the $^{13}$CO ($1-0$) integrated intensities toward the Polaris cloud observed with the FCRAO 14~m telescope \citep{2001A&A...366..636B,2003ApJ...591.1013B} and the Delingha 13.7~m telescope (this work; see Sect.~\ref{sec_codistri} for the large-scale distribution of the Polaris Flare).
Both datasets were convolved to a common angular resolution of $2\arcmin$.
The resulting gas morphologies and line intensities show excellent agreement, with minor small-scale discrepancies likely arising from beam sidelobe structures, scanning strategies, or gridding artifacts.
Although these datasets were acquired more than a decade apart, significant structural or kinematic evolution of the cloud is highly unlikely at the spatial resolutions probed by single-dish observations \citep[e.g.,][]{2018MNRAS.476.3688J}, validating the temporal reliability of this cross-calibration.
A pixel-by-pixel linear regression analysis (panel c of Figure~\ref{fig:compare_fcrao}) yields an intensity ratio averaging near unity with a scatter of $\sim$10\%, demonstrating global consistency and point-to-point fidelity.
This result aligns with previous benchmarking between the MWISP survey (conducted with the Delingha 13.7~m telescope) and the GRS survey (conducted with the FCRAO 14~m telescope), which similarly demonstrated high consistency in $^{13}$CO intensities \citep{2021ApJS..256...32S}.
Because the FCRAO 14~m and Delingha 13.7~m telescopes possess comparable apertures and operate under similar calibration frameworks, their close agreement provides a direct validation of our flux calibration.
Taken together with the IRAM 30~m comparisons, these cross-checks across multiple independent facilities demonstrate that both the velocity and intensity scales of our survey are robust.
Consequently, we adopt a conservative systematic intensity uncertainty of $\sim$10\% to encompass calibration offsets.


\section{A glimpse of the Polaris Flare} \label{sec_glimpse}

In this section, we take a first look at the PPCOS data cubes, offering an unprecedented 100~deg$^2$ view of the molecular gas within the Polaris Flare, one of the most prominent high-latitude molecular clouds in the sky. In Sect.~\ref{sec_codistri}, we examine the spatial distributions of $^{12}$CO and $^{13}$CO emission, revealing the intricate large-scale morphology and structural variations of the diffuse-to-dense molecular gas.
Sect.~\ref{sec_C18O} focuses on the highly localized C$^{18}$O emission, tracing the densest clumps and embedded compact cores within the cloud matrix. Sect.~\ref{sec_highmoms} presents the higher-order moment maps of $^{12}$CO, characterizing the occurrence of multiple velocity components, centroid velocities, line widths, and peak radiation temperatures to provide insight into the internal cloud dynamics. Finally, in Sect.~\ref{sec_pvandoverall}, we inspect the $^{12}$CO data cube in the position--velocity plane and through three-dimensional rendering, offering an assessment of the complex kinematic structure and its variations across different spatial scales.

\subsection{Distribution of $^{12}$CO and $^{13}$CO}\label{sec_codistri}
\subsubsection{$^{12}$CO and $^{13}$CO moment-0 maps}\label{sec_12co_dist}
Almost all of the CO flux originating from the Polaris Flare within our survey footprint is confined to the velocity interval between $-8$ and $+6$~km~s$^{-1}$ (Figure~\ref{fig:avespec}).
The $^{12}$CO and $^{13}$CO integrated intensity (Moment 0) maps generated over this velocity range are displayed in Figure~\ref{fig:mom0_12_13}.
In contrast, the C$^{18}$O emission is exceptionally weak and typically exhibits narrow line widths comparable to the native spectrometer channel width.
Direct integration over a wide velocity interval would drastically degrade the signal-to-noise ratio, obscuring most of the faint C$^{18}$O signal; a carefully optimized Moment 0 map designed to maximize sensitivity while preserving real emission is presented and discussed in Sect.~\ref{sec_C18O}.

The spatial distribution of the $^{12}$CO emission reveals a striking morphology that broadly traces the features of the \textit{Planck} 353~GHz dust continuum map (Figure~\ref{fig:planck}), underscoring the intimate coupling between molecular gas and dust at high Galactic latitudes.
Nevertheless, this comparison also highlights several intriguing structural discrepancies.
On large scales, certain diffuse features apparent in the dust map lack corresponding $^{12}$CO emission, a trend likely attributable to local sensitivity limitations, or the presence of diffuse atomic components and CO-dark molecular gas \citep[e.g.,][]{2010ApJ...716.1191W}.
This issue will be comprehensively addressed in an upcoming study utilizing linear decomposition to assign the dust optical depth to distinct atomic and molecular phases (Liu et al., in prep).
On smaller scales, narrow, elongated structures that are entirely unresolved by \textit{Planck} stand out clearly in our $^{12}$CO data, such as the prominent, slender filamentary feature surrounding the North Celestial Pole (left panel of Figure~\ref{fig:mom0_12_13}).

Earlier legacy surveys, such as DHT16 \citep{1990ApJ...353L..49H,1993A&A...268..265H}, primarily captured clumpy, macroscopic CO structures due to their coarse 8.5\arcmin\ angular resolution, with the spatial Nyquist frequency effectively limited to scales of $\sim$15\arcmin\ by the coarse half-Nyquist grid.
In contrast, our sub-arcminute resolution mapping resolves a wealth of complex substructures, including narrow ripples, striations, filaments, and fine fibers, that were previously blended or entirely unresolved.
The ubiquity of these features demonstrates that even highly quiescent, high-latitude molecular clouds like the Polaris Flare possess an intricate filamentary network.
These morphological properties are likely governed by a combination of magnetohydrodynamic turbulence and ambient magnetic fields \citep{2023ASPC..534..193P,2025arXiv250220458L,2025NatAs.tmp..148L,2025RAA....25b5020L}, forces that may also dominate the large-scale envelope dynamics of the complex (Sect.~\ref{sec_lvbv}).

The $^{13}$CO Moment 0 map maps a significantly more compact and dense component of the cloud, effectively delineating the high-density "skeleton" that forms the structural backbone of the Polaris Flare.
Given that this high-latitude environment is remarkably quiescent and entirely devoid of active star formation \citep{2010A&A...518L.104M}, these data offer a pristine, unperturbed view of how dense gas is structurally organized prior to the onset of stellar feedback.

\begin{figure}
    \centering
    \includegraphics[width=0.99\linewidth]{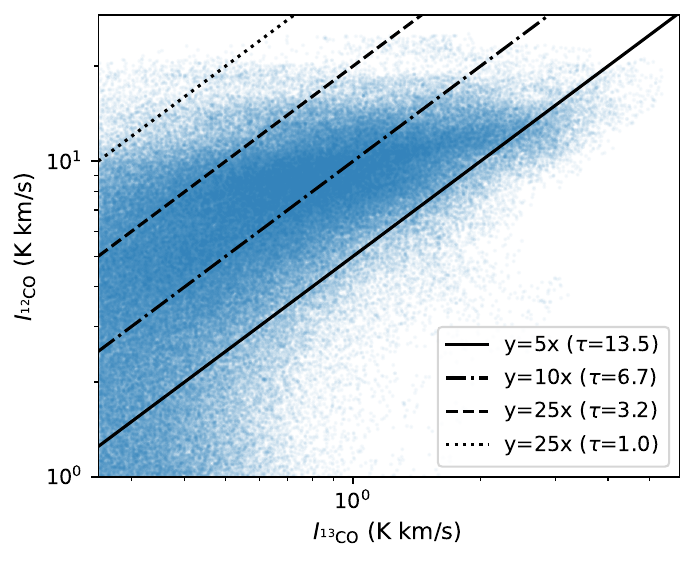}
    \caption{Distribution of $I_{\rm ^{12}CO}$ and $I_{\rm ^{13}CO}$
in the $^{13}$CO-emitting region (Sect.~\ref{sec_codistri}).
The black lines denote different intensity ratios, with their values and the corresponding optical depths ($\tau$) of $^{12}$CO
calculated using Eq.~\ref{eq:tau12}  labeled in the legend.
Note that the optical depth is obtained under the uniform assumption (Sect.~\ref{sec_codistri}). }
    \label{fig:12_13COratio}
\end{figure}

\begin{figure*}[!t]
    \centering
    \includegraphics[width=0.99\linewidth]{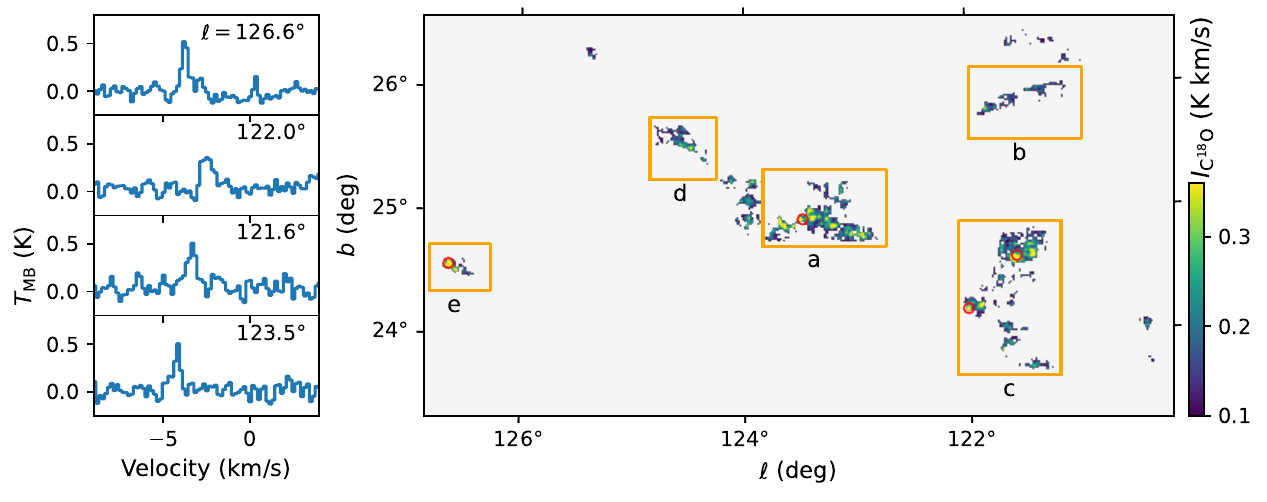}
    \caption{Left: Example C$^{18}$O (1--0) spectra at the positions marked by small red circles in the right panel. The approximate longitude of each spectrum is indicated in the upper right corner of each panel. 
Right: Moment-0 map of C$^{18}$O (1--0), showing only the region with detectable emission. Integrated intensities below 0.1~K~km~s$^{-1}$ are blanked. The emission region is divided into five parts, enclosed by orange boxes and labeled a--e (Sect.~\ref{sec_C18O}). See Fig.~\ref{fig:mom0_12_13} for the locations of these C$^{18}$O emission regions within the Polaris Flare, and Fig.~\ref{fig:regiona_c18o} for a zoom-in view of region a (Polaris cloud).}
    \label{fig:mom0_18}
\end{figure*}

\subsubsection{Subregions}
The combination of wide-field coverage, high angular resolution, and kinematic information (Sect.~\ref{sec_globalgradient}; Figure~\ref{fig:subpart}) allows us to partition the Polaris Flare into several discrete subregions, hereafter referred to as complexes. We identify four primary complexes within the CO emission footprint, designated A through D.
Complex~A corresponds to the central and most prominent structure of the Polaris Flare, harboring the bulk of the total molecular gas mass. Complex~B extends to the southeast of Complex~A and represents the next most substantial concentration of gas. Complexes~C and D trace supplementary molecular structures in the southern portion of the Flare, contributing to the globally extended and filamentary morphology of the cloud.

Complex~A can be further subdivided into three distinct components based on its physical and kinematic properties. The northern component (AN) is characterized by slightly redshifted velocities of approximately $\sim$1~km~s$^{-1}$.
The central component (AC) hosts the well-known Polaris cloud, which contains the majority of the dense molecular material in the region. The western component (AW) extends toward the southwest, bridging toward the Cepheus Flare \citep{2001ApJ...547..792D} and representing a more diffuse extension of the cloud. Complex~B is similarly divided into an eastern component (BE) and a western component (BW), which are segregated by a prominent spatial discontinuity in the CO Moment~0 map and a velocity offset of approximately 1.5~km~s$^{-1}$. These structural subdivisions emphasize the highly fragmented and non-uniform nature of the molecular gas in the Polaris Flare.

To characterize the kinematics of these subregions quantitatively, we present the area-averaged $^{12}$CO ($1-0$) spectra for all seven components in the right panel of Figure~\ref{fig:subpart}. In general, the line profiles exhibit a single dominant peak, and the velocity centroids derived from Gaussian fitting are compiled in Table~\ref{tab:subregions}.
A notable exception is Complex~D, which displays two clearly resolved velocity components, indicative of overlapping or interacting gas layers along the line of sight. We discuss these distinct kinematic features of Complex~D in detail in Sect.~\ref{sec_complexD}.

This systematic partitioning of the Polaris Flare into specific complexes and subcomponents provides a robust framework for analyzing the spatial and kinematic architecture of the molecular gas.
Specifically, it facilitates rigorous cross-comparisons between local filamentary alignments, large-scale velocity gradients, and localized dynamical processes across the cloud complex (e.g., Sects.~\ref{sec_highmoms} and \ref{sec_lvbv}).

\subsubsection{Intensity ratios of $^{12}$CO/$^{13}$CO} \label{sec_Iratio}

In the solar neighborhood, the abundance ratio of $^{12}$CO to $^{13}$CO, denoted as $X_{12;13}$, is generally assumed to follow the elemental $^{12}$C/$^{13}$C isotopic ratio, with a representative value of $\sim$68 \citep{2005ApJ...634.1126M}. If both isotopologues were optically thin and co-spatial, the observed integrated intensity ratio $I_{\rm ^{12}CO}/I_{\rm ^{13}CO}$ would asymptotically approach $X_{12;13}$. Deviations from this value therefore carry important information regarding optical depth, excitation conditions, and the spatial distribution of the emitting gas.

Figure~\ref{fig:12_13COratio} shows the distribution of $^{12}$CO ($I_{\rm ^{12}CO}$) and $^{13}$CO ($I_{\rm ^{13}CO}$) integrated intensities, restricted to positions with significant $^{13}$CO detections ($I_{\rm ^{13}CO} > 0.25$~K~km~s$^{-1}$). The measured intensity ratios fall within the range of 5--25.
These values are much lower than the intrinsic isotopic abundance ratio, which requires that the $^{12}$CO ($1-0$) transition is optically thick over a substantial fraction of the beam area. Ratios of this order are consistent with those found in giant molecular clouds and diffuse cirrus clouds \citep{1979ApJ...232L..89S,1995A&A...301..873S}.
However, these ratios are higher than the very low values ($\sim$2) reported for \textit{Planck} Galactic Cold Clumps \citep[PGCCs;][]{2012ApJ...756...76W,2016A&A...594A..28P}, which are considered to represent some of the coldest and least evolved molecular environments \citep[e.g.,][]{2019A&A...622A..32L}. This comparison suggests that the Polaris Flare occupies a transitional regime: it is still chemically young, but already contains material sufficiently dense and shielded for $^{12}$CO to become optically thick, preventing the ratio from reaching the optically thin isotopic limit.

To quantify the impact of optical depth, we adopt a simplified framework in which both $^{12}$CO and $^{13}$CO emissions are assumed to be uniformly distributed within the telescope beam. Under this uniform assumption, the optical depth of the $^{12}$CO ($1-0$) transition can be expressed as\footnote{This relation follows from inverting the standard radiative transfer expression $\tau / (1 - e^{-\tau}) = 1/A$, where $\tau$ represents the line-center optical depth assuming a Gaussian profile.}
\begin{equation}
\tau_{12} = \frac{1 + A W_{0}\!\left(-\dfrac{1}{A} e^{-1/A}\right)}{A}, 
\qquad 0 < A < 1,
\label{eq:tau12}
\end{equation}
where $W_{0}$ denotes the principal branch of the Lambert $W$ function\footnote{We evaluate $W$ using the {\tt scipy.special.lambertw} implementation \citep{2020NatMe..17..261V}.}. The dimensionless parameter $A$ is defined as
\begin{equation}
A = \frac{I_{\rm ^{12}CO}}{I_{\rm ^{13}CO} \, X_{12;13}}.
\end{equation}
For our observed integrated intensity ratios of 5--25, we find $\tau_{12} \gtrsim 1$ (Figure~\ref{fig:12_13COratio}), confirming that the $^{12}$CO emission is heavily saturated while $^{13}$CO remains predominantly optically thin.

This behavior can be explained by two physical scenarios that can occur simultaneously. First, the $^{13}$CO emission may originate from compact, unresolved substructures embedded within a more widespread $^{12}$CO envelope where the main isotopologue has already saturated, meaning that different area filling factors can significantly depress the observed beam-averaged $I_{\rm ^{12}CO}/I_{\rm ^{13}CO}$ ratio \citep[e.g.,][]{2002A&A...391..295O}. Second, the $^{12}$CO emission may be genuinely optically thick on beam-averaged scales, but its low peak radiation temperatures ($\lesssim$10~K; Sect.~\ref{sec_peakintensity}) indicate subthermal excitation, implying that the local gas density is below the critical density required to fully thermalize the transition to the kinetic temperature of the cloud. Both mechanisms are highly compatible with a dynamically young molecular cloud environment, where ultraviolet shielding is sufficient to facilitate a high abundance of $^{12}$CO, yet local densities remain too low to efficiently thermalize the lower rotational states.

Additional chemical and radiative transfer effects may further influence the observed ratios. Selective photodissociation can reduce the abundance of $^{13}$CO in the outer, more weakly shielded layers of molecular gas, since the rarer isotopologue lacks the self-shielding capability of $^{12}$CO and is more easily destroyed by ambient ultraviolet photons \citep{2009A&A...503..323V}. This process would elevate the observed $I_{\rm ^{12}CO}/I_{\rm ^{13}CO}$ ratio above the expected baseline. Conversely, in cold gas environments, isotope-exchange reactions can enhance $^{13}$CO abundances through chemical fractionation \citep{1980ApJ...235L..39L}, thereby depressing the ratio. The delicate balance between these competing astrochemical processes depends on the local column density, shielding geometry, and radiation field intensity, which likely drives the substantial scatter in the $I_{\rm ^{12}CO}/I_{\rm ^{13}CO}$ distribution shown in Figure~\ref{fig:12_13COratio}.

Taken together, the moderate $^{12}$CO/$^{13}$CO ratios, widespread $^{12}$CO optical thickness, low radiation temperatures, and the competing signatures of selective photodissociation and chemical fractionation demonstrate that the Polaris Flare occupies an intermediate developmental regime. While its structural properties resemble those of giant molecular clouds on large scales, the subthermal excitation conditions and relatively faint $^{13}$CO emission indicate that the gas has not yet reached the dense, fully thermalized states typical of mature GMC interiors. These characteristics firmly establish the Polaris Flare as a young molecular cloud complex that is transitional between diffuse atomic-dominated cirrus and massive, evolved star-forming environments.

\begin{figure}[!t]
    \centering
    \includegraphics[width=0.99\linewidth]{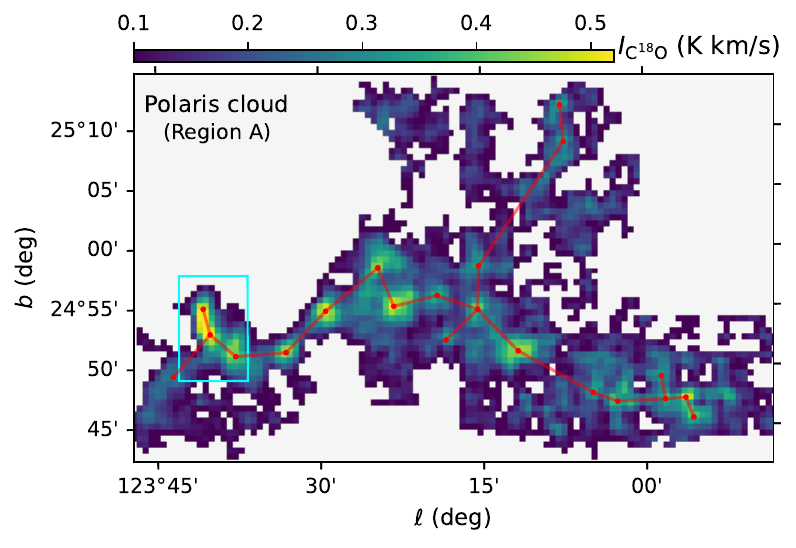}
    \caption{Zoom-in of the C$^{18}$O moment 0 map of Polaris cloud (region $a$ in Figure~\ref{fig:mom0_18}). Red lines indicate the minimal spanning tree (MST) constructed from C$^{18}$O spots (prestellar core candidates; Sect.~\ref{sec_C18O}), which are marked by small red dots.
    The cyan box denotes MCLD 123.5+24.9, the densest part of the Polaris cloud.}
    \label{fig:regiona_c18o}
\end{figure}

\subsection{C$^{18}$O emission}\label{sec_C18O}
\subsubsection{Moment 0 map}
The emission of C$^{18}$O is generally weak and only marginally detected in a few localized regions, reflecting its lower abundance and column density compared to $^{12}$CO and $^{13}$CO. A straightforward integration over the wide velocity intervals commonly used for $^{12}$CO and $^{13}$CO would produce a Moment~0 map dominated by noise and obscure the faint C$^{18}$O features. To extract reliable emission, we first inspected spectra in regions of relatively strong $^{13}$CO emission (Fig.~\ref{fig:mom0_12_13}) and identified positions with detectable C$^{18}$O, which serve as anchor points for the subsequent analysis (examples are shown in the left panels of Fig.~\ref{fig:mom0_18}). The C$^{18}$O ($1-0$) lines are typically narrow and are heavily blended or under-resolved at our current velocity resolution. For each pixel in the data cube, we therefore adopted a narrow integration window of $\pm$0.5~km~s$^{-1}$ centered on the velocity of its nearest anchor point, effectively isolating genuine line emission from surrounding baseline noise. This tailored masking approach recovers the faintest detectable C$^{18}$O structures while suppressing noise fluctuations, resulting in the optimized Moment~0 map shown in the right panel of Fig.~\ref{fig:mom0_18}.

The resulting distribution of C$^{18}$O is far more localized than that of $^{13}$CO, appearing only in compact, isolated pockets rather than forming extended filamentary networks. The emission is primarily concentrated at the junctions, crests, and densest nodes of the $^{13}$CO filaments, tracing the highest-density substructures within the cloud. In contrast to the relatively widespread and continuous $^{13}$CO emission, C$^{18}$O selectively highlights only those dense peaks where the column density and excitation conditions are sufficient to overcome our detection threshold. This stratified emission pattern provides a complementary perspective on the molecular cloud architecture: while $^{13}$CO outlines the extended filamentary framework, C$^{18}$O emphasizes the densest embedded condensations. Together, these tracers reveal the hierarchical structure of the cloud, in which compact, high-density regions are embedded within a more diffuse molecular envelope traced by the more abundant CO isotopologues.

\begin{table}
\centering
\caption{Parameters of C$^{18}$O spots in the Polaris cloud (Figure \ref{fig:regiona_c18o}). \label{tab:C180}}
\begin{tabular}{ccccc}
\hline
dex & l & b & diameter & peak$^{(1)}$ \\
 & $\degr$ & $\degr$ & $\arcmin$ & K km s$^{-1}$ \\
\hline
1 & 123.6814 & 24.9231 & 2.5 & 0.55 \\
2 & 123.6716 & 24.8855 & 2.3 & 0.63 \\
3 & 123.7237 & 24.8273 & 2.5 & 0.28 \\
4 & 123.6341 & 24.8526 & 2.5 & 0.46 \\
5 & 123.5521 & 24.8570 & 2.8 & 0.45 \\
6 & 123.4917 & 24.9108 & 2.9 & 0.50 \\
7 & 123.4127 & 24.9761 & 2.6 & 0.45 \\
8 & 123.3875 & 24.9199 & 2.8 & 0.51 \\
9 & 123.3201 & 24.9327 & 2.4 & 0.35 \\
10 & 123.2569 & 24.9772 & 2.5 & 0.32 \\
11 & 123.2595 & 24.9170 & 2.8 & 0.42 \\
12 & 123.3024 & 24.8740 & 2.0 & 0.34 \\
13 & 123.2013 & 24.8588 & 3.1 & 0.46 \\
14 & 123.0784 & 24.7992 & 1.9 & 0.35 \\
15 & 123.0452 & 24.7871 & 2.2 & 0.39 \\
16 & 122.9779 & 24.8225 & 2.0 & 0.32 \\
17 & 122.9708 & 24.7851 & 2.0 & 0.40 \\
18 & 122.9386 & 24.7905 & 2.1 & 0.50 \\
19 & 122.9249 & 24.7601 & 2.2 & 0.41 \\
20 & 123.1219 & 25.1449 & 2.2 & 0.32 \\
21 & 123.1269 & 25.1986 & 2.5 & 0.33 \\
\hline
\end{tabular}\\
\vspace{0.5em}
\footnotesize{
$^{(1)}$ The uncertainty of the peak integrated intensity of C$^{18}$O is estimated to be 0.05 K km s$^{-1}$. 
}
\end{table}

\begin{figure*}[!t]
    \centering
    \includegraphics[width=0.9\linewidth]{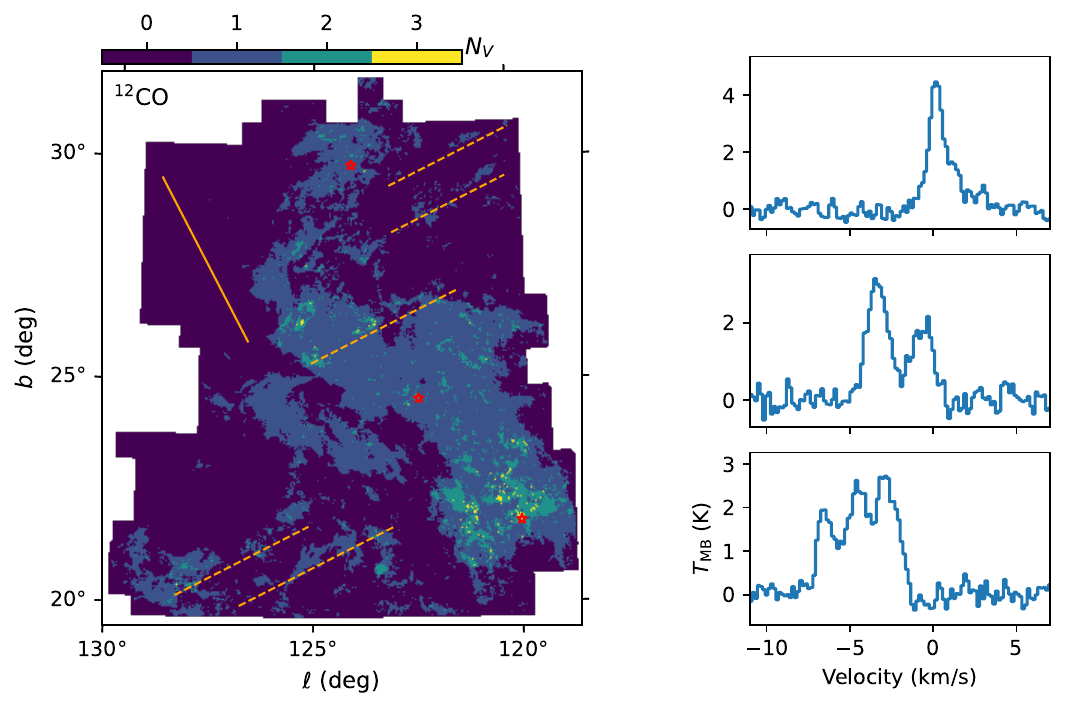}
    \caption{Left: Distribution of the number of $^{12}$CO velocity components ($N_{V}$) identified in each pixel (Sect.~\ref{sec_multivelos}).  
The three red pentastars, from top to bottom, mark example positions with single, double, and triple velocity components.  
The solid orange line indicates the direction of the global velocity gradient (Sect. \ref{sec_globalgradient}), which is also shown by a gray line in the left panel of Figure~\ref{fig:mom_highorder}.  
The five dashed orange lines mark linear structures perpendicular to the global velocity gradient direction (Sect. \ref{sec_lvbv}).  
Right: $^{12}$CO spectra at the three example positions indicated in the left panel.
\label{fig:ncomp}}
\end{figure*}

\subsubsection{C$^{18}$O spots -- prestellar cores in the Polaris cloud?}
The emission of C$^{18}$O is partitioned into five distinct regions, designated as $a$ through $e$ (Fig.~\ref{fig:mom0_18}). Region~$a$ in Fig.~\ref{fig:mom0_18} corresponds to the Polaris cloud, which is the densest structure within the Polaris Flare (Fig.~\ref{fig:compare_fcrao}). It serves as a central hub\footnote{A structural arrangement commonly observed in both atomic clouds dominated by supersonic turbulence and molecular clouds dominated by self-gravity \citep[e.g.,][]{2023ASPC..534..153H,2025NatAs.tmp..148L}.} for dense substructures, from which three C$^{18}$O branches extend discontinuously toward regions~$b$, $c$, and $d$. Within the Polaris cloud, the brightest C$^{18}$O emission is associated with MCLD~123.5+24.9 (indicated by the cyan box in Fig.~\ref{fig:compare_fcrao}a), a well-studied pre-stellar core region. In addition to this main MCLD structure, localized bright C$^{18}$O peaks ($I_{\rm C^{18}O} > 0.3$~K~km~s$^{-1}$) are distributed across the surrounding clump environment (Fig.~\ref{fig:regiona_c18o}). These bright spots likely trace a population of pre-stellar cores \citep[potentially in virial equilibrium; e.g.,][]{2012ApJ...745..195S} forming along the central skeletal bone (Fig.~\ref{fig:regiona_c18o}) within a host clump that is itself likely not purely bound by self-gravity \citep{2025ApJ...981..158S}.

The cores within region~$a$ were initially identified via visual inspection of the C$^{18}$O integrated intensity map, after which each emission peak was fitted with a two-dimensional ellipse to determine the spatial size of the core structure. The resulting C$^{18}$O peaks (compiled in Table~\ref{tab:C180}) exhibit a typical angular size of $\sim$2.5\arcmin\ (FWHM), corresponding to a physical scale of $\sim$0.1~pc at the adopted distance of 150~pc (Sect.~\ref{sec_intro}). A Minimum Spanning Tree (MST\footnote{Constructed using the \texttt{MiSTree} Python package (\url{https://knaidoo29.github.io/mistreedoc/}).}; see Fig.~\ref{fig:regiona_c18o}) analysis yields a characteristic nearest-neighbor separation of $0.16 \pm 0.06$~pc. These dense structures remain beam-diluted and are not fully resolved at our current angular resolution (Sect.~\ref{sec_dr}), implying that their true deconvolved sizes are smaller than 0.1~pc. This physical scale is in excellent agreement with the characteristic core sizes of $\sim$0.05--0.1~pc derived from \textit{Herschel} observations of both the Polaris Flare and other nearby molecular clouds \citep{2010A&A...518L.103M,2010A&A...518L.106K}.

Assuming a conversion factor between the C$^{18}$O integrated intensity and H$_2$ column density of $X_{\mathrm{C^{18}O}} = 5\times10^{21}~\mathrm{cm^{-2}\,(K\,km\,s^{-1})^{-1}}$ \citep{1996ApJ...465..815O}, the total H$_2$ column density can be estimated via
\begin{equation}
N(\mathrm{H_2}) = X_{\mathrm{C^{18}O}} \times I_{\rm C^{18}O}.
\end{equation}
For a typical C$^{18}$O peak with $I_{\rm C^{18}O} \sim 0.3~\mathrm{K\,km\,s^{-1}}$ and a nominal angular size of $3\arcmin$, this formulation yields an aggregate H$_2$ mass of $\sim$0.5~$M_\odot$, assuming a mean molecular weight of $\mu = 2.8$ and uniform integration over the core area. This mass estimate aligns with the characteristic scales of pre-stellar cores identified by \textit{Herschel} in nearby molecular clouds \citep{2010A&A...518L.102A,2010A&A...518L.106K}. It also agrees well with the predictions of turbulent fragmentation models in filamentary structures \citep{2025RAA....25b5020L}, supporting the scenario that these C$^{18}$O structures trace dense cores at an early evolutionary stage.

The dynamical state of a typical C$^{18}$O condensation can be quantified by evaluating its virial parameter, defined as \citep{1992ApJ...395..140B}
\begin{equation}
\alpha_{\rm vir} = \frac{5 \sigma_v^2 R}{G M},
\end{equation}
where $\sigma_v$ is the one-dimensional velocity dispersion, $R$ is the core radius, and $M$ is the core mass. Adopting a typical velocity dispersion of $\sigma_v \sim 0.1~\mathrm{km\,s^{-1}}$ (estimated from the channel width, consistent with the narrow-line nature of these features), a radius of $R \sim 0.05$~pc, and a mass of $M \sim 0.5~M_\odot$, we obtain $\alpha_{\rm vir} \sim 1$. This unity value indicates that these localized bright C$^{18}$O structures are close to a state of virial equilibrium. Taken together, their coupled mass, spatial dimensions, and gravitationally bound virial states suggest that these bright C$^{18}$O emission peaks are highly promising pre-stellar core candidates.

\subsubsection{C$^{18}$O in region $e$ as a filament head}
Region~$e$ in Figure~\ref{fig:mom0_18} is a relatively isolated structure exhibiting C$^{18}$O emission, which corresponds to the southeastern bright head of a narrow filament traced in $^{13}$CO (see Sect.~\ref{sec_multivelos} and Figure~\ref{fig:mom0_12_13}). This filament, a constituent of complex BE, is elongated perpendicular to the global velocity gradient (Sect.~\ref{sec_globalgradient}). Such slender, perpendicular filaments are common across the Polaris Flare (Figure~\ref{fig:ncomp}) and will be discussed extensively in Sect.~\ref{sec_lvbv}. Among these structures, the filament hosting region~$e$ stands out due to its well-defined C$^{18}$O emission profiles. Its morphology aligns with the end-dominated evolution of filaments, where preferential mass accumulation occurs at the boundaries due to longitudinal accretion flows and gravitational focusing \citep[e.g.,][]{1983A&A...119..109B,2020A&A...637A..67Y}.

The narrow C$^{18}$O line widths indicate that region~$e$ is dynamically quiescent, while the total absence of embedded protostars points toward a pre-stellar or early evolutionary stage. Alternatively, the observed properties of this filament may also reflect dynamical interactions with the surrounding natal clump BE, which could actively shape its localized structure and internal kinematics.


\begin{figure*}[!t]
    \centering
    \includegraphics[width=0.99\linewidth]{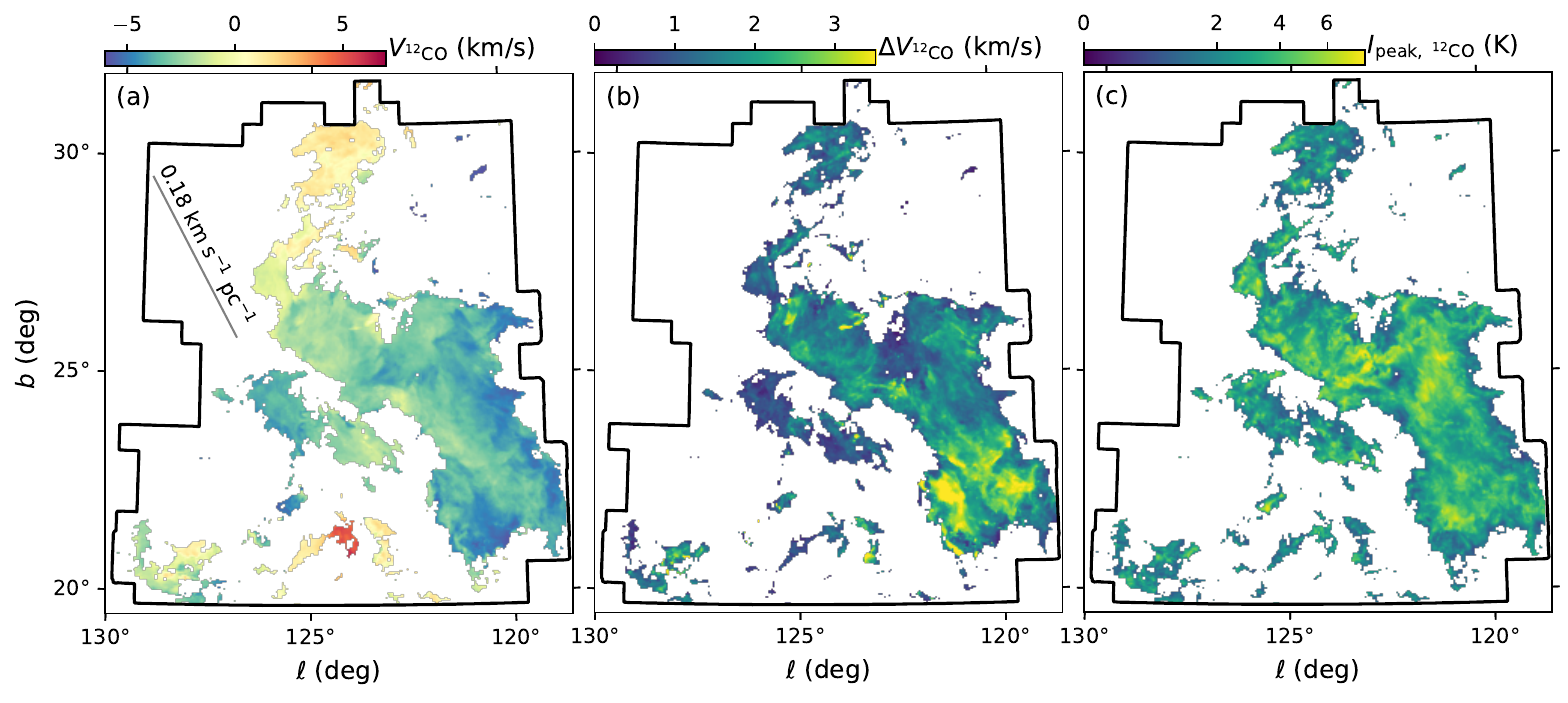}
    \caption{Moment 1 (a), line width (b), and spectral peak (c) maps of $^{12}$CO (Sect. \ref{sec_momenteqs}). The gray line in panel (a) indicates the direction of the global velocity gradient (Sect.~\ref{sec_globalgradient}). \label{fig:mom_highorder}
    }
\end{figure*}

\subsubsection{Hierarchical structure traced by CO isotopologues}\label{sec_hiera}
The Polaris Flare exhibits a hierarchical structure that is clearly revealed by the stratified emission of multiple CO isotopologues. C$^{18}$O selectively traces the densest and most compact structures, corresponding to gravitationally bound pre-stellar cores and marking sites where fragmentation driven by self-gravity and converging flows is most advanced. In contrast, $^{13}$CO delineates a more extended, partially bound skeleton that forms a relatively stable intermediate kernel, embedding these dense cores and organizing the overall architectural framework.

The $^{12}$CO emission traces the more diffuse outskirts or turbulent envelope, where the gas is less bound and highly dynamically active, likely undergoing simultaneous assembly and dispersal driven by interstellar turbulence, waves, and large-scale macroscopic motions \citep[e.g.,][]{2003ApJ...583..280B,2022ApJ...932...47W,2025ApJ...989...25Y}. Together, these three tracers provide a coherent view of the Polaris Flare, transitioning from dense, gravitationally confined nodes (C$^{18}$O) to a partially bound, stable skeleton ($^{13}$CO), and finally to a dynamically evolving periphery ($^{12}$CO). This multi-line perspective beautifully highlights the delicate interplay between gravity and turbulence in shaping the initial conditions of filamentary star formation.

\subsection{Higher-order moment maps of $^{12}$CO}\label{sec_highmoms}
In addition to the integrated intensity (Moment~0) map, higher-order velocity moments provide valuable insights into the detailed kinematics and physical interactions of different gas components in position--position--velocity (PPV) space. In this section, we present maps of the number of discrete velocity components, centroid velocities, line widths, and peak intensities. In Sect.~\ref{sec_momenteqs}, we outline a unified formalism demonstrating how each of these quantities can be rigorously expressed within the framework of moment parameters. The dynamical information revealed by these respective mappings is subsequently discussed in Sects.~\ref{sec_multivelos}--\ref{sec_peakintensity}.

\subsubsection{Kinematic parameters}\label{sec_momenteqs}

For a given spectrum $I$, the standard $n$-th order velocity moments are calculated numerically across the channel array as
\begin{equation}
    M_n = \frac{\sum_j \left( V_j - V_0 \right)^n I_j}{\sum_j I_j},\label{eq_mom}
\end{equation}
where $j$ is the velocity channel index, and the reference velocity $V_0$ is set to 0 for $n \le 1$ and to the centroid velocity $M_1$ for $n > 1$. Under the assumption of a symmetric Gaussian profile, the line full width at half maximum ($\Delta V$) can be directly estimated from the second-order moment ($M_2$) via
\begin{equation}
    \Delta V = \sqrt{8 \ln(2) \, M_2}. \label{eq_calDV}
\end{equation}
We apply Eq.~\ref{eq_calDV} across the entire map as an effective line-width indicator, even in regions where the profiles deviate from a Gaussian or contain multiple velocity components. Because higher-order moments ($n \ge 1$) are inherently sensitive to baseline fluctuations, only channels with intensities exceeding a $3\sigma$ threshold are included in these calculations.

In addition to the velocity moments, the peak radiation temperature of each spectrum is extracted directly from the raw data cube via
\begin{equation}
    I_{\rm peak} = \max_j I_j.
\end{equation}
In data reduction software such as CASA, this parameter is historically referred to as ``moment 8,'' although it represents a simple maximum intensity filter rather than a true statistical moment.

To robustly determine the number of discrete velocity components ($N_{V}$) at each spatial pixel, we implement a practical peak-identification algorithm rather than an analytical moment formalism. We utilize the \texttt{find\_peaks} routine from the Python package \texttt{SciPy} \citep{2020NatMe..17..261V}, applying a strict set of topological constraints to filter out noise and baseline artifacts. Specifically, we enforce a minimum signal-to-noise ratio threshold of 5 to eliminate spurious low-intensity features, a minimum channel separation of 5 channels to merge tightly blended neighboring local maxima, and a prominence threshold of $4\sigma$ to discard insignificant side lobes or noise spikes near primary emission components. This numerical approach provides a robust and reproducible census of the cloud's multi-component velocity structure.

\begin{figure*}[!t]
    \centering
    \includegraphics[width=0.98\linewidth]{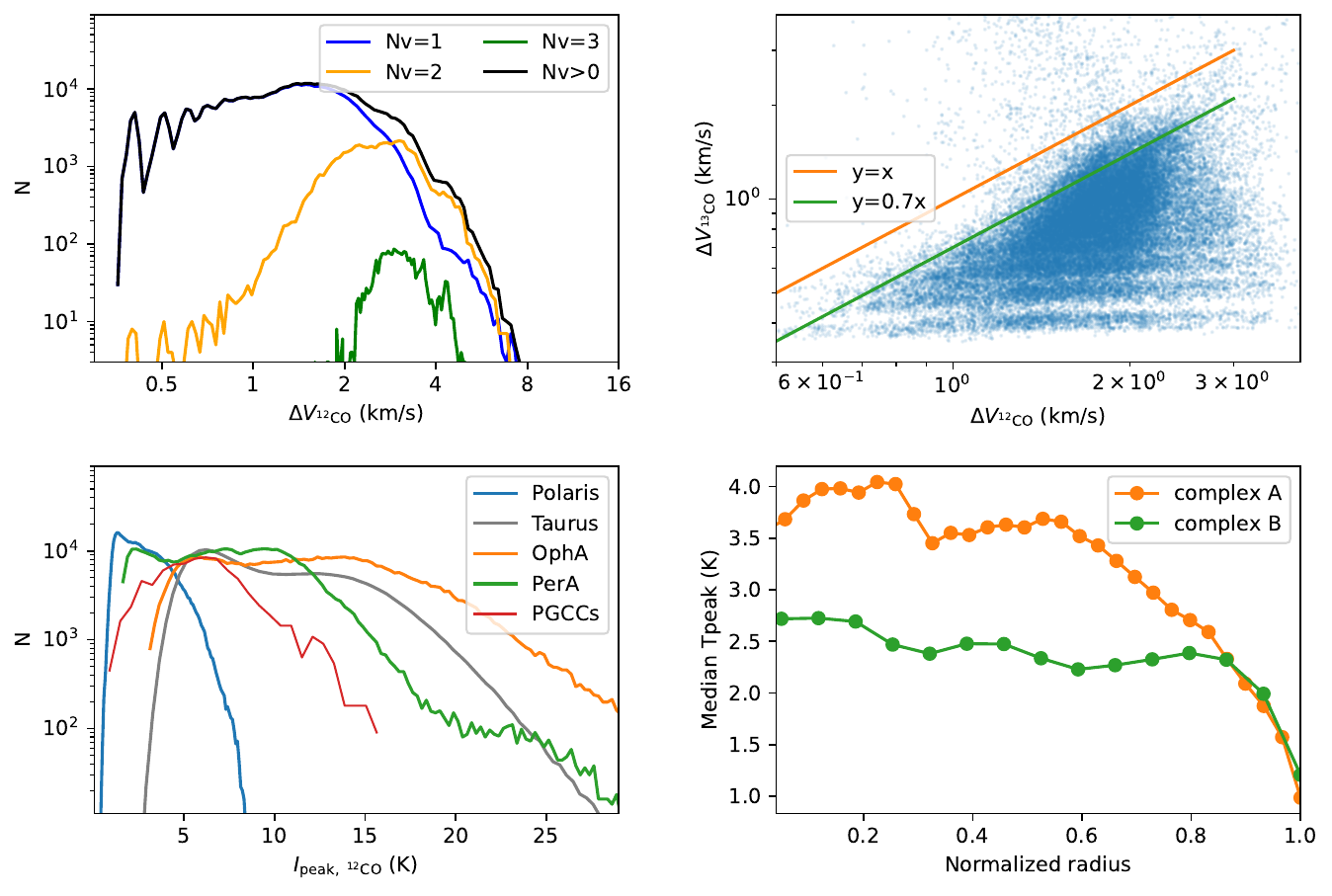}
\caption{Upper left: Pixel-based histogram of the $^{12}$CO line widths in the Polaris Flare, derived from the second-order velocity moments using Eq.~\ref{eq_calDV} and partitioned into different groups based on the number of velocity components ($N_V$, Sect.~\ref{sec_highmoms}). Upper right: Correlation between the single-component line widths of $^{12}$CO and $^{13}$CO, restricted strictly to pixel populations with $N_V = 1$. The orange solid line indicates $y = x$, while the green solid line marks the upper envelope defined by $y = 0.7x$.
Lower left: Distribution of peak brightness temperatures ($I_{\rm peak,\ ^{12}CO}$) for the Polaris Flare alongside other target groups. The comparison datasets include Taurus, Perseus, and Ophiuchus observed with FCRAO \citep{2008ApJS..177..341N,2006AJ....131.2921R}, as well as PGCCs mapped with the Delingha 13.7~m telescope \citep{2012ApJ...756...76W}. For the PGCC sample, the values correspond to the individual clump-peak spectra. The total pixel counts for each group have been normalized to match the distribution of Polaris to facilitate direct comparison. Lower right: Median value of $I_{\rm peak,\ ^{12}CO}$ as a function of the normalized radius $R$ (see Sect.~\ref{sec_peakintensity}) computed for complex~A (orange) and complex~B (yellow).}

    \label{fig:DVTpeak}
\end{figure*}

\subsubsection{Multiple velocity components}\label{sec_multivelos}
The resulting map displaying the spatial distribution of the number of velocity components ($N_{V}$) is shown in Figure~\ref{fig:ncomp}. The vast majority of pixels exhibit single-peaked $^{12}$CO spectra, with only approximately 10\% of the detected pixels above the $5\sigma$ threshold displaying two or three discrete velocity components. The Polaris cloud, in particular, is characterized by a single velocity component across its entire spatial extent. Multi-component CO spectra are a common feature of molecular clouds, typically tracing overlapping, velocity-coherent substructures or fibers along the line of sight, as observed in complexes like Taurus and Orion \citep[e.g.,][]{2013A&A...554A..55H,2021RAA....21...24Y}. For comparison, systematic Gaussian decompositions in the Musca filament revealed that roughly 20\% of $^{13}$CO profiles require at least two distinct components \citep{2016A&A...587A..97H}. The relatively small fraction of multi-component spectra in the Polaris Flare, even when traced using the more diffuse $^{12}$CO emission, suggests that Polaris is an exceptionally quiescent and dynamically young cloud complex, where the gas has not yet developed the complex, overlapping kinematic structures typical of more evolved filamentary networks.

Nevertheless, regions exhibiting multiple velocity components provide crucial dynamical clues. The multi-velocity structures identified here, taken together with the sharp velocity discontinuities observed between different complexes (Sects.~\ref{sec_globalgradient} and \ref{sec_lvbv}), point to potential cloud--cloud interactions or localized sheet-like collisions \citep[e.g.,][]{2021PASJ...73S...1F}, occurring both between and within the individual complexes. Owing to its otherwise simple and unconfused environment, the Polaris cloud provides an ideal astrophysical laboratory for investigating the earliest stages of such gas interactions. Notably, the $N_{V}$ map also highlights several slender filaments elongated perpendicular to the global velocity gradient (indicated by the dashed orange lines in Figure~\ref{fig:ncomp}), which will be analyzed in detail in Sect.~\ref{sec_lvbv}.

The right panels of Figure~\ref{fig:ncomp} present representative spectra displaying single, double, and triple peaks, verifying the empirical robustness of our automated \texttt{find\_peaks} extraction method. These profile variations also reveal a clear large-scale systematic trend: line profiles in the southwestern sector of Polaris are systematically blueshifted, whereas those in the northern sector shift toward redshifted velocities. This systematic kinematic behavior motivates a rigorous examination of the large-scale velocity gradient using the first-order moment (centroid velocity) map, as presented in Sect.~\ref{sec_globalgradient}.

\subsubsection{Global velocity gradient}\label{sec_globalgradient}

The first-order velocity moment (Moment~1) map clearly reveals a coherent global velocity gradient across the entire expanse of the Polaris Flare (Figure~\ref{fig:mom_highorder}). To robustly quantify this kinematic feature, we apply a two-dimensional linear regression to the Moment~1 map over the detected area. Let $(x_i, y_i)$ denote the spatial position of the $i$-th pixel in Cartesian coordinates. For a trial projection angle $\theta$ (defined counterclockwise from north), we introduce the rotated spatial coordinate
\begin{equation}
    x'_i = y_i \cos\theta - x_i \sin\theta.
\end{equation}
The principal direction of the velocity gradient, $\theta_{\rm grad}$, and its absolute magnitude, $V_{\rm grad}$, are subsequently determined by solving the minimization problem
\begin{equation}
    (\theta_{\rm grad},\,V_{\rm grad}) = \underset{\theta,\,a}{\mathrm{argmin}} \left[ \min_{b} \sum_i \bigl(a x_i' + b - V_i\bigr)^2 \right],
\end{equation}
where $V_i$ represents the intensity-weighted line-of-sight velocity at pixel $i$, $a$ is the gradient slope along the $x'$ axis, and $b$ is the systemic velocity intercept.

Fitting the $^{12}$CO Moment~1 map via this framework yields a principal direction of 
\begin{equation}
    \theta_{\rm grad} = 27.3\degr,
\end{equation} 
and a gradient magnitude of 
\begin{equation}
    V_{\rm grad} = 0.18~\mathrm{km~s^{-1}~pc^{-1}}
\end{equation} 
(panel a of Figure~\ref{fig:mom_highorder}), with statistical uncertainties below 1\%. This measured shear is notably larger than the gradients typically found across giant molecular clouds ($\sim$0.01--0.1~km~s$^{-1}$~pc$^{-1}$; \citealt{2003ApJ...599..258R,2011ApJ...732...78I}). Instead, it is highly comparable to values observed in localized, quiescent molecular filaments ($\sim$0.1--0.3~km~s$^{-1}$~pc$^{-1}$; \citealt{2013A&A...554A..55H,2014ApJ...790L..19F}), while remaining significantly smaller than the steep gradients characteristic of dense, star-forming infrared dark clouds ($\sim$0.5--2~km~s$^{-1}$~pc$^{-1}$; \citealt{2014MNRAS.440.2860H,2014A&A...561A..83P}). These comparative results demonstrate that while the Polaris Flare is more dynamically structured than a standard, monolithic GMC, its kinematics remain firmly within the regime of non-star-forming filaments—consistent with its uncollapsed, quiescent nature.

\begin{figure}
    \centering
    \includegraphics[width=0.99\linewidth]{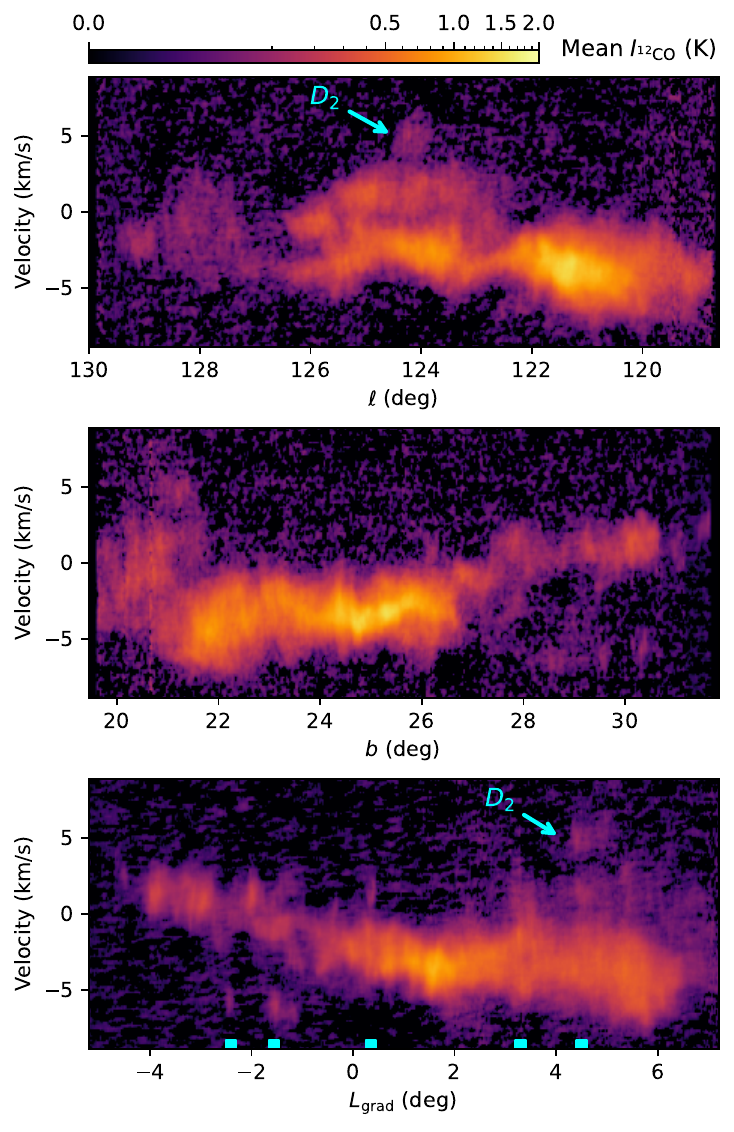}
    \caption{Images were obtained by averaging the $^{12}$CO PPV cube perpendicular to the 
$l$ (upper), $b$ (middle), and $L_{\rm grad}$ (lower) directions 
(Sect.~\ref{sec_lvbv}).  
Here, $L_{\rm grad}$ denotes the angular offset along the direction of 
$V_{\rm grad}$ (Figure~\ref{fig:mom_highorder}), measured from the northeast corner.  
All three panels share the same color scale, indicated by the color bar above the upper panel.  
The red component (D$_2$) of complex~D (Sect.~\ref{sec_complexD}) is marked by a blue arrow.  
In the lower panel, the cyan bars along the x-axis highlight the narrow vertical strips 
identified in the map (see Figure~\ref{fig:ncomp} for their corresponding 
linear structures perpendicular to the global velocity gradient).
    } \label{fig:ppvcube}
\end{figure}

\subsubsection{Line widths} \label{sec_linewidths}

The spatial distribution of line widths (middle panel of Figure~\ref{fig:mom_highorder}) derived from Eq.~\ref{eq_calDV} closely mimics that of $N_{V}$ (left panel of Figure \ref{fig:ncomp}). This structural correspondence suggests that single-component spectra generally maintain a relatively uniform and narrow width, whereas regions containing multiple velocity components contribute significantly to the observed line broadening. A notable exception is the Polaris cloud (Figure \ref{fig:planck}), which, despite exclusively harboring a single velocity component, exhibits broader line profiles than its surrounding ambient environment (Figure \ref{fig:mom_highorder}).

The line widths ($\Delta V$) of the $^{12}$CO emission (see the upper left panel of Figure~\ref{fig:DVTpeak}) are systematically broader than those of its less abundant isotopologues. For pixels containing a single velocity component ($N_{V}=1$; Figure~\ref{fig:ncomp}), $\Delta V$($^{12}$CO) typically averages $1.2 \pm 0.6$~km~s$^{-1}$. In contrast, pixels with multiple velocity components reach a 90th-percentile value of approximately 2~km~s$^{-1}$, with extreme profile wings extending up to 8~km~s$^{-1}$. For physical context, a kinetic temperature of 10~K corresponds to a purely thermal line width of only $\sim$0.13~km~s$^{-1}$ for a $^{12}$CO molecule, which is far narrower than our observed values.

Line broadening driven solely by high optical depth cannot account for the magnitude of these profiles. Even an optical depth as high as $\tau=10$ increases the second-order moment width of an intrinsic Gaussian profile by only about 50\%. When evaluated via the full width at half maximum, the optical depth broadening factor at $\tau=10$ can be quantified analytically as \citep[e.g.,][]{1979ApJ...231..720P}
\begin{equation}
    \beta_\tau = \sqrt{\frac{\ln(\tau)}{\ln(2)}},
\end{equation}
which yields $\beta_\tau \sim 1.8$, corresponding to an 80\% increase in the apparent FWHM. Consequently, the observed line profiles must be heavily dominated by supersonic, non-thermal motions, at least within the extended structures mapped by the $^{12}$CO emission.

Although the $^{12}$CO transitions are generally broad, the corresponding $^{13}$CO spectra are considerably narrower. In the upper right panel of Figure~\ref{fig:DVTpeak}, we explore the correlation between $\Delta V$($^{12}$CO) and $\Delta V$($^{13}$CO) within regions where $^{13}$CO is detected and characterized by a single velocity component. Among the standard velocity moments, the second-order moment ($M_2$) is inherently the most sensitive to baseline noise fluctuations and artifacts. To derive a robust $M_2$ value for the $^{13}$CO spectra, we excluded all channels with radiation intensities below a $3\sigma$ threshold. For exceptionally narrow profiles where only a few channels survive this mask, the resulting $M_2$ values become discretely quantized, clustering around integer multiples of the channel width, $n\delta V$. Occasional random noise spikes migrating above the threshold can also artificially inflate the calculated line width. This finite spectral sampling explains the appearance of striping artifacts at the low-value end in the upper right panel of Figure~\ref{fig:DVTpeak}, as well as the unphysical outliers that display $\Delta V({\rm ^{13}CO}) > \Delta V({\rm ^{12}CO})$.

It is clear that $\Delta V$($^{13}$CO) is systematically narrower than $\Delta V$($^{12}$CO), bounded by an approximate upper envelope defined as
\begin{equation}
    \Delta V^{\rm upper}({\rm ^{13}CO}) = 0.7 \,\Delta V({\rm ^{12}CO}),
\end{equation}
which is shown as a green solid line in the upper right panel of Figure~\ref{fig:DVTpeak}. The median and standard deviation of these single-component $^{13}$CO widths are $0.8 \pm 0.3$~km~s$^{-1}$, representing roughly 70\% of the typical $^{12}$CO line width. Importantly, the discrete sampling effects discussed above do not alter this underlying physical trend, which reflects a true kinematic transition within the cloud hierarchy.

\begin{figure}[!t]
    \centering
    \includegraphics[width=0.99\linewidth]{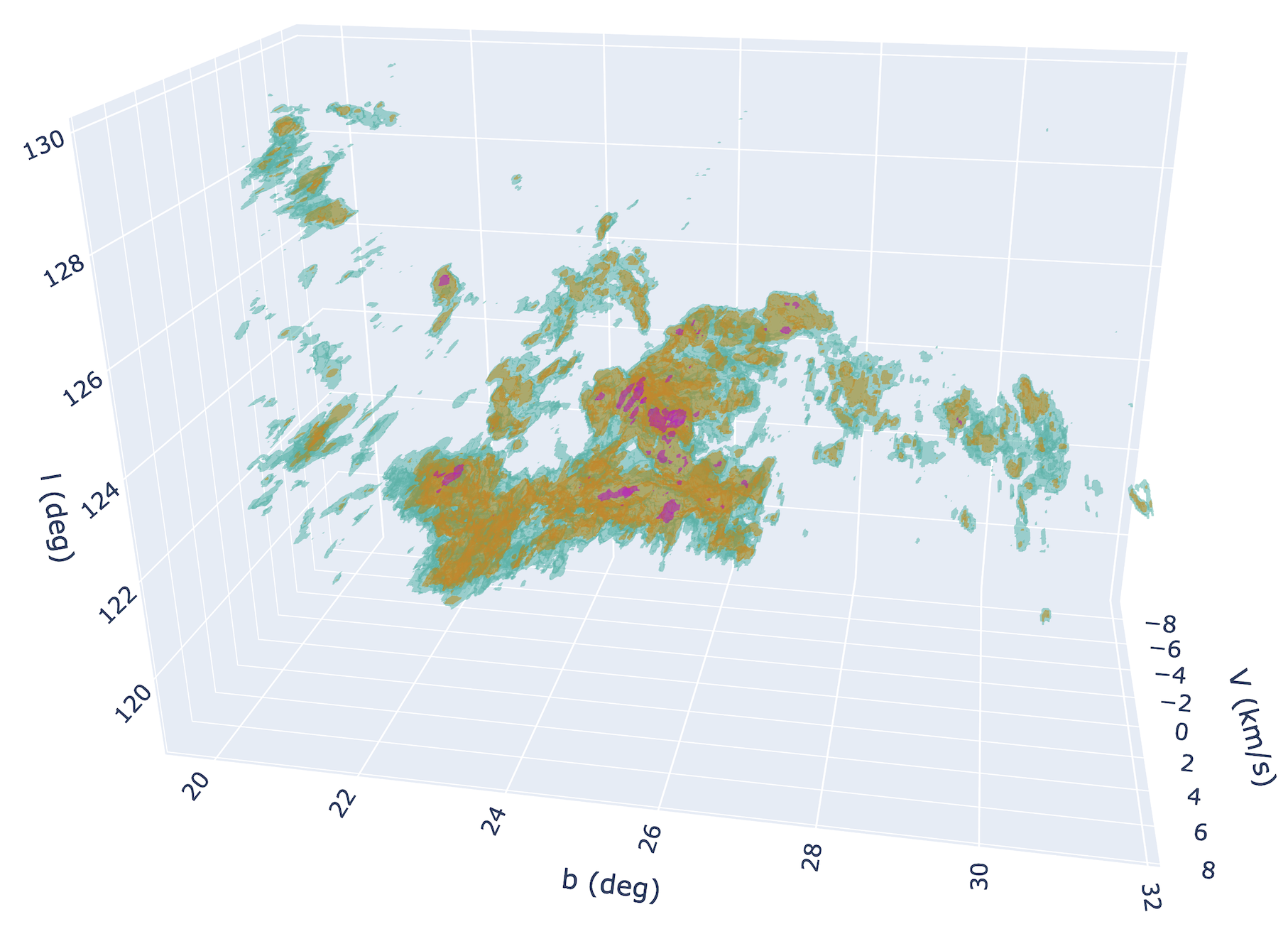}
    \includegraphics[width=0.99\linewidth]{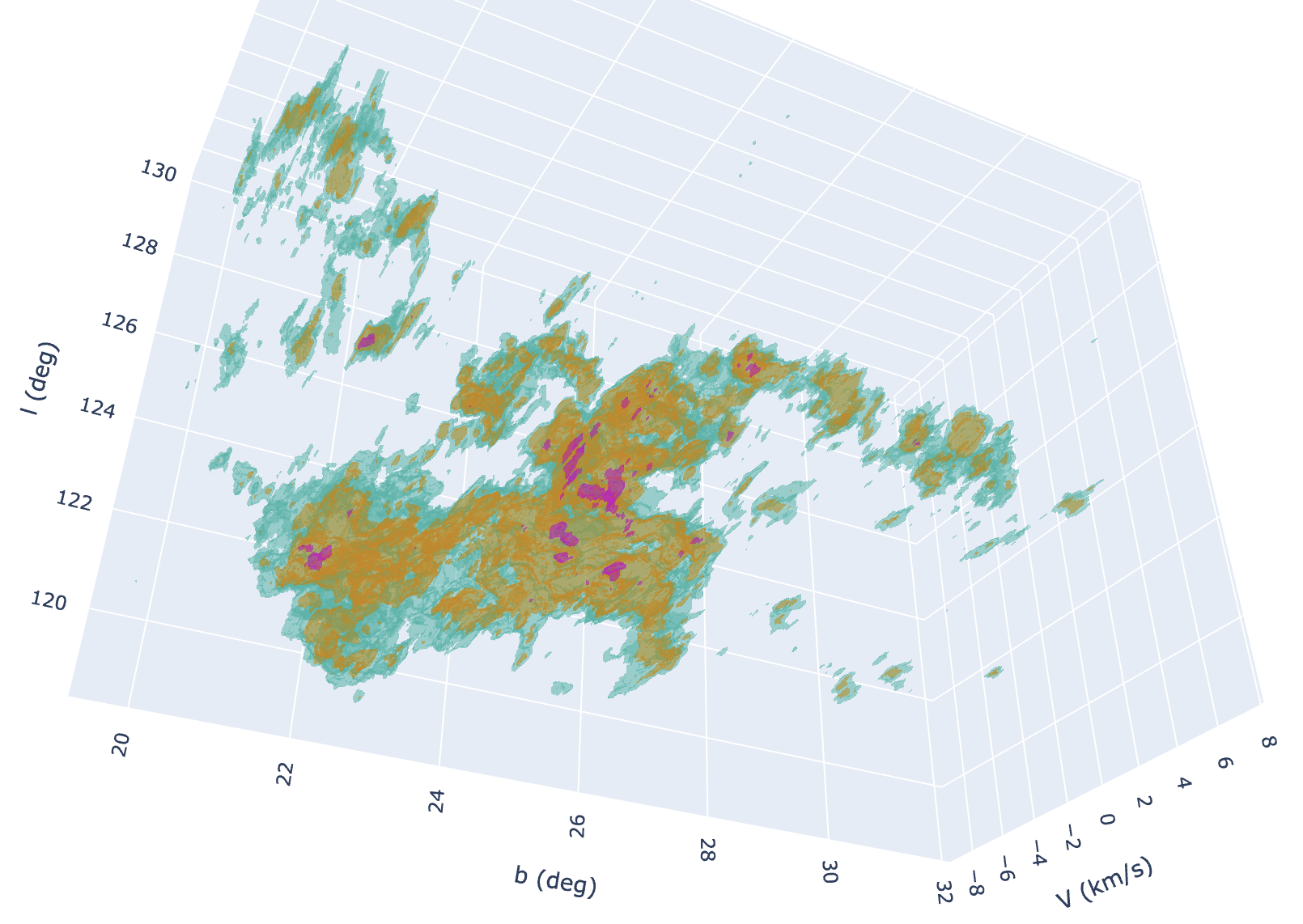}
    \caption{ Three dimensional rendering of the PPV cube (Sect. \ref{sec_ppvcube}), viewed from
    different viewing angles, showing isosurfaces at 2, 4, and 6 K in turquoise, orange, and magenta, respectively. }
    \label{fig:isosurfaces}
\end{figure}

\subsubsection{Spectral peak intensities}\label{sec_peakintensity}

In the lower left panel of Figure~\ref{fig:DVTpeak}, we present the histogram of the spectral peak intensity $I_{\rm peak,\ ^{12}CO}$ for the Polaris Flare alongside those of Taurus, Perseus, and Ophiuchus observed with FCRAO \citep{2008ApJS..177..341N,2006AJ....131.2921R}, as well as a sample of PGCCs observed with the Delingha 13.7~m telescope \citep{2012ApJ...756...76W}. The high-intensity tails of the distributions for the active star-forming clouds (Taurus, Perseus, and Ophiuchus) extend well beyond 20~K, reflecting thermal heating associated with ongoing stellar feedback and protostellar activity. In contrast, the Polaris Flare exhibits significantly lower peak temperatures—even when compared directly to the PGCC sample—with the vast majority of pixels constrained to $I_{\rm peak,\ ^{12}CO} < 8$~K.
This property aligns with the integrated intensity ratio analysis (Sect.~\ref{sec_Iratio}), which revealed $I_{\rm ^{12}CO}/I_{\rm ^{13}CO}$ systematically larger than those characteristic of PGCCs. Overall, these results imply that the Polaris Flare may be younger than the PGCCs, which in turn are less evolved than the nearby star-forming regions.

The 60th, 90th, and 95th percentiles of $I_{\rm peak,\ ^{12}CO}$ in Polaris are 2.6, 4.7, and 5.3~K, respectively. Assuming an excitation temperature of 10~K, these values correspond to optical depths of 0.13, 0.28, and 0.33, suggesting that the beam-averaged $^{12}$CO emission is largely optically thin. This appears somewhat at odds with the relatively low $I_{\rm ^{12}CO}/I_{\rm ^{13}CO}$ ratio (Sect.~\ref{sec_Iratio}), which implies that $^{12}$CO may not be fully thermalized or that density fluctuations within the telescope beam are significant. A comprehensive assessment of CO excitation in Polaris would require detailed radiative transfer calculations \citep[e.g.,][]{2020A&A...644A.151J,2020A&A...644A..27B} on the extended 3D structures inferred from the 2D maps \citep[e.g.,][]{2025arXiv250319259L}, which is beyond the scope of this work.

Complexes~A and B are the most extended molecular structures in the Polaris Flare, enabling us to systematically examine the spatial architecture of their CO excitation. The value of $I_{\rm peak,\ ^{12}CO}$ in complex~A is significantly higher than in complex~B and the other surrounding complexes. To quantify this spatial variation, the lower right panel of Figure~\ref{fig:DVTpeak} presents the median value of $I_{\rm peak,\ ^{12}CO}$ as a function of the normalized radius for Complexes~A and B. The distance $D$ of each pixel from its nearest complex boundary is first computed using a distance transform technique\footnote{The distance transform of a region was implemented using the \texttt{distance\_transform\_edt} function from the \texttt{SciPy} library \citep{2020NatMe..17..261V}.}
, and the normalized radius $R$ is then defined via
\begin{equation}
    R = \frac{\max(D) - D}{\max(D)}.
\end{equation}
Defined this way, $R = 1$ corresponds to the outer boundary of the complex, while $R = 0$ identifies the centermost pixels with the largest distance from the edge. For $R \gtrsim 0.85$, both complexes exhibit similar median peak intensities of $\sim$2.4~K, indicating that their outermost peripheries share comparable gas properties.
For $R < 0.85$, the median peak intensity in complex~B remains roughly constant at $\sim$2.5~K, while in complex~A it increases steadily toward the core, reaching $\sim$4~K as $R$ approaches 0. This distinct behavioral divergence suggests that Complex~A contains denser or more highly excited gas at its internal kernel, whereas Complex~B is internally more uniform. Overall, these radial trends imply that both complexes likely originated from a common parental structure, with complex~A having experienced a higher degree of localized evolution or accumulation of dense gas. This structural interpretation is highly consistent with the framework of a dynamically assembling and dispersing $^{12}$CO periphery proposed in Sect.~\ref{sec_hiera}.

\subsection{PV maps and 3D rendering}\label{sec_pvandoverall}
\subsubsection{PV maps and perpendicular structures}\label{sec_lvbv}

The projections of the $^{12}$CO data cube onto the $l$--$V$ and $b$--$V$ planes (upper and middle panels of Figure~\ref{fig:ppvcube}) reveal the comprehensive velocity distribution across the cloud. A clear velocity gradient is visible in the $b$--$V$ map, indicating systematic motion along the Galactic latitude direction. This trend reflects large-scale ordered motions that may be associated with shear flows in the diffuse envelope or with global turbulence cascading across the region.

In the $l$--$V$ map, the northern ($b > 27\degr$) and southern ($b < 27\degr$) regions within the longitude range $123\degr < l < 126\degr$ contribute to two separated redshifted and blueshifted components, corresponding to complexes~AN and AC (see Figure~\ref{fig:subpart}). Yet, these two components merge smoothly into a coherent structure in the $b$--$V$ plane (middle panel of Figure~\ref{fig:ppvcube}). This implies that both components follow a similar overall monotonic behavior along Galactic latitude, possibly governed by this same velocity trend in PPV space.

Such coherence suggests that the cloud should be regarded as a single turbulent system rather than a set of discrete sub-clouds, although their molecular components are spatially separated. The velocity separation between complexes~AN and AC likely arises from local kinematic distortions, possibly driven by shear, partial compression, or filamentary streaming motions, superposed on the global velocity gradient. This raises the possibility of cloud–cloud interactions or large-scale converging flows, which could contribute to the observed fragmentation of the complex. We suggest that the global velocity field in Polaris is not random, but organized on large scales, while still showing signs of turbulent mixing and local interaction.

\begin{figure}[!t]
    \centering
    \includegraphics[width=0.99\linewidth]{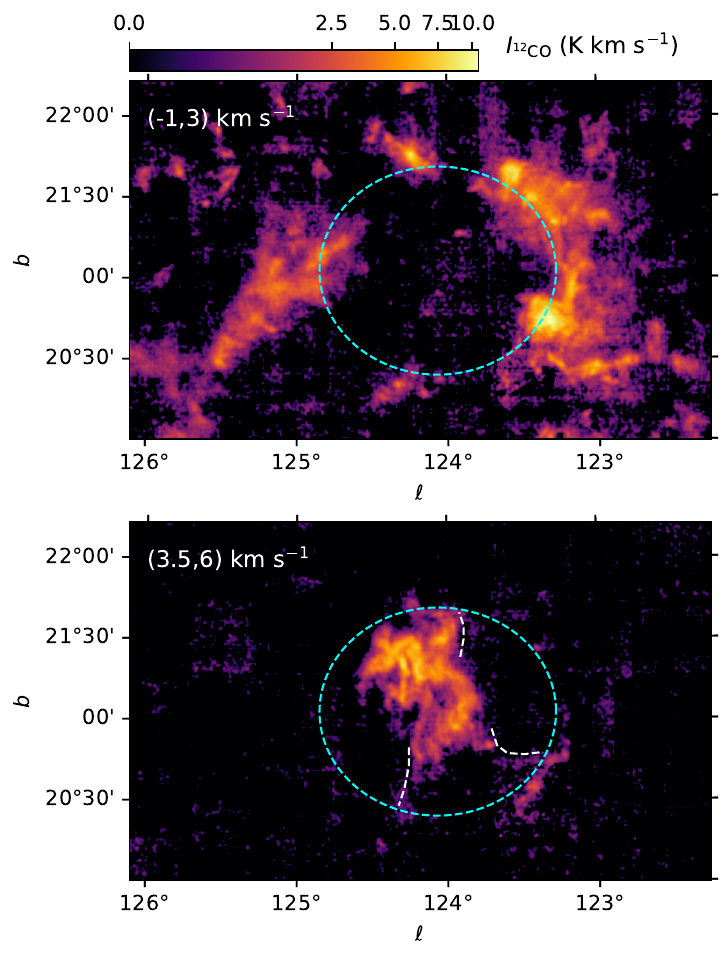}
    \caption{Moment-0 maps of the blue (upper) and red (lower) velocity components of Complex D (Sect. \ref{sec_complexD}). The cyan ellipse denotes the inner boundary traced by the blue-component emission. The white dashed lines highlight the filamentary structure of the red component, showing strands that connect to the shell (cyan ellipse).}
    \label{fig:Dmom0}
\end{figure}

The projections of the $^{12}$CO data cube onto the $L_{\rm grad}$--$V$ plane, where $L_{\rm grad}$ is defined along the direction of the velocity gradient (Sect.~\ref{sec_globalgradient}), reveal enhanced emission concentrated along a compact ridge. A striking morphological feature is the presence of vertical strips on this plane (indicated by cyan bars), which appear much clearer and narrower than in the $l$--$V$ and $b$--$V$ planes. These strips, identified through careful inspection of the data cube, correspond to linear structures oriented perpendicular to the velocity gradient, five of which are marked by orange dashed lines in Figure~\ref{fig:ncomp}.
The two strips in the northwest corner trace elongated, slender filaments that are not assigned to any of the seven complexes (Figure~\ref{fig:subpart}). The central strip, representing the third of these marked linear structures, corresponds to a filament that spatially overlaps with complex~AC but shows a distinct velocity offset, while the two southern strips correspond to complexes~C and D.

The global morphology and kinematics suggest that the Polaris Flare can be divided into two main structural components: a primary component, comprising complexes~A and B, which extend parallel to the global velocity gradient; and a secondary component, consisting of slender filaments and complexes~C and D, which form features elongated perpendicular to this gradient. This arrangement indicates that all identified complexes and surrounding flaring gas are part of a single, dynamically regulated system.
The physical origin of these perpendicular structures remains an open question. They may represent large-scale wavelike features, such as striations generated by magnetohydrodynamic wave modes \citep[e.g.,][]{2016MNRAS.462.3602T,2019FrASS...6....5H,2023A&A...673A..76S}, or they could be driven by local shear, compressive flows, shock fronts, or other macro-scale dynamical processes shaping the cloud complex \citep{2016MNRAS.461.3918H,2019MNRAS.485.4509L,2025NatAs.tmp..148L}. Such large-scale wavelike modes may also be physically linked to the small-scale striations resolved within the individual complexes.

\subsubsection{3D view of the cube}\label{sec_ppvcube}

To complement the two-dimensional projections, we construct a three-dimensional isosurface rendering of the $^{12}$CO PPV cube at intensity levels of 1.5, 3, and 6~K, enabling visualization from multiple viewing angles (Figure~\ref{fig:isosurfaces}). This rendering was generated using an isosurface extraction approach based on the marching-cubes algorithm \citep{Lorensen87}, implemented with the \texttt{scikit-image} package\footnote{\url{https://scikit-image.org}} for surface extraction and the \texttt{Plotly} library\footnote{\url{https://plotly.com/python/}} for interactive visualization. Prior to rendering, the data cube was smoothed and downsampled onto a spatial grid with a step size of 1.5\arcmin.

This three-dimensional representation provides a global perspective on the emission distribution in PPV space, allowing structural coherence to be assessed simultaneously across spatial and kinematic dimensions. The isosurface rendering emphasizes the extended and interconnected nature of the emission features, which are less apparent in individual $l$--$b$, $l$--$V$, or $b$--$V$ slices. Unlike conventional two-dimensional projections, the three-dimensional view reveals how emission structures are nested and intertwined, with lower-brightness envelopes surrounding brighter, more compact concentrations.

Complexes~AC and AW are dominated by compact emission condensations with brightness temperatures above 3~K, where the 1.5~K isosurface tightly encloses the 3~K structures. This nested morphology points to steep brightness (and likely density or excitation) gradients at the boundaries of these complexes, consistent with the picture of spatially coherent substructures embedded within more extended emission
(Sects.~\ref{sec_hiera} and \ref{sec_peakintensity}).
In contrast, across complexes~AN and B (including components~BW and BE), the 3~K isosurface occupies only about half the volume enclosed by the 1.5~K surface, suggesting that these regions are more diffuse and may represent flaring or extended gas enveloping the denser condensations of AC and AW. The relative absence of high-brightness emission indicates that their excitation conditions are lower, and that they lack prominent compact substructures.

The 6~K isosurface highlights only a small fraction of the total volume, corresponding closely to the C$^{18}$O detection region (Figure~\ref{fig:mom0_18}). This tight spatial correlation indicates that the brightest $^{12}$CO features coincide with the densest gas in the cloud, where C$^{18}$O remains abundant and shielded against photodissociation. Taken together, the isosurface view reveals a nested hierarchy: diffuse envelopes at $\sim$1.5~K, compact condensations at $>$3~K, and dense cores at $>$6~K. This three-dimensional morphology directly complements the isotopologue-based hierarchy described in Sect.~\ref{sec_hiera}, in which $^{12}$CO, $^{13}$CO, and C$^{18}$O trace the periphery, kernel, and core of the cloud, respectively.

\begin{figure}[!t]
    \centering
    \includegraphics[width=0.99\linewidth]{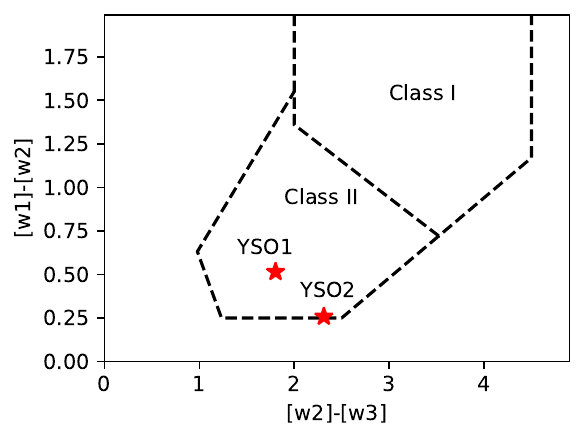}
    \caption{WISE color–color diagram \citep{2014ApJ...791..131K} showing the two YSO candidates (see left panel of Figure~\ref{fig:subpart}) for further classification (Sect. \ref{sec_starless}).}
    \label{fig:yso}
\end{figure}

\section{Discussion}\label{sec_discuss}
\subsection{Nested structures of complex D}\label{sec_complexD}

In this work, we primarily explore the global properties of the molecular gas within the Polaris Flare through statistical analyses and comparative studies of different subregions, rather than focusing on local emission. Nevertheless, the nested structures uncovered within complex~D exhibit unique kinematic properties that warrant immediate discussion. The Moment~0 maps of complex~D (Figure~\ref{fig:Dmom0}) reveal, for the first time, its intricate internal morphology, showing two kinematically distinct yet spatially connected components of comparable brightness. The bluer component ($V \sim -1$ to 3~km~s$^{-1}$; designated as D$_{1}$; see Figure~\ref{fig:subpart}) forms a ring-like shell that completely surrounds a central cavity. An ellipse traced along the inner boundary of this shell delineates the curvature of the cavity and defines the spatial extent of the surrounding gas, spanning approximately 45\arcmin\ (or $\sim$2~pc) in diameter. Similar shell--cavity morphologies have been widely observed in other molecular cloud complexes and are frequently linked to localized gas motions or past dynamical events, which may drive the formation of filamentary structures at the interaction interfaces \citep[e.g.,][]{2018PASJ...70S..46F,2020A&A...644A...5B}.

The redder component ($V \sim 3.5$ to 6~km~s$^{-1}$; designated as D$_2$) occupies this central cavity and exhibits a highly structured filamentary morphology, with its inner region showing greater structural complexity than the surrounding shell. Multiple filaments belonging to this redshifted component extend in different directions, reaching and anchoring directly to the inner boundary of the bluer shell, indicating that the two components constitute a single, spatially coherent system. While interacting molecular filaments and nested systems have been reported in more evolved regions \citep[e.g.,][]{2018PASJ...70S..46F}, the pronounced internal complexity and clear embedding of the red component within the shell observed here appear unique, offering a direct view of such nested filamentary configurations within a diffuse, high-latitude cloud.

Although several filaments of the red component (D$_2$) connect to the shell, it fills only a portion of the cavity volume, and direct kinematically continuous connections between the two components remain localized. We speculate that this red component may have physically impacted the natal blue cloud, a process that could have initiated before the gas fully transitioned into the molecular phase. Under this plausible interpretation, the red cloud might have collided with—or been expelled from—the region that later formed the bluer shell, potentially helping to carve out the cavity observed today. At the localized interfaces where the red filaments appear to contact the shell, localized compression could enhance the gas density, reinforcing the intricate filamentary network that remains clearly resolved in our Delingha $^{12}$CO observations. Similar filament--shell interactions have been characterized in dense star-forming regions such as Sgr~B2 \citep{2020MNRAS.499.4918A} and in diffuse atomic clouds associated with supershells like GS040 \citep{2016ApJ...827L..27P}; however, capturing this specific configuration in a young, quiescent molecular environment is unprecedented.

We suggest that complex~D can be viewed as a nested kinematic system where the intruding red cloud appears to actively push against and shape the interior of the surrounding bluer shell. These observations provide a detailed view of the internal structure of complex~D, illustrating how early-phase cloud collisions or macroscopic gas dynamics might generate complex filamentary patterns in the interstellar medium prior to the onset of star formation. A comprehensive numerical analysis to evaluate the exact driving mechanisms behind this structure is beyond the scope of the present work.

\subsection{Starless nature of Polaris Flare}\label{sec_starless}
From the AllWISE catalog of young stellar object (YSO) candidates \citep{2016MNRAS.458.3479M}, we identified two Class~I/II sources within the Polaris Flare region, hereafter designated YSO1 and YSO2 (Figure~\ref{fig:subpart}). To further assess their evolutionary stages, we placed these sources on the WISE color--color diagram \citep{2014ApJ...791..131K}, using the standard [3.4]--[4.6] versus [4.6]--[12] color scheme to distinguish between Class~I and Class~II YSOs (Figure~\ref{fig:yso}). YSO1 falls firmly within the Class~II region, suggesting a relatively evolved object with a substantial circumstellar disk, while YSO2 lies close to the Class~I/II boundary, such that its classification as a YSO is itself uncertain.  
YSO1 is projected onto a CO-bright portion of Complex~AW, coinciding with regions of dense molecular gas, whereas YSO2 lies near the periphery of the CO emission, where the molecular material is more diffuse. Based on their colors and spatial distribution, one might initially consider them as candidate members of the Polaris Flare.  
However, Gaia~DR3 data \citep{2023A&A...674A...1G} indicate a parallax of 2.85~mas\footnote{Parallax value from \url{https://gea.esac.esa.int/archive/}} for YSO1, corresponding to a distance of approximately 350~pc. Given that the Polaris Flare is located at $\sim$150~pc, YSO1 is clearly a background object. For YSO2, its marginal color--color position, together with the lack of spatial or kinematic association with CO, argues against its being either a bona fide YSO or a Polaris member. We therefore conclude that both YSO1 and YSO2 are very likely unrelated background sources projected along the line of sight, and that the Polaris Flare shows no convincing evidence of embedded YSOs.  

\begin{figure}
    \centering
    \includegraphics[width=0.95\linewidth]{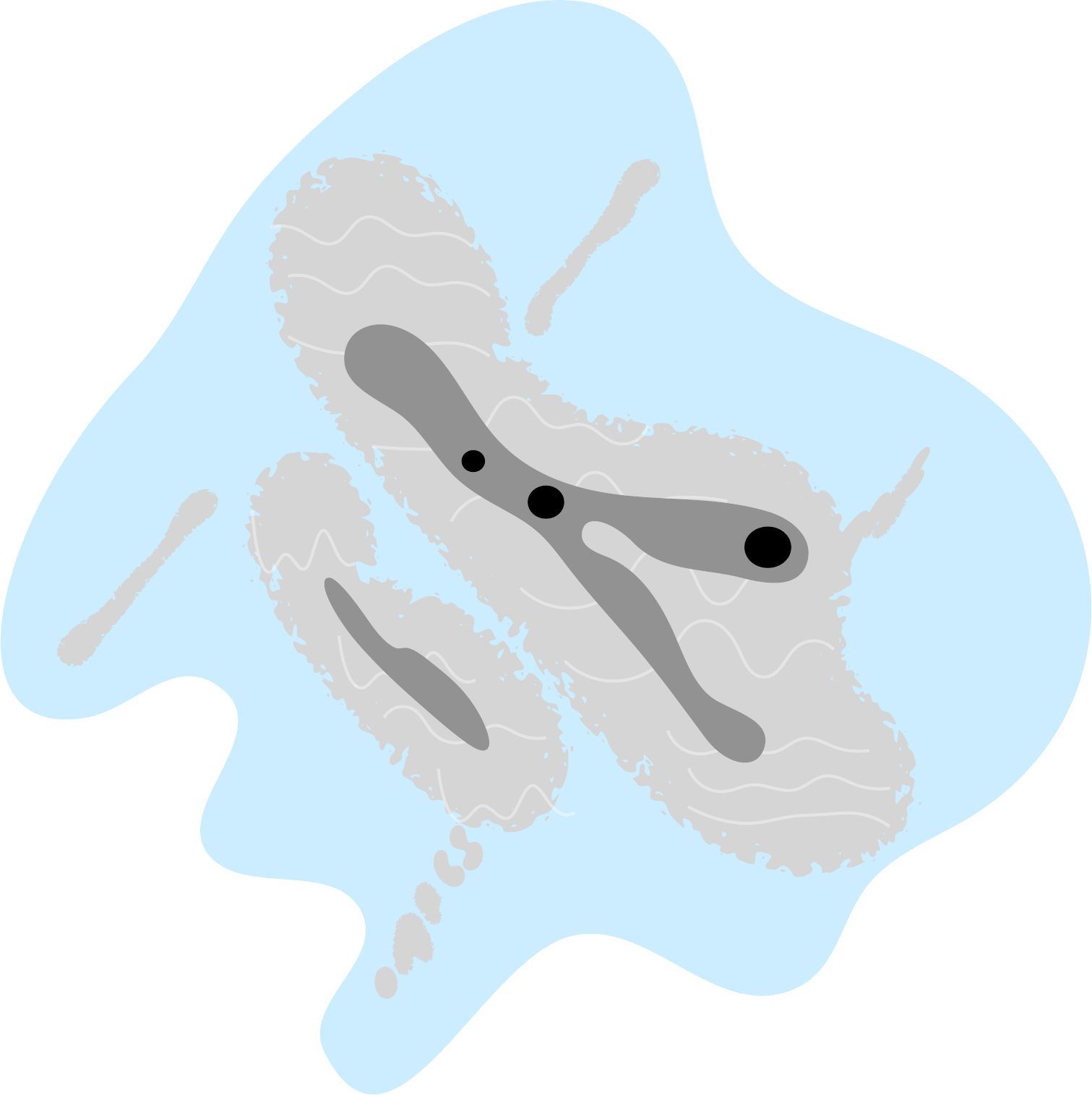}
\caption{Hierarchical structure of a young molecular cloud, inspired by the Polaris Flare analysis (Sect.~\ref{sec_glimpse}). A common natal atomic cloud is represented in blue. The $^{12}$CO periphery (light gray), developing within the natal atomic medium, represents a dynamically assembling and dispersing envelope layer (Sect.~\ref{sec_peakintensity}).
The relatively stable kernel of the molecular cloud is traced by $^{13}$CO, forming a prominent filamentary backbone (Sect.~\ref{sec_codistri}), while localized, gravitationally bound pre-stellar cores are traced by C$^{18}$O (Sect.~\ref{sec_C18O}). White curves highlight ripples and fine striations resolved within the $^{12}$CO periphery. Slender $^{12}$CO structures oriented perpendicular to the main complexes are also indicated, which are potentially shaped by large-scale, overarching dynamic processes (Sect.~\ref{sec_lvbv}).}
    \label{fig:diagram}
\end{figure}

The unmatching of YSOs is consistent with the overall starless nature of the Polaris Flare. Previous studies have consistently reported that the region exhibits little to no current star formation activity. Observations by the \textit{Herschel} Space Observatory revealed a network of filamentary structures with over 300 compact, mostly gravitationally unbound starless cores, indicating that star formation has not yet commenced \citep{2010A&A...518L..92W,2010A&A...518L.102A}. These findings underscore the quiescent nature of the Polaris Flare and are consistent with our Gaia-based conclusion that the candidate YSOs are likely background sources.
High-resolution single-dish observations further support this picture. Mapping of the densest region of the Polaris cloud with the IRAM 30-meter telescope revealed a network of filamentary structures without any gravitationally bound starless cores. Similarly, Nobeyama 45-meter telescope observations \citep{2025ApJ...981..158S} identified dense clumps, but these structures do not exhibit the characteristics of starless cores, such as gravitational binding or signs of active star formation. Together, these results confirm that the Polaris Flare is a quiescent molecular cloud, lacking the conditions necessary for ongoing star formation.
However, a distinct structural hierarchy is already well-established within such a young molecular cloud, as schematically depicted in Figure~\ref{fig:diagram} based on our comprehensive analysis of the Polaris Flare CO data cubes (Sect.~\ref{sec_glimpse}).

\subsection{Scientific topics enabled by PPCOS}\label{sec_topics}
The PMO Polaris CO survey, with its significantly improved spectral line information compared to previous surveys towards Polaris Flare that were constrained by limited spectral and spatial resolution as well as sparse spatial sampling \citep{1990ApJ...353L..49H,1993A&A...268..265H}, offers a valuable opportunity to advance our understanding of the Polaris Flare. Among the possible further scientific explorations enabled by this dataset, investigations of turbulent dynamics and the early stages of cloud evolution are of particular interest.

The 100 deg$^2$ coverage of this survey provides a unique opportunity to systematically explore turbulence-related statistics. In particular, analyses such as the $\Delta$-variance, column density PDFs, and velocity increment PDFs characterized by Normal Inverse Gaussian (NIG) profiles \citep[e.g.,][]{1998A&A...336..697S,2004EPJB...41..345B,2016A&A...590A.104O,2025arXiv250220458L} will enable a quantitative investigation of the turbulent properties of the Polaris Flare. These tools can help assess the role of possible supersonic shocks and turbulent intermittency in shaping the structure of high-Galactic-latitude, early-phase clouds. Related topics include understanding how turbulent energy in such a starless system is maintained and cascades to scales below 0.1 pc—the possible universal scale of filaments—as well as how the molecular regions of the Polaris Flare are confined, whether by external pressure, magnetic fields, turbulence, or converging flows.

By comparing the CO distribution from this survey with dust emission traced by the \textit{Planck} continuum \citep{2016A&A...594A..28P} and \ion{H}{I} data from surveys such as HI4PI \citep{2016A&A...594A.116H}, it is possible to quantitatively separate the dust components associated with atomic and molecular gas. This, in turn, enables a more precise determination of the $X_{\rm CO}$ factor in high-Galactic-latitude clouds \citep[e.g.,][]{2000ApJ...535..211I,2022A&A...668L...9M}. Importantly, the present work constitutes the first large-area CO survey of a high-Galactic-latitude cloud at sub-arcminute resolution. Previous CO surveys have concentrated mainly on the Galactic plane, with Taurus being the only exception covering $\sim$100~deg$^2$. However, Taurus lies at relatively low Galactic latitude ($-20\degr \lesssim b \lesssim -10\degr$) and may represent an intermediate case between the Polaris Flare and typical Galactic plane clouds\footnote{Adopting an off-Galactic-plane vertical height of the Sun of 20~pc \citep{10.1093/mnras/sty2813}, the vertical height $Z$ of Taurus is $\sim -15$~pc, much smaller in absolute value than that of the Polaris Flare ($\sim 85$~pc).}. By jointly comparing the Polaris Flare, Taurus, and Galactic plane molecular clouds, it becomes possible to characterize the conversion of \ion{H}{I} into the earliest molecular structures.

In addition to the topics discussed above, another advantage of this survey is that it provides information for both $^{12}$CO and $^{13}$CO, enabling studies of the excitation state and optical depth of CO, as well as more precise calculations of the CO column density. By adopting methods such as 3D reconstruction and radiative transfer solutions \citep{2020A&A...644A.151J,2020A&A...644A..27B,2025arXiv250319259L}, it becomes possible to model the excitation conditions within the Polaris Flare accurately. Consequently, the column density PDF and $X_{\rm CO}$ factor can be better constrained, offering improved insights into the turbulent structures and early evolution of high-Galactic-latitude clouds.

\section{Summary}\label{sec_summary}
We present the PMO Polaris CO survey (PPCOS), the first sub-arcminute, large-area CO survey, covering 100~deg$^2$ of the Polaris Flare in the $J=1\!-\!0$ transitions of $^{12}$CO, $^{13}$CO, and C$^{18}$O with the Delingha 13.7~m telescope, providing unprecedented, high-quality wide-field data at sub-arcminute angular resolution. Observations, conducted from 2012 over 10 years, cover more than 400 cells of $30\arcmin\times 30\arcmin$ each through over 2000 OTF scans, totaling more than 2000 hours of telescope time including overhead. The survey achieves sensitivities of $\sim0.46$~K for $^{12}$CO and $\sim0.23$~K for $^{13}$CO and C$^{18}$O at a spectral resolution of $\sim0.16$~km~s$^{-1}$ and spatial resolution of $\sim50\arcsec$. 

The CO flux measured by PPCOS is $\sim$30\% higher than that reported by DHT16 using the CfA 1.2\,m telescope, but is consistent within $\sim$10\% with measurements from IRAM and FCRAO over the Polaris cloud. The $\sim$30\% discrepancy with CfA likely arises from factors such as the larger beam sidelobes of the 1.2\,m telescope, differences in main-beam efficiency calibration, or variations in atmospheric correction. The average line profiles show general agreement across different instruments. We conclude that the PPCOS flux is well calibrated, with a systematic intensity uncertainty of at most $\sim$10\%.

By smoothing to 1.5\arcmin, we obtained merged cubes with even lower noise (0.21~K for $^{12}$CO and 0.11~K for $^{13}$CO and C$^{18}$O), enabling rapid, large-scale exploration of the cloud.
The main findings, based on the merged cubes, include:
\begin{itemize}

    \item[1.] $^{12}$CO unveils intricate large-scale striations and wavy substructures across the cloud, while $^{13}$CO and C$^{18}$O trace the main cloud skeleton and highlight dense regions that may host prestellar cores. The bright C$^{18}$O spots likely lie close to virial equilibrium, embedded within a turbulent natal cloud that may not be gravitationally bound, consistent with previous studies.
    No WISE young stars are firmly associated with the Polaris Flare, as the closest YSO candidate is excluded by Gaia distance measurements.
    \item[2.] The cloud is divided into seven complexes (denoted as AN, AC, AW, BE, BW, C and D); most show single-peaked spectra, except Complex D, which exhibits a shell-like blue component and a red component within the cavity. 
    We suggest that Complex D represents a nested system in which the red cloud has actively pushed against and impacted the bluer shell.
    Different complexes are morphologically and kinematically connected, forming a single turbulent system
    \item[3]
    Only about 10\% of the pixels in the Polaris Flare exhibit multiple velocity components, and the global velocity gradient is measured to be 0.18~km~s$^{-1}$~pc$^{-1}$.  
The $^{12}$CO line widths (FWHM) are typically $1.2 \pm 0.6$~km~s$^{-1}$, while $^{13}$CO lines are significantly narrower, generally $\le 0.7\,\Delta V_{\rm ^{12}CO}$.  
\item[4]
Both the complexes and the surrounding flare gas appear to be regulated by the global velocity gradient and can be divided into two groups: a major group, comprising complexes~A and B, which extend along the global velocity gradient; and a secondary group, consisting of slim filaments and complexes~C and D, which form vertical structures elongated perpendicular to the global velocity gradient.  
This suggests that large-scale dynamical processes play a dominant role in shaping the morphology of the Polaris Flare.
\item[5] 
The $^{12}$CO peak intensities in Polaris reach 2.6, 4.7, and 5.3~K at the 60th, 90th, and 95th percentiles. Complexes AC and AW are dominated by compact condensations with $T_{\rm MB} > 3$~K, tightly enclosed within the 1.5~K isosurface, whereas complexes AN and B are more diffuse, with the 3~K isosurface covering only about half the volume enclosed by the 1.5~K surface, indicating extended gas surrounding the denser regions. The 6~K isosurface highlights the densest condensations and corresponds closely to the C$^{18}$O detection region.

\item[6] Based on a comprehensive analysis of morphology, kinematics, and isotopologue stratification, we propose that molecular clouds such as the Polaris Flare follow a three-layer hierarchy: a dynamically assembling and dispersing periphery traced by $^{12}$CO, a more stable and partly bound kernel traced by $^{13}$CO, and gravitationally bound compact cores traced by C$^{18}$O. Different complexes may have originated from the same parental cloud, sharing similar peripheries.

\end{itemize}

Overall, Polaris Flare is a young, simple, and quiescent cloud, ideal for studying turbulence and early cloud evolution.
PPCOS provides an unprecedented large-area view of the Polaris Flare, with dynamic information, establishing it as a benchmark for statistical studies of these topics.

\begin{acknowledgement}
X.L. acknowledges the support of the Strategic Priority Research Program of the Chinese Academy of Sciences  under Grant No. XDB0800303.
T.Z. acknowledges the Leading Innovation and Entrepreneurship Team of Zhejiang Province of China (Grant No. 2023R01008).
This research was carried out in part at the Jet Propulsion Laboratory, which is operated by the California Institute of Technology under a contract with the National Aeronautics and Space Administration (80NM0018D0004).
We thank the staff of the Delingha Observatory for their continuous efforts in carrying out the observations.
We thank Volker Ossenkopf-Okada for kindly sharing the 
$^{13}$CO moment~0 map of the Polaris cloud obtained with FCRAO, 
for comparison with the Delingha observations.
\end{acknowledgement}

\bibliography{PMOPolarisSurvey}

@BOOK{Spitzer1978,
  author    = {Spitzer, Lyman},
  title     = {Physical Processes in the Interstellar Medium},
  publisher = {Wiley-Interscience},
  address   = {New York},
  year      = {1978},
  doi       = {10.1002/9783527617722},
  url       = {https://ui.adsabs.harvard.edu/abs/1978ppim.book.....S},
  note      = {Cited by over 5,000 times as of 2025}
}

@ARTICLE{1979ApJ...232L..89S,
       author = {{Solomon}, P.~M. and {Scoville}, N.~Z. and {Sanders}, D.~B.},
        title = "{Giant molecular clouds in the Galaxy: the distribution of $^{13}$CO emission in the galactic plane.}",
      journal = {\apjl},
         year = 1979,
        month = sep,
       volume = {232},
        pages = {L89-L93},
          doi = {10.1086/183042},
       adsurl = {https://ui.adsabs.harvard.edu/abs/1979ApJ...232L..89S}
}

@ARTICLE{1998ApJ...498..541K,
       author = {{Kennicutt}, Jr., Robert C.},
        title = "{The Global Schmidt Law in Star-forming Galaxies}",
      journal = {\apj},
         year = 1998,
        month = may,
       volume = {498},
       number = {2},
        pages = {541-552},
          doi = {10.1086/305588},
archivePrefix = {arXiv},
       eprint = {astro-ph/9712213},
 primaryClass = {astro-ph},
       adsurl = {https://ui.adsabs.harvard.edu/abs/1998ApJ...498..541K}
}

@ARTICLE{1977ApJ...216..291S,
       author = {{Savage}, B.~D. and {Bohlin}, R.~C. and {Drake}, J.~F. and {Budich}, W.},
        title = "{A survey of interstellar molecular hydrogen. I.}",
      journal = {\apj},
         year = 1977,
        month = aug,
       volume = {216},
        pages = {291-307},
          doi = {10.1086/155471},
       adsurl = {https://ui.adsabs.harvard.edu/abs/1977ApJ...216..291S}
}

@ARTICLE{2017ApJ...841...25G,
       author = {{Goldsmith}, Paul F. and {Kauffmann}, Jens},
        title = "{Electron Excitation of High Dipole Moment Molecules Re-examined}",
      journal = {\apj},
         year = 2017,
        month = may,
       volume = {841},
       number = {1},
          eid = {25},
        pages = {25},
          doi = {10.3847/1538-4357/aa6f12},
archivePrefix = {arXiv},
       eprint = {1708.07553},
 primaryClass = {astro-ph.GA},
       adsurl = {https://ui.adsabs.harvard.edu/abs/2017ApJ...841...25G}
}

@ARTICLE{2009ApJ...693..216K,
       author = {{Krumholz}, Mark R. and {McKee}, Christopher F. and {Tumlinson}, Jason},
        title = "{The Atomic-to-Molecular Transition in Galaxies. II: H I and H$_{2}$ Column Densities}",
      journal = {\apj},
         year = 2009,
        month = mar,
       volume = {693},
       number = {1},
        pages = {216-235},
          doi = {10.1088/0004-637X/693/1/216},
archivePrefix = {arXiv},
       eprint = {0811.0004},
 primaryClass = {astro-ph},
       adsurl = {https://ui.adsabs.harvard.edu/abs/2009ApJ...693..216K}
}

@ARTICLE{2001ApJ...547..792D,
       author = {{Dame}, T.~M. and {Hartmann}, Dap and {Thaddeus}, P.},
        title = "{The Milky Way in Molecular Clouds: A New Complete CO Survey}",
      journal = {\apj},
         year = 2001,
        month = feb,
       volume = {547},
       number = {2},
        pages = {792-813},
          doi = {10.1086/318388},
archivePrefix = {arXiv},
       eprint = {astro-ph/0009217},
 primaryClass = {astro-ph},
       adsurl = {https://ui.adsabs.harvard.edu/abs/2001ApJ...547..792D}
}

@ARTICLE{2015ARA&A..53..583H,
       author = {{Heyer}, Mark and {Dame}, T.~M.},
        title = "{Molecular Clouds in the Milky Way}",
      journal = {\araa},
         year = 2015,
        month = aug,
       volume = {53},
        pages = {583-629},
          doi = {10.1146/annurev-astro-082214-122324},
       adsurl = {https://ui.adsabs.harvard.edu/abs/2015ARA&A..53..583H}
}

@INPROCEEDINGS{1993AIPC..278..279H,
       author = {{Hartmann}, Dap and {Burton}, W.~B.},
        title = "{The Leiden/Dwingeloo survey of galactic HI at {\ensuremath{\delta}}>=-30{\textdegree}}",
    booktitle = {Back to the Galaxy},
         year = 1993,
       editor = {{Holt}, Stephen S. and {Verter}, Francis},
       series = {American Institute of Physics Conference Series},
       volume = {278},
        month = jan,
        pages = {279-282},
          doi = {10.1063/1.43986},
       adsurl = {https://ui.adsabs.harvard.edu/abs/1993AIPC..278..279H}
}

@ARTICLE{2016A&A...594A.116H,
       author = {{HI4PI Collaboration} and {Ben Bekhti}, N. and {Fl{\"o}er}, L. and {Keller}, R. and {Kerp}, J. and {Lenz}, D. and {Winkel}, B. and {Bailin}, J. and {Calabretta}, M.~R. and {Dedes}, L. and {Ford}, H.~A. and {Gibson}, B.~K. and {Haud}, U. and {Janowiecki}, S. and {Kalberla}, P.~M.~W. and {Lockman}, F.~J. and {McClure-Griffiths}, N.~M. and {Murphy}, T. and {Nakanishi}, H. and {Pisano}, D.~J. and {Staveley-Smith}, L.},
        title = "{HI4PI: A full-sky H I survey based on EBHIS and GASS}",
      journal = {\aap},
         year = 2016,
        month = oct,
       volume = {594},
          eid = {A116},
        pages = {A116},
          doi = {10.1051/0004-6361/201629178},
archivePrefix = {arXiv},
       eprint = {1610.06175},
 primaryClass = {astro-ph.GA},
       adsurl = {https://ui.adsabs.harvard.edu/abs/2016A&A...594A.116H}
}

@ARTICLE{1954BAN....12..117V,
       author = {{van de Hulst}, H.~C. and {Muller}, C.~A. and {Oort}, J.~H.},
        title = "{The spiral structure of the outer part of the Galactic System derived from the hydrogen emission at 21 cm wavelength}",
      journal = {\bain},
         year = 1954,
        month = may,
       volume = {12},
        pages = {117},
       adsurl = {https://ui.adsabs.harvard.edu/abs/1954BAN....12..117V}
}

@ARTICLE{2020A&A...639A..26K,
       author = {{Kalberla}, P.~M.~W. and {Kerp}, J. and {Haud}, U.},
        title = "{H I filaments are cold and associated with dark molecular gas. HI4PI-based estimates of the local diffuse CO-dark H$_{2}$ distribution}",
      journal = {\aap},
         year = 2020,
        month = jul,
       volume = {639},
          eid = {A26},
        pages = {A26},
          doi = {10.1051/0004-6361/202037602},
archivePrefix = {arXiv},
       eprint = {2004.14630},
 primaryClass = {astro-ph.GA},
       adsurl = {https://ui.adsabs.harvard.edu/abs/2020A&A...639A..26K}
}

@ARTICLE{2003ApJ...586.1067H,
       author = {{Heiles}, Carl and {Troland}, T.~H.},
        title = "{The Millennium Arecibo 21 Centimeter Absorption-Line Survey. II. Properties of the Warm and Cold Neutral Media}",
      journal = {\apj},
         year = 2003,
        month = apr,
       volume = {586},
       number = {2},
        pages = {1067-1093},
          doi = {10.1086/367828},
archivePrefix = {arXiv},
       eprint = {astro-ph/0207105},
 primaryClass = {astro-ph},
       adsurl = {https://ui.adsabs.harvard.edu/abs/2003ApJ...586.1067H}
}

@ARTICLE{2025NatAs.tmp..148L,
       author = {{Liu}, Xunchuan and {Liu}, Tie and {Li}, Pak-Shing and {Mai}, Xiaofeng and {Henkel}, Christian and {Goldsmith}, Paul F. and {Qin}, Sheng-Li and {Gong}, Yan and {Lu}, Xing and {Xu}, Fengwei and {Luo}, Qiuyi and {Liu}, Hong-Li and {Zhang}, Tianwei and {Cheng}, Yu and {Di}, Yihuan and {Wu}, Yuefang and {Gu}, Qilao and {Tang}, Ningyu and {Yang}, Aiyuan and {Shen}, Zhiqiang},
        title = "{A network of velocity-coherent filaments formed by supersonic turbulence in a very-high-velocity H I cloud}",
      journal = {Nature Astronomy},
         year = 2025,
        month = jul,
          doi = {10.1038/s41550-025-02605-8},
archivePrefix = {arXiv},
       eprint = {2502.10897},
 primaryClass = {astro-ph.GA},
       adsurl = {https://ui.adsabs.harvard.edu/abs/2025NatAs.tmp..148L}
}

@ARTICLE{2018MNRAS.474..289W,
       author = {{Westmeier}, Tobias},
        title = "{A new all-sky map of Galactic high-velocity clouds from the 21-cm HI4PI survey}",
      journal = {\mnras},
         year = 2018,
        month = feb,
       volume = {474},
       number = {1},
        pages = {289-299},
          doi = {10.1093/mnras/stx2757},
archivePrefix = {arXiv},
       eprint = {1712.00909},
 primaryClass = {astro-ph.GA},
       adsurl = {https://ui.adsabs.harvard.edu/abs/2018MNRAS.474..289W}
}

@ARTICLE{2006ApJS..163..145J,
       author = {{Jackson}, J.~M. and {Rathborne}, J.~M. and {Shah}, R.~Y. and {Simon}, R. and {Bania}, T.~M. and {Clemens}, D.~P. and {Chambers}, E.~T. and {Johnson}, A.~M. and {Dormody}, M. and {Lavoie}, R. and {Heyer}, M.~H.},
        title = "{The Boston University-Five College Radio Astronomy Observatory Galactic Ring Survey}",
      journal = {\apjs},
         year = 2006,
        month = mar,
       volume = {163},
       number = {1},
        pages = {145-159},
          doi = {10.1086/500091},
archivePrefix = {arXiv},
       eprint = {astro-ph/0602160},
 primaryClass = {astro-ph},
       adsurl = {https://ui.adsabs.harvard.edu/abs/2006ApJS..163..145J}
}

@ARTICLE{2015ApJ...812....6B,
       author = {{Barnes}, Peter J. and {Muller}, Erik and {Indermuehle}, Balthasar and {O'Dougherty}, Stefan N. and {Lowe}, Vicki and {Cunningham}, Maria and {Hernandez}, Audra K. and {Fuller}, Gary A.},
        title = "{The Three-mm Ultimate Mopra Milky Way Survey. I. Survey Overview, Initial Data Releases, and First Results}",
      journal = {\apj},
         year = 2015,
        month = oct,
       volume = {812},
       number = {1},
          eid = {6},
        pages = {6},
          doi = {10.1088/0004-637X/812/1/6},
archivePrefix = {arXiv},
       eprint = {1507.05095},
 primaryClass = {astro-ph.GA},
       adsurl = {https://ui.adsabs.harvard.edu/abs/2015ApJ...812....6B}
}

@ARTICLE{2019ApJS..240....9S,
       author = {{Su}, Yang and {Yang}, Ji and {Zhang}, Shaobo and {Gong}, Yan and {Wang}, Hongchi and {Zhou}, Xin and {Wang}, Min and {Chen}, Zhiwei and {Sun}, Yan and {Chen}, Xuepeng and {Xu}, Ye and {Jiang}, Zhibo},
        title = "{The Milky Way Imaging Scroll Painting (MWISP): Project Details and Initial Results from the Galactic Longitudes of 25.{\textdegree}8-49.{\textdegree}7}",
      journal = {\apjs},
         year = 2019,
        month = jan,
       volume = {240},
       number = {1},
          eid = {9},
        pages = {9},
          doi = {10.3847/1538-4365/aaf1c8},
archivePrefix = {arXiv},
       eprint = {1901.00285},
 primaryClass = {astro-ph.GA},
       adsurl = {https://ui.adsabs.harvard.edu/abs/2019ApJS..240....9S}
}

@ARTICLE{2017PASJ...69...78U,
       author = {{Umemoto}, Tomofumi and {Minamidani}, Tetsuhiro and {Kuno}, Nario and {Fujita}, Shinji and {Matsuo}, Mitsuhiro and {Nishimura}, Atsushi and {Torii}, Kazufumi and {Tosaki}, Tomoka and {Kohno}, Mikito and {Kuriki}, Mika and {Tsuda}, Yuya and {Hirota}, Akihiko and {Ohashi}, Satoshi and {Yamagishi}, Mitsuyoshi and {Handa}, Toshihiro and {Nakanishi}, Hiroyuki and {Omodaka}, Toshihiro and {Koide}, Nagito and {Matsumoto}, Naoko and {Onishi}, Toshikazu and {Tokuda}, Kazuki and {Seta}, Masumichi and {Kobayashi}, Yukinori and {Tachihara}, Kengo and {Sano}, Hidetoshi and {Hattori}, Yusuke and {Onodera}, Sachiko and {Oasa}, Yumiko and {Kamegai}, Kazuhisa and {Tsuboi}, Masato and {Sofue}, Yoshiaki and {Higuchi}, Aya E. and {Chibueze}, James O. and {Mizuno}, Norikazu and {Honma}, Mareki and {Muller}, Erik and {Inoue}, Tsuyoshi and {Morokuma-Matsui}, Kana and {Shinnaga}, Hiroko and {Ozawa}, Takeaki and {Takahashi}, Ryo and {Yoshiike}, Satoshi and {Costes}, Jean and {Kuwahara}, Sho},
        title = "{FOREST unbiased Galactic plane imaging survey with the Nobeyama 45 m telescope (FUGIN). I. Project overview and initial results}",
      journal = {\pasj},
         year = 2017,
        month = oct,
       volume = {69},
       number = {5},
          eid = {78},
        pages = {78},
          doi = {10.1093/pasj/psx061},
archivePrefix = {arXiv},
       eprint = {1707.05981},
 primaryClass = {astro-ph.GA},
       adsurl = {https://ui.adsabs.harvard.edu/abs/2017PASJ...69...78U}
}

@ARTICLE{1985ApJ...295..402M,
       author = {{Magnani}, L. and {Blitz}, L. and {Mundy}, L.},
        title = "{Molecular gas at high galactic latitudes.}",
      journal = {\apj},
         year = 1985,
        month = aug,
       volume = {295},
        pages = {402-421},
          doi = {10.1086/163385},
       adsurl = {https://ui.adsabs.harvard.edu/abs/1985ApJ...295..402M}
}

@ARTICLE{2008ApJ...680..428G,
       author = {{Goldsmith}, Paul F. and {Heyer}, Mark and {Narayanan}, Gopal and {Snell}, Ronald and {Li}, Di and {Brunt}, Chris},
        title = "{Large-Scale Structure of the Molecular Gas in Taurus Revealed by High Linear Dynamic Range Spectral Line Mapping}",
      journal = {\apj},
         year = 2008,
        month = jun,
       volume = {680},
       number = {1},
        pages = {428-445},
          doi = {10.1086/587166},
archivePrefix = {arXiv},
       eprint = {0802.2206},
 primaryClass = {astro-ph},
       adsurl = {https://ui.adsabs.harvard.edu/abs/2008ApJ...680..428G}
}

@ARTICLE{1990ApJ...353L..49H,
       author = {{Heithausen}, Andreas and {Thaddeus}, Patrick},
        title = "{The Polaris Flare: Extensive Molecular Gas near the North Celestial Pole}",
      journal = {\apjl},
         year = 1990,
        month = apr,
       volume = {353},
        pages = {L49},
          doi = {10.1086/185705},
       adsurl = {https://ui.adsabs.harvard.edu/abs/1990ApJ...353L..49H}
}

@ARTICLE{2010A&A...518L.104M,
       author = {{Miville-Desch{\^e}nes}, M. -A. and {Martin}, P.~G. and {Abergel}, A. and {Bernard}, J. -P. and {Boulanger}, F. and {Lagache}, G. and {Anderson}, L.~D. and {Andr{\'e}}, P. and {Arab}, H. and {Baluteau}, J. -P. and {Blagrave}, K. and {Bontemps}, S. and {Cohen}, M. and {Compiegne}, M. and {Cox}, P. and {Dartois}, E. and {Davis}, G. and {Emery}, R. and {Fulton}, T. and {Gry}, C. and {Habart}, E. and {Huang}, M. and {Joblin}, C. and {Jones}, S.~C. and {Kirk}, J. and {Lim}, T. and {Madden}, S. and {Makiwa}, G. and {Menshchikov}, A. and {Molinari}, S. and {Moseley}, H. and {Motte}, F. and {Naylor}, D.~A. and {Okumura}, K. and {Pinheiro Gon{\c{c}}alves}, D. and {Polehampton}, E. and {Rod{\'o}n}, J.~A. and {Russeil}, D. and {Saraceno}, P. and {Schneider}, N. and {Sidher}, S. and {Spencer}, L. and {Swinyard}, B. and {Ward-Thompson}, D. and {White}, G.~J. and {Zavagno}, A.},
        title = "{Herschel-SPIRE observations of the Polaris flare: Structure of the diffuse interstellar medium at the sub-parsec scale}",
      journal = {\aap},
         year = 2010,
        month = jul,
       volume = {518},
          eid = {L104},
        pages = {L104},
          doi = {10.1051/0004-6361/201014678},
archivePrefix = {arXiv},
       eprint = {1005.2746},
 primaryClass = {astro-ph.GA},
       adsurl = {https://ui.adsabs.harvard.edu/abs/2010A&A...518L.104M}
}

@ARTICLE{2010A&A...518L.102A,
       author = {{Andr{\'e}}, Ph. and {Men'shchikov}, A. and {Bontemps}, S. and {K{\"o}nyves}, V. and {Motte}, F. and {Schneider}, N. and {Didelon}, P. and {Minier}, V. and {Saraceno}, P. and {Ward-Thompson}, D. and {di Francesco}, J. and {White}, G. and {Molinari}, S. and {Testi}, L. and {Abergel}, A. and {Griffin}, M. and {Henning}, Th. and {Royer}, P. and {Mer{\'\i}n}, B. and {Vavrek}, R. and {Attard}, M. and {Arzoumanian}, D. and {Wilson}, C.~D. and {Ade}, P. and {Aussel}, H. and {Baluteau}, J. -P. and {Benedettini}, M. and {Bernard}, J. -Ph. and {Blommaert}, J.~A.~D.~L. and {Cambr{\'e}sy}, L. and {Cox}, P. and {di Giorgio}, A. and {Hargrave}, P. and {Hennemann}, M. and {Huang}, M. and {Kirk}, J. and {Krause}, O. and {Launhardt}, R. and {Leeks}, S. and {Le Pennec}, J. and {Li}, J.~Z. and {Martin}, P.~G. and {Maury}, A. and {Olofsson}, G. and {Omont}, A. and {Peretto}, N. and {Pezzuto}, S. and {Prusti}, T. and {Roussel}, H. and {Russeil}, D. and {Sauvage}, M. and {Sibthorpe}, B. and {Sicilia-Aguilar}, A. and {Spinoglio}, L. and {Waelkens}, C. and {Woodcraft}, A. and {Zavagno}, A.},
        title = "{From filamentary clouds to prestellar cores to the stellar IMF: Initial highlights from the Herschel Gould Belt Survey}",
      journal = {\aap},
         year = 2010,
        month = jul,
       volume = {518},
          eid = {L102},
        pages = {L102},
          doi = {10.1051/0004-6361/201014666},
archivePrefix = {arXiv},
       eprint = {1005.2618},
 primaryClass = {astro-ph.GA},
       adsurl = {https://ui.adsabs.harvard.edu/abs/2010A&A...518L.102A}
}

@ARTICLE{1998A&A...331..669F,
       author = {{Falgarone}, E. and {Panis}, J. -F. and {Heithausen}, A. and {Perault}, M. and {Stutzki}, J. and {Puget}, J. -L. and {Bensch}, F.},
        title = "{The IRAM key-project: small-scale structure of pre-star-forming regions. I. Observational results}",
      journal = {\aap},
         year = 1998,
        month = mar,
       volume = {331},
        pages = {669-696},
       adsurl = {https://ui.adsabs.harvard.edu/abs/1998A&A...331..669F}
}

@ARTICLE{1996A&A...313..929M,
       author = {{Meyerdierks}, H. and {Heithausen}, A.},
        title = "{Diffuse molecular gas in the Polaris flare.}",
      journal = {\aap},
         year = 1996,
        month = sep,
       volume = {313},
        pages = {929-937},
       adsurl = {https://ui.adsabs.harvard.edu/abs/1996A&A...313..929M}
}

@ARTICLE{1998A&A...333..709L,
       author = {{Lagache}, G. and {Abergel}, A. and {Boulanger}, F. and {Puget}, J. -L.},
        title = "{The interstellar cold dust observed by COBE}",
      journal = {\aap},
         year = 1998,
        month = may,
       volume = {333},
        pages = {709-720},
          doi = {10.48550/arXiv.astro-ph/9812474},
archivePrefix = {arXiv},
       eprint = {astro-ph/9812474},
 primaryClass = {astro-ph},
       adsurl = {https://ui.adsabs.harvard.edu/abs/1998A&A...333..709L}
}

@ARTICLE{1999A&A...347..640B,
       author = {{Bernard}, J.~P. and {Abergel}, A. and {Ristorcelli}, I. and {Pajot}, F. and {Torre}, J.~P. and {Boulanger}, F. and {Giard}, M. and {Lagache}, G. and {Serra}, G. and {Lamarre}, J.~M. and {Puget}, J.~L. and {Lepeintre}, F. and {Cambr{\'e}sy}, L.},
        title = "{PRONAOS observations of MCLD 123.5 + 24.9: cold dust in the Polaris cirrus cloud}",
      journal = {\aap},
         year = 1999,
        month = jul,
       volume = {347},
        pages = {640-649},
       adsurl = {https://ui.adsabs.harvard.edu/abs/1999A&A...347..640B}
}

@ARTICLE{2025ApJ...981..158S,
       author = {{Shimoikura}, Tomomi and {Dobashi}, Kazuhito and {Nakamura}, Fumitaka and {Taniguchi}, Kotomi},
        title = "{Velocity Structure and Molecular Formation in the Polaris Molecular Cloud}",
      journal = {\apj},
         year = 2025,
        month = mar,
       volume = {981},
       number = {2},
          eid = {158},
        pages = {158},
          doi = {10.3847/1538-4357/adb418},
archivePrefix = {arXiv},
       eprint = {2502.10668},
 primaryClass = {astro-ph.GA},
       adsurl = {https://ui.adsabs.harvard.edu/abs/2025ApJ...981..158S}
}

@ARTICLE{1999A&A...352..645Z,
       author = {{Zagury}, F. and {Boulanger}, F. and {Banchet}, V.},
        title = "{Optical images of MCLD123.5+24.9: a cloud illuminated by the North star?}",
      journal = {\aap},
         year = 1999,
        month = dec,
       volume = {352},
        pages = {645-658},
       adsurl = {https://ui.adsabs.harvard.edu/abs/1999A&A...352..645Z}
}

@ARTICLE{2002A&A...390..307O,
       author = {{Ossenkopf}, V. and {Mac Low}, M. -M.},
        title = "{Turbulent velocity structure in molecular clouds}",
      journal = {\aap},
         year = 2002,
        month = jul,
       volume = {390},
        pages = {307-326},
          doi = {10.1051/0004-6361:20020629},
       adsurl = {https://ui.adsabs.harvard.edu/abs/2002A&A...390..307O}
}

@ARTICLE{2006JKAS...39....9C,
       author = {{Chi}, Seung-Youp and {Park}, Yong-Sun},
        title = "{Molecular Line Observation Toward Polaris Flare}",
      journal = {Journal of Korean Astronomical Society},
         year = 2006,
        month = mar,
       volume = {39},
       number = {1},
        pages = {9-17},
          doi = {10.5303/JKAS.2006.39.1.009},
       adsurl = {https://ui.adsabs.harvard.edu/abs/2006JKAS...39....9C}
}

@ARTICLE{2010A&A...518L..92W,
       author = {{Ward-Thompson}, D. and {Kirk}, J.~M. and {Andr{\'e}}, P. and {Saraceno}, P. and {Didelon}, P. and {K{\"o}nyves}, V. and {Schneider}, N. and {Abergel}, A. and {Baluteau}, J. -P. and {Bernard}, J. -Ph. and {Bontemps}, S. and {Cambr{\'e}sy}, L. and {Cox}, P. and {di Francesco}, J. and {di Giorgio}, A.~M. and {Griffin}, M. and {Hargrave}, P. and {Huang}, M. and {Li}, J.~Z. and {Martin}, P. and {Men'shchikov}, A. and {Minier}, V. and {Molinari}, S. and {Motte}, F. and {Olofsson}, G. and {Pezzuto}, S. and {Russeil}, D. and {Sauvage}, M. and {Sibthorpe}, B. and {Spinoglio}, L. and {Testi}, L. and {White}, G. and {Wilson}, C. and {Woodcraft}, A. and {Zavagno}, A.},
        title = "{A Herschel study of the properties of starless cores in the Polaris Flare dark cloud region using PACS and SPIRE}",
      journal = {\aap},
         year = 2010,
        month = jul,
       volume = {518},
          eid = {L92},
        pages = {L92},
          doi = {10.1051/0004-6361/201014618},
archivePrefix = {arXiv},
       eprint = {1005.2519},
 primaryClass = {astro-ph.GA},
       adsurl = {https://ui.adsabs.harvard.edu/abs/2010A&A...518L..92W}
}

@ARTICLE{1993A&A...268..265H,
       author = {{Heithausen}, A. and {Stacy}, J.~G. and {de Vries}, H.~W. and {Mebold}, U. and {Thaddeus}, P.},
        title = "{A composite large-scale CO survey at high galactic latitudes in the second quadrant.}",
      journal = {\aap},
         year = 1993,
        month = feb,
       volume = {268},
        pages = {265-275},
       adsurl = {https://ui.adsabs.harvard.edu/abs/1993A&A...268..265H}
}

@ARTICLE{2014A&A...571A..11P,
       author = {{Planck Collaboration} and {Abergel}, A. and {Ade}, P.~A.~R. and {Aghanim}, N. and {Alves}, M.~I.~R. and {Aniano}, G. and {Armitage-Caplan}, C. and {Arnaud}, M. and {Ashdown}, M. and {Atrio-Barandela}, F. and {Aumont}, J. and {Baccigalupi}, C. and {Banday}, A.~J. and {Barreiro}, R.~B. and {Bartlett}, J.~G. and {Battaner}, E. and {Benabed}, K. and {Beno{\^\i}t}, A. and {Benoit-L{\'e}vy}, A. and {Bernard}, J. -P. and {Bersanelli}, M. and {Bielewicz}, P. and {Bobin}, J. and {Bock}, J.~J. and {Bonaldi}, A. and {Bond}, J.~R. and {Borrill}, J. and {Bouchet}, F.~R. and {Boulanger}, F. and {Bridges}, M. and {Bucher}, M. and {Burigana}, C. and {Butler}, R.~C. and {Cardoso}, J. -F. and {Catalano}, A. and {Chamballu}, A. and {Chary}, R. -R. and {Chiang}, H.~C. and {Chiang}, L. -Y. and {Christensen}, P.~R. and {Church}, S. and {Clemens}, M. and {Clements}, D.~L. and {Colombi}, S. and {Colombo}, L.~P.~L. and {Combet}, C. and {Couchot}, F. and {Coulais}, A. and {Crill}, B.~P. and {Curto}, A. and {Cuttaia}, F. and {Danese}, L. and {Davies}, R.~D. and {Davis}, R.~J. and {de Bernardis}, P. and {de Rosa}, A. and {de Zotti}, G. and {Delabrouille}, J. and {Delouis}, J. -M. and {D{\'e}sert}, F. -X. and {Dickinson}, C. and {Diego}, J.~M. and {Dole}, H. and {Donzelli}, S. and {Dor{\'e}}, O. and {Douspis}, M. and {Draine}, B.~T. and {Dupac}, X. and {Efstathiou}, G. and {En{\ss}lin}, T.~A. and {Eriksen}, H.~K. and {Falgarone}, E. and {Finelli}, F. and {Forni}, O. and {Frailis}, M. and {Fraisse}, A.~A. and {Franceschi}, E. and {Galeotta}, S. and {Ganga}, K. and {Ghosh}, T. and {Giard}, M. and {Giardino}, G. and {Giraud-H{\'e}raud}, Y. and {Gonz{\'a}lez-Nuevo}, J. and {G{\'o}rski}, K.~M. and {Gratton}, S. and {Gregorio}, A. and {Grenier}, I.~A. and {Gruppuso}, A. and {Guillet}, V. and {Hansen}, F.~K. and {Hanson}, D. and {Harrison}, D.~L. and {Helou}, G. and {Henrot-Versill{\'e}}, S. and {Hern{\'a}ndez-Monteagudo}, C. and {Herranz}, D. and {Hildebrandt}, S.~R. and {Hivon}, E. and {Hobson}, M. and {Holmes}, W.~A. and {Hornstrup}, A. and {Hovest}, W. and {Huffenberger}, K.~M. and {Jaffe}, A.~H. and {Jaffe}, T.~R. and {Jewell}, J. and {Joncas}, G. and {Jones}, W.~C. and {Juvela}, M. and {Keih{\"a}nen}, E. and {Keskitalo}, R. and {Kisner}, T.~S. and {Knoche}, J. and {Knox}, L. and {Kunz}, M. and {Kurki-Suonio}, H. and {Lagache}, G. and {L{\"a}hteenm{\"a}ki}, A. and {Lamarre}, J. -M. and {Lasenby}, A. and {Laureijs}, R.~J. and {Lawrence}, C.~R. and {Leonardi}, R. and {Le{\'o}n-Tavares}, J. and {Lesgourgues}, J. and {Levrier}, F. and {Liguori}, M. and {Lilje}, P.~B. and {Linden-V{\o}rnle}, M. and {L{\'o}pez-Caniego}, M. and {Lubin}, P.~M. and {Mac{\'\i}as-P{\'e}rez}, J.~F. and {Maffei}, B. and {Maino}, D. and {Mandolesi}, N. and {Maris}, M. and {Marshall}, D.~J. and {Martin}, P.~G. and {Mart{\'\i}nez-Gonz{\'a}lez}, E. and {Masi}, S. and {Massardi}, M. and {Matarrese}, S. and {Matthai}, F. and {Mazzotta}, P. and {McGehee}, P. and {Melchiorri}, A. and {Mendes}, L. and {Mennella}, A. and {Migliaccio}, M. and {Mitra}, S. and {Miville-Desch{\^e}nes}, M. -A. and {Moneti}, A. and {Montier}, L. and {Morgante}, G. and {Mortlock}, D. and {Munshi}, D. and {Murphy}, J.~A. and {Naselsky}, P. and {Nati}, F. and {Natoli}, P. and {Netterfield}, C.~B. and {N{\o}rgaard-Nielsen}, H.~U. and {Noviello}, F. and {Novikov}, D. and {Novikov}, I. and {Osborne}, S. and {Oxborrow}, C.~A. and {Paci}, F. and {Pagano}, L. and {Pajot}, F. and {Paladini}, R. and {Paoletti}, D. and {Pasian}, F. and {Patanchon}, G. and {Perdereau}, O. and {Perotto}, L. and {Perrotta}, F. and {Piacentini}, F. and {Piat}, M. and {Pierpaoli}, E. and {Pietrobon}, D. and {Plaszczynski}, S. and {Pointecouteau}, E. and {Polenta}, G. and {Ponthieu}, N. and {Popa}, L. and {Poutanen}, T. and {Pratt}, G.~W. and {Pr{\'e}zeau}, G. and {Prunet}, S. and {Puget}, J. -L. and {Rachen}, J.~P. and {Reach}, W.~T. and {Rebolo}, R. and {Reinecke}, M. and {Remazeilles}, M. and {Renault}, C. and {Ricciardi}, S. and {Riller}, T.},
        title = "{Planck 2013 results. XI. All-sky model of thermal dust emission}",
      journal = {\aap},
         year = 2014,
        month = nov,
       volume = {571},
          eid = {A11},
        pages = {A11},
          doi = {10.1051/0004-6361/201323195},
archivePrefix = {arXiv},
       eprint = {1312.1300},
 primaryClass = {astro-ph.GA},
       adsurl = {https://ui.adsabs.harvard.edu/abs/2014A&A...571A..11P}
}

@ARTICLE{2001A&A...366..636B,
       author = {{Bensch}, F. and {Stutzki}, J. and {Ossenkopf}, V.},
        title = "{Quantification of molecular cloud structure using the Delta -variance}",
      journal = {\aap},
         year = 2001,
        month = feb,
       volume = {366},
        pages = {636-650},
          doi = {10.1051/0004-6361:20000292},
       adsurl = {https://ui.adsabs.harvard.edu/abs/2001A&A...366..636B}
}

@ARTICLE{2016PASP..128j5003T,
       author = {{Tian}, J.~F. and {Deng}, L.~C. and {Zhang}, X.~B. and {Lu}, X.~M. and {Sun}, J.~J. and {Liu}, Q.~L. and {Zhou}, Q. and {Yan}, Z.~Z. and {Xin}, Y. and {Wang}, K. and {Jiang}, X.~J. and {Luo}, Z.~Q. and {Yang}, J.},
        title = "{Optical Observing Conditions at Delingha Station}",
      journal = {\pasp},
         year = 2016,
        month = oct,
       volume = {128},
       number = {968},
        pages = {105003},
          doi = {10.1088/1538-3873/128/968/105003},
archivePrefix = {arXiv},
       eprint = {1602.00838},
 primaryClass = {astro-ph.IM},
       adsurl = {https://ui.adsabs.harvard.edu/abs/2016PASP..128j5003T}
}

@ARTICLE{2004ChJAA...4..390Z,
       author = {{Zuo}, Ying-Xi and {Yang}, Ji and {Shi}, Sheng-Cai and {Chen}, Shan-Huai and {Pei}, Li-Ben and {Yao}, Qi-Jun and {Sun}, Jin-Jiang and {Lin}, Zhen-Hui},
        title = "{Upgrade Procedure for the Delingha 13.7-m Telescope}",
      journal = {\cjaa},
         year = 2004,
        month = aug,
       volume = {4},
        pages = {390-396},
          doi = {10.1088/1009-9271/4/4/390},
       adsurl = {https://ui.adsabs.harvard.edu/abs/2004ChJAA...4..390Z}
}

@INPROCEEDINGS{2012SPIE.8444E..4BY,
       author = {{Yang}, Dehua and {Zhang}, Yong and {Zhou}, Guohua and {Li}, Aihua and {Chen}, Kunxin and {Zhang}, Zhenchao and {Li}, Guoping and {Zuo}, Yingxi and {Xu}, Ye},
        title = "{An active surface upgrade for the Delingha 13.7-m Radio Telescope}",
    booktitle = {Ground-based and Airborne Telescopes IV},
         year = 2012,
       editor = {{Stepp}, Larry M. and {Gilmozzi}, Roberto and {Hall}, Helen J.},
       series = {Society of Photo-Optical Instrumentation Engineers (SPIE) Conference Series},
       volume = {8444},
        month = sep,
          eid = {84444B},
        pages = {84444B},
          doi = {10.1117/12.925789},
       adsurl = {https://ui.adsabs.harvard.edu/abs/2012SPIE.8444E..4BY}
}

@ARTICLE{2012ITTST...2..593S,
       author = {{Shan}, Wenlei and {Yang}, Ji and {Shi}, Shengcai and {Yao}, Qijun and {Zuo}, Yingxi and {Lin}, Zhenhui and {Chen}, Shanhuai and {Zhang}, Xuguo and {Duan}, Wenying and {Cao}, Aiqing and {Li}, Sheng and {Li}, Zhenqiang and {Liu}, Jie and {Zhong}, Jiaqiang},
        title = "{Development of Superconducting Spectroscopic Array Receiver: A Multibeam 2SB SIS Receiver for Millimeter-Wave Radio Astronomy}",
      journal = {IEEE Transactions on Terahertz Science and Technology},
         year = 2012,
        month = nov,
       volume = {2},
       number = {6},
        pages = {593-604},
          doi = {10.1109/TTHZ.2012.2213818},
       adsurl = {https://ui.adsabs.harvard.edu/abs/2012ITTST...2..593S}
}

@article{ZHENQIANG2021559,
  title = {The Operation of Multi-beam Receiver of Delingha 13.7m Telescope},
  journal = {Chinese Astronomy and Astrophysics},
  volume = {45},
  number = {4},
  pages = {559-586},
  year = {2021},
  issn = {0275-1062},
  doi = {https://doi.org/10.1016/j.chinastron.2021.11.008},
  url = {https://www.sciencedirect.com/science/article/pii/S0275106221000953},
  author = {{Li}, Zhenqiang and {Zhang}, Xuguo and {Li}, Jibin and {Sun}, Jixian and {Lu}, Dengrong and {Dong}, Shouhao and {Li}, Yapeng and {Li}, Shengxue}
}

@ARTICLE{2018AcASn..59....3S,
       author = {{Sun}, J.~X. and {Lu}, D.~R. and {Yang}, J. and {Su}, Y. and {Zhang}, S.~B. and {Zhou}, X. and {Lin}, Z.~H.},
        title = "{Spectral Line On-The-Fly Observing System of the Delingha 13.7 m Telescope}",
      journal = {Acta Astronomica Sinica},
         year = 2018,
        month = jan,
       volume = {59},
       number = {1},
          eid = {3},
        pages = {3},
       adsurl = {https://ui.adsabs.harvard.edu/abs/2018AcASn..59....3S}
}

@INPROCEEDINGS{2005sf2a.conf..721P,
       author = {{Pety}, J.},
        title = "{Successes of and Challenges to GILDAS, a State-of-the-Art Radioastronomy Toolkit}",
    booktitle = {SF2A-2005: Semaine de l'Astrophysique Francaise},
         year = 2005,
       editor = {{Casoli}, F. and {Contini}, T. and {Hameury}, J.~M. and {Pagani}, L.},
        month = dec,
        pages = {721},
       adsurl = {https://ui.adsabs.harvard.edu/abs/2005sf2a.conf..721P}
}

@ARTICLE{2022ApJ...935..167A,
       author = {{Astropy Collaboration} and {Price-Whelan}, Adrian M. and {Lim}, Pey Lian and {Earl}, Nicholas and {Starkman}, Nathaniel and {Bradley}, Larry and {Shupe}, David L. and {Patil}, Aarya A. and {Corrales}, Lia and {Brasseur}, C.~E. and {N{\"o}the}, Maximilian and {Donath}, Axel and {Tollerud}, Erik and {Morris}, Brett M. and {Ginsburg}, Adam and {Vaher}, Eero and {Weaver}, Benjamin A. and {Tocknell}, James and {Jamieson}, William and {van Kerkwijk}, Marten H. and {Robitaille}, Thomas P. and {Merry}, Bruce and {Bachetti}, Matteo and {G{\"u}nther}, H. Moritz and {Aldcroft}, Thomas L. and {Alvarado-Montes}, Jaime A. and {Archibald}, Anne M. and {B{\'o}di}, Attila and {Bapat}, Shreyas and {Barentsen}, Geert and {Baz{\'a}n}, Juanjo and {Biswas}, Manish and {Boquien}, M{\'e}d{\'e}ric and {Burke}, D.~J. and {Cara}, Daria and {Cara}, Mihai and {Conroy}, Kyle E. and {Conseil}, Simon and {Craig}, Matthew W. and {Cross}, Robert M. and {Cruz}, Kelle L. and {D'Eugenio}, Francesco and {Dencheva}, Nadia and {Devillepoix}, Hadrien A.~R. and {Dietrich}, J{\"o}rg P. and {Eigenbrot}, Arthur Davis and {Erben}, Thomas and {Ferreira}, Leonardo and {Foreman-Mackey}, Daniel and {Fox}, Ryan and {Freij}, Nabil and {Garg}, Suyog and {Geda}, Robel and {Glattly}, Lauren and {Gondhalekar}, Yash and {Gordon}, Karl D. and {Grant}, David and {Greenfield}, Perry and {Groener}, Austen M. and {Guest}, Steve and {Gurovich}, Sebastian and {Handberg}, Rasmus and {Hart}, Akeem and {Hatfield-Dodds}, Zac and {Homeier}, Derek and {Hosseinzadeh}, Griffin and {Jenness}, Tim and {Jones}, Craig K. and {Joseph}, Prajwel and {Kalmbach}, J. Bryce and {Karamehmetoglu}, Emir and {Ka{\l}uszy{\'n}ski}, Miko{\l}aj and {Kelley}, Michael S.~P. and {Kern}, Nicholas and {Kerzendorf}, Wolfgang E. and {Koch}, Eric W. and {Kulumani}, Shankar and {Lee}, Antony and {Ly}, Chun and {Ma}, Zhiyuan and {MacBride}, Conor and {Maljaars}, Jakob M. and {Muna}, Demitri and {Murphy}, N.~A. and {Norman}, Henrik and {O'Steen}, Richard and {Oman}, Kyle A. and {Pacifici}, Camilla and {Pascual}, Sergio and {Pascual-Granado}, J. and {Patil}, Rohit R. and {Perren}, Gabriel I. and {Pickering}, Timothy E. and {Rastogi}, Tanuj and {Roulston}, Benjamin R. and {Ryan}, Daniel F. and {Rykoff}, Eli S. and {Sabater}, Jose and {Sakurikar}, Parikshit and {Salgado}, Jes{\'u}s and {Sanghi}, Aniket and {Saunders}, Nicholas and {Savchenko}, Volodymyr and {Schwardt}, Ludwig and {Seifert-Eckert}, Michael and {Shih}, Albert Y. and {Jain}, Anany Shrey and {Shukla}, Gyanendra and {Sick}, Jonathan and {Simpson}, Chris and {Singanamalla}, Sudheesh and {Singer}, Leo P. and {Singhal}, Jaladh and {Sinha}, Manodeep and {Sip{\H{o}}cz}, Brigitta M. and {Spitler}, Lee R. and {Stansby}, David and {Streicher}, Ole and {{\v{S}}umak}, Jani and {Swinbank}, John D. and {Taranu}, Dan S. and {Tewary}, Nikita and {Tremblay}, Grant R. and {de Val-Borro}, Miguel and {Van Kooten}, Samuel J. and {Vasovi{\'c}}, Zlatan and {Verma}, Shresth and {de Miranda Cardoso}, Jos{\'e} Vin{\'\i}cius and {Williams}, Peter K.~G. and {Wilson}, Tom J. and {Winkel}, Benjamin and {Wood-Vasey}, W.~M. and {Xue}, Rui and {Yoachim}, Peter and {Zhang}, Chen and {Zonca}, Andrea and {Astropy Project Contributors}},
        title = "{The Astropy Project: Sustaining and Growing a Community-oriented Open-source Project and the Latest Major Release (v5.0) of the Core Package}",
      journal = {\apj},
         year = 2022,
        month = aug,
       volume = {935},
       number = {2},
          eid = {167},
        pages = {167},
          doi = {10.3847/1538-4357/ac7c74},
archivePrefix = {arXiv},
       eprint = {2206.14220},
 primaryClass = {astro-ph.IM},
       adsurl = {https://ui.adsabs.harvard.edu/abs/2022ApJ...935..167A}
}

@ARTICLE{2010MNRAS.406.2713B,
       author = {{Barriault}, L. and {Joncas}, G. and {Falgarone}, E. and {Marshall}, D.~J. and {Heyer}, M. and {Boulanger}, F. and {Foster}, T. and {Brunt}, C. and {Miville-Desch{\^e}nes}, M. -A. and {Blagrave}, K. and {Kothes}, R. and {Landecker}, T.~L. and {Martin}, P.~G. and {Scott}, D. and {Stil}, J.~M. and {Taylor}, A.~R.},
        title = "{Multiwavelength observations of cirrus clouds in the North Celestial Loop: the transition from atomic to molecular gas}",
      journal = {\mnras},
         year = 2010,
        month = aug,
       volume = {406},
       number = {4},
        pages = {2713-2731},
          doi = {10.1111/j.1365-2966.2010.16871.x},
       adsurl = {https://ui.adsabs.harvard.edu/abs/2010MNRAS.406.2713B}
}

@ARTICLE{2001A&A...365..285B,
       author = {{Bensch}, F. and {Stutzki}, J. and {Heithausen}, A.},
        title = "{Methods and constraints for the correction of the error beam pick-up in single dish radio observations}",
      journal = {\aap},
         year = 2001,
        month = jan,
       volume = {365},
        pages = {285-293},
          doi = {10.1051/0004-6361:20000485},
       adsurl = {https://ui.adsabs.harvard.edu/abs/2001A&A...365..285B}
}

@INPROCEEDINGS{2023ASPC..534..193P,
       author = {{Pattle}, K. and {Fissel}, L. and {Tahani}, M. and {Liu}, T. and {Ntormousi}, E.},
        title = "{Magnetic Fields in Star Formation: from Clouds to Cores}",
    booktitle = {Protostars and Planets VII},
         year = 2023,
       editor = {{Inutsuka}, S. and {Aikawa}, Y. and {Muto}, T. and {Tomida}, K. and {Tamura}, M.},
       series = {Astronomical Society of the Pacific Conference Series},
       volume = {534},
        month = jul,
        pages = {193},
          doi = {10.48550/arXiv.2203.11179},
archivePrefix = {arXiv},
       eprint = {2203.11179},
 primaryClass = {astro-ph.GA},
       adsurl = {https://ui.adsabs.harvard.edu/abs/2023ASPC..534..193P}
}

@ARTICLE{2010ApJ...716.1191W,
       author = {{Wolfire}, Mark G. and {Hollenbach}, David and {McKee}, Christopher F.},
        title = "{The Dark Molecular Gas}",
      journal = {\apj},
         year = 2010,
        month = jun,
       volume = {716},
       number = {2},
        pages = {1191-1207},
          doi = {10.1088/0004-637X/716/2/1191},
archivePrefix = {arXiv},
       eprint = {1004.5401},
 primaryClass = {astro-ph.GA},
       adsurl = {https://ui.adsabs.harvard.edu/abs/2010ApJ...716.1191W}
}

@ARTICLE{2025RAA....25b5020L,
       author = {{Liu}, Xunchuan and {Liu}, Tie and {Mai}, Xiaofeng and {Cheng}, Yu and {Jiao}, Sihan and {Jiao}, Wenyu and {Liu}, Hongli and {Zhang}, Siju},
        title = "{Core Mass Function in View of Fractal and Turbulent Filaments and Fibers}",
      journal = {Research in Astronomy and Astrophysics},
         year = 2025,
        month = feb,
       volume = {25},
       number = {2},
          eid = {025020},
        pages = {025020},
          doi = {10.1088/1674-4527/adb15a},
archivePrefix = {arXiv},
       eprint = {2501.17502},
 primaryClass = {astro-ph.GA},
       adsurl = {https://ui.adsabs.harvard.edu/abs/2025RAA....25b5020L}
}

@ARTICLE{2005ApJ...634.1126M,
       author = {{Milam}, S.~N. and {Savage}, C. and {Brewster}, M.~A. and {Ziurys}, L.~M. and {Wyckoff}, S.},
        title = "{The $^{12}$C/$^{13}$C Isotope Gradient Derived from Millimeter Transitions of CN: The Case for Galactic Chemical Evolution}",
      journal = {\apj},
         year = 2005,
        month = dec,
       volume = {634},
       number = {2},
        pages = {1126-1132},
          doi = {10.1086/497123},
       adsurl = {https://ui.adsabs.harvard.edu/abs/2005ApJ...634.1126M}
}

@ARTICLE{2002A&A...391..295O,
       author = {{Ossenkopf}, V.},
        title = "{Molecular line emission from turbulent clouds}",
      journal = {\aap},
         year = 2002,
        month = aug,
       volume = {391},
        pages = {295-315},
          doi = {10.1051/0004-6361:20020812},
       adsurl = {https://ui.adsabs.harvard.edu/abs/2002A&A...391..295O}
}

@ARTICLE{2025arXiv250220458L,
       author = {{Liu}, Xunchuan},
        title = "{Turbulence in virtual: Origin of the variance and skewness of density function}",
      journal = {arXiv e-prints},
         year = 2025,
        month = feb,
          eid = {arXiv:2502.20458},
        pages = {arXiv:2502.20458},
          doi = {10.48550/arXiv.2502.20458},
archivePrefix = {arXiv},
       eprint = {2502.20458},
 primaryClass = {astro-ph.GA},
       adsurl = {https://ui.adsabs.harvard.edu/abs/2025arXiv250220458L}
}

@ARTICLE{1995A&A...301..873S,
       author = {{Stark}, R.},
        title = "{Diffuse molecular cirrus clouds at high galactic latitude.}",
      journal = {\aap},
         year = 1995,
        month = sep,
       volume = {301},
        pages = {873},
       adsurl = {https://ui.adsabs.harvard.edu/abs/1995A&A...301..873S}
}

@ARTICLE{2012ApJ...756...76W,
       author = {{Wu}, Yuefang and {Liu}, Tie and {Meng}, Fanyi and {Li}, Di and {Qin}, Sheng-Li and {Ju}, Bing-Gang},
        title = "{Gas Emissions in Planck Cold Dust Clumps{\textemdash}A Survey of the J = 1-0 Transitions of $^{12}$CO, $^{13}$CO, and C$^{18}$O}",
      journal = {\apj},
         year = 2012,
        month = sep,
       volume = {756},
       number = {1},
          eid = {76},
        pages = {76},
          doi = {10.1088/0004-637X/756/1/76},
archivePrefix = {arXiv},
       eprint = {1206.7027},
 primaryClass = {astro-ph.SR},
       adsurl = {https://ui.adsabs.harvard.edu/abs/2012ApJ...756...76W}
}

@ARTICLE{2016A&A...594A..28P,
       author = {{Planck Collaboration} and {Ade}, P.~A.~R. and {Aghanim}, N. and {Arnaud}, M. and {Ashdown}, M. and {Aumont}, J. and {Baccigalupi}, C. and {Banday}, A.~J. and {Barreiro}, R.~B. and {Bartolo}, N. and {Battaner}, E. and {Benabed}, K. and {Beno{\^\i}t}, A. and {Benoit-L{\'e}vy}, A. and {Bernard}, J. -P. and {Bersanelli}, M. and {Bielewicz}, P. and {Bonaldi}, A. and {Bonavera}, L. and {Bond}, J.~R. and {Borrill}, J. and {Bouchet}, F.~R. and {Boulanger}, F. and {Bucher}, M. and {Burigana}, C. and {Butler}, R.~C. and {Calabrese}, E. and {Catalano}, A. and {Chamballu}, A. and {Chiang}, H.~C. and {Christensen}, P.~R. and {Clements}, D.~L. and {Colombi}, S. and {Colombo}, L.~P.~L. and {Combet}, C. and {Couchot}, F. and {Coulais}, A. and {Crill}, B.~P. and {Curto}, A. and {Cuttaia}, F. and {Danese}, L. and {Davies}, R.~D. and {Davis}, R.~J. and {de Bernardis}, P. and {de Rosa}, A. and {de Zotti}, G. and {Delabrouille}, J. and {D{\'e}sert}, F. -X. and {Dickinson}, C. and {Diego}, J.~M. and {Dole}, H. and {Donzelli}, S. and {Dor{\'e}}, O. and {Douspis}, M. and {Ducout}, A. and {Dupac}, X. and {Efstathiou}, G. and {Elsner}, F. and {En{\ss}lin}, T.~A. and {Eriksen}, H.~K. and {Falgarone}, E. and {Fergusson}, J. and {Finelli}, F. and {Forni}, O. and {Frailis}, M. and {Fraisse}, A.~A. and {Franceschi}, E. and {Frejsel}, A. and {Galeotta}, S. and {Galli}, S. and {Ganga}, K. and {Giard}, M. and {Giraud-H{\'e}raud}, Y. and {Gjerl{\o}w}, E. and {Gonz{\'a}lez-Nuevo}, J. and {G{\'o}rski}, K.~M. and {Gratton}, S. and {Gregorio}, A. and {Gruppuso}, A. and {Gudmundsson}, J.~E. and {Hansen}, F.~K. and {Hanson}, D. and {Harrison}, D.~L. and {Helou}, G. and {Henrot-Versill{\'e}}, S. and {Hern{\'a}ndez-Monteagudo}, C. and {Herranz}, D. and {Hildebrandt}, S.~R. and {Hivon}, E. and {Hobson}, M. and {Holmes}, W.~A. and {Hornstrup}, A. and {Hovest}, W. and {Huffenberger}, K.~M. and {Hurier}, G. and {Jaffe}, A.~H. and {Jaffe}, T.~R. and {Jones}, W.~C. and {Juvela}, M. and {Keih{\"a}nen}, E. and {Keskitalo}, R. and {Kisner}, T.~S. and {Knoche}, J. and {Kunz}, M. and {Kurki-Suonio}, H. and {Lagache}, G. and {Lamarre}, J. -M. and {Lasenby}, A. and {Lattanzi}, M. and {Lawrence}, C.~R. and {Leonardi}, R. and {Lesgourgues}, J. and {Levrier}, F. and {Liguori}, M. and {Lilje}, P.~B. and {Linden-V{\o}rnle}, M. and {L{\'o}pez-Caniego}, M. and {Lubin}, P.~M. and {Mac{\'\i}as-P{\'e}rez}, J.~F. and {Maggio}, G. and {Maino}, D. and {Mandolesi}, N. and {Mangilli}, A. and {Marshall}, D.~J. and {Martin}, P.~G. and {Mart{\'\i}nez-Gonz{\'a}lez}, E. and {Masi}, S. and {Matarrese}, S. and {Mazzotta}, P. and {McGehee}, P. and {Melchiorri}, A. and {Mendes}, L. and {Mennella}, A. and {Migliaccio}, M. and {Mitra}, S. and {Miville-Desch{\^e}nes}, M. -A. and {Moneti}, A. and {Montier}, L. and {Morgante}, G. and {Mortlock}, D. and {Moss}, A. and {Munshi}, D. and {Murphy}, J.~A. and {Naselsky}, P. and {Nati}, F. and {Natoli}, P. and {Netterfield}, C.~B. and {N{\o}rgaard-Nielsen}, H.~U. and {Noviello}, F. and {Novikov}, D. and {Novikov}, I. and {Oxborrow}, C.~A. and {Paci}, F. and {Pagano}, L. and {Pajot}, F. and {Paladini}, R. and {Paoletti}, D. and {Pasian}, F. and {Patanchon}, G. and {Pearson}, T.~J. and {Pelkonen}, V. -M. and {Perdereau}, O. and {Perotto}, L. and {Perrotta}, F. and {Pettorino}, V. and {Piacentini}, F. and {Piat}, M. and {Pierpaoli}, E. and {Pietrobon}, D. and {Plaszczynski}, S. and {Pointecouteau}, E. and {Polenta}, G. and {Pratt}, G.~W. and {Pr{\'e}zeau}, G. and {Prunet}, S. and {Puget}, J. -L. and {Rachen}, J.~P. and {Reach}, W.~T. and {Rebolo}, R. and {Reinecke}, M. and {Remazeilles}, M. and {Renault}, C. and {Renzi}, A. and {Ristorcelli}, I. and {Rocha}, G. and {Rosset}, C. and {Rossetti}, M. and {Roudier}, G. and {Rubi{\~n}o-Mart{\'\i}n}, J.~A. and {Rusholme}, B. and {Sandri}, M. and {Santos}, D. and {Savelainen}, M. and {Savini}, G. and {Scott}, D. and {Seiffert}, M.~D. and {Shellard}, E.~P.~S. and {Spencer}, L.~D. and {Stolyarov}, V. and {Sudiwala}, R.},
        title = "{Planck 2015 results. XXVIII. The Planck Catalogue of Galactic cold clumps}",
      journal = {\aap},
         year = 2016,
        month = sep,
       volume = {594},
          eid = {A28},
        pages = {A28},
          doi = {10.1051/0004-6361/201525819},
archivePrefix = {arXiv},
       eprint = {1502.01599},
 primaryClass = {astro-ph.GA},
       adsurl = {https://ui.adsabs.harvard.edu/abs/2016A&A...594A..28P}
}

@ARTICLE{2020NatMe..17..261V,
       author = {{Virtanen}, Pauli and {Gommers}, Ralf and {Oliphant}, Travis E. and {Haberland}, Matt and {Reddy}, Tyler and {Cournapeau}, David and {Burovski}, Evgeni and {Peterson}, Pearu and {Weckesser}, Warren and {Bright}, Jonathan and {van der Walt}, St{\'e}fan J. and {Brett}, Matthew and {Wilson}, Joshua and {Millman}, K. Jarrod and {Mayorov}, Nikolay and {Nelson}, Andrew R.~J. and {Jones}, Eric and {Kern}, Robert and {Larson}, Eric and {Carey}, C.~J. and {Polat}, {\.I}lhan and {Feng}, Yu and {Moore}, Eric W. and {VanderPlas}, Jake and {Laxalde}, Denis and {Perktold}, Josef and {Cimrman}, Robert and {Henriksen}, Ian and {Quintero}, E.~A. and {Harris}, Charles R. and {Archibald}, Anne M. and {Ribeiro}, Ant{\^o}nio H. and {Pedregosa}, Fabian and {van Mulbregt}, Paul and {SciPy 1. 0 Contributors}},
        title = "{SciPy 1.0: fundamental algorithms for scientific computing in Python}",
      journal = {Nature Methods},
         year = 2020,
        month = feb,
       volume = {17},
        pages = {261-272},
          doi = {10.1038/s41592-019-0686-2},
archivePrefix = {arXiv},
       eprint = {1907.10121},
 primaryClass = {cs.MS},
       adsurl = {https://ui.adsabs.harvard.edu/abs/2020NatMe..17..261V}
}

@ARTICLE{2003ApJ...591.1013B,
       author = {{Bensch}, F. and {Leuenhagen}, U. and {Stutzki}, J. and {Schieder}, R.},
        title = "{[C I] 492 GHz Mapping Observations of the High-Latitude Translucent Cloud MCLD 123.5+24.9}",
      journal = {\apj},
         year = 2003,
        month = jul,
       volume = {591},
       number = {2},
        pages = {1013-1024},
          doi = {10.1086/375393},
       adsurl = {https://ui.adsabs.harvard.edu/abs/2003ApJ...591.1013B}
}

@ARTICLE{2018MNRAS.476.3688J,
       author = {{Jeffreson}, Sarah M.~R. and {Kruijssen}, J.~M. Diederik},
        title = "{A general theory for the lifetimes of giant molecular clouds under the influence of galactic dynamics}",
      journal = {\mnras},
         year = 2018,
        month = may,
       volume = {476},
       number = {3},
        pages = {3688-3715},
          doi = {10.1093/mnras/sty594},
archivePrefix = {arXiv},
       eprint = {1803.01850},
 primaryClass = {astro-ph.GA},
       adsurl = {https://ui.adsabs.harvard.edu/abs/2018MNRAS.476.3688J}
}

@ARTICLE{2020A&A...637A..67Y,
       author = {{Yuan}, Lixia and {Li}, Guang-Xing and {Zhu}, Ming and {Liu}, Tie and {Wang}, Ke and {Liu}, Xunchuan and {Kim}, Kee-Tae and {Tatematsu}, Ken'ichi and {Yuan}, Jinghua and {Wu}, Yuefang},
        title = "{Edge collapse and subsequent longitudinal accretion in filament S242}",
      journal = {\aap},
         year = 2020,
        month = may,
       volume = {637},
          eid = {A67},
        pages = {A67},
          doi = {10.1051/0004-6361/201936625},
archivePrefix = {arXiv},
       eprint = {2004.04886},
 primaryClass = {astro-ph.GA},
       adsurl = {https://ui.adsabs.harvard.edu/abs/2020A&A...637A..67Y}
}

@ARTICLE{1983A&A...119..109B,
       author = {{Bastien}, P.},
        title = "{Gravitational collapse and fragmentation of isothermal, non-rotating, cylindrical clouds}",
      journal = {\aap},
         year = 1983,
        month = mar,
       volume = {119},
       number = {1},
        pages = {109-116},
       adsurl = {https://ui.adsabs.harvard.edu/abs/1983A&A...119..109B}
}

@INPROCEEDINGS{2023ASPC..534..153H,
       author = {{Hacar}, A. and {Clark}, S.~E. and {Heitsch}, F. and {Kainulainen}, J. and {Panopoulou}, G.~V. and {Seifried}, D. and {Smith}, R.},
        title = "{Initial Conditions for Star Formation: a Physical Description of the Filamentary ISM}",
    booktitle = {Protostars and Planets VII},
         year = 2023,
       editor = {{Inutsuka}, S. and {Aikawa}, Y. and {Muto}, T. and {Tomida}, K. and {Tamura}, M.},
       series = {Astronomical Society of the Pacific Conference Series},
       volume = {534},
        month = jul,
        pages = {153},
          doi = {10.48550/arXiv.2203.09562},
archivePrefix = {arXiv},
       eprint = {2203.09562},
 primaryClass = {astro-ph.GA},
       adsurl = {https://ui.adsabs.harvard.edu/abs/2023ASPC..534..153H}
}

@ARTICLE{2002A&A...383..591H,
       author = {{Heithausen}, A. and {Bertoldi}, F. and {Bensch}, F.},
        title = "{Gravitationally bound cores in a molecular cirrus cloud}",
      journal = {\aap},
         year = 2002,
        month = feb,
       volume = {383},
        pages = {591-597},
          doi = {10.1051/0004-6361:20011806},
       adsurl = {https://ui.adsabs.harvard.edu/abs/2002A&A...383..591H}
}

@ARTICLE{2012ApJ...745..195S,
       author = {{Shimoikura}, Tomomi and {Dobashi}, Kazuhito and {Sakurai}, Tohko and {Takano}, Shuro and {Nishiura}, Shingo and {Hirota}, Tomoya},
        title = "{Molecular Line Observations of MCLD 123.5+24.9 in the Polaris Cirrus}",
      journal = {\apj},
         year = 2012,
        month = feb,
       volume = {745},
       number = {2},
          eid = {195},
        pages = {195},
          doi = {10.1088/0004-637X/745/2/195},
       adsurl = {https://ui.adsabs.harvard.edu/abs/2012ApJ...745..195S}
}

@ARTICLE{2010A&A...518L.103M,
       author = {{Men'shchikov}, A. and {Andr{\'e}}, Ph. and {Didelon}, P. and {K{\"o}nyves}, V. and {Schneider}, N. and {Motte}, F. and {Bontemps}, S. and {Arzoumanian}, D. and {Attard}, M. and {Abergel}, A. and {Baluteau}, J. -P. and {Bernard}, J. -Ph. and {Cambr{\'e}sy}, L. and {Cox}, P. and {di Francesco}, J. and {di Giorgio}, A.~M. and {Griffin}, M. and {Hargrave}, P. and {Huang}, M. and {Kirk}, J. and {Li}, J.~Z. and {Martin}, P. and {Minier}, V. and {Miville-Desch{\^e}nes}, M. -A. and {Molinari}, S. and {Olofsson}, G. and {Pezzuto}, S. and {Roussel}, H. and {Russeil}, D. and {Saraceno}, P. and {Sauvage}, M. and {Sibthorpe}, B. and {Spinoglio}, L. and {Testi}, L. and {Ward-Thompson}, D. and {White}, G. and {Wilson}, C.~D. and {Woodcraft}, A. and {Zavagno}, A.},
        title = "{Filamentary structures and compact objects in the Aquila and Polaris clouds observed by Herschel}",
      journal = {\aap},
         year = 2010,
        month = jul,
       volume = {518},
          eid = {L103},
        pages = {L103},
          doi = {10.1051/0004-6361/201014668},
archivePrefix = {arXiv},
       eprint = {1005.3115},
 primaryClass = {astro-ph.GA},
       adsurl = {https://ui.adsabs.harvard.edu/abs/2010A&A...518L.103M}
}

@ARTICLE{2010A&A...518L.106K,
       author = {{K{\"o}nyves}, V. and {Andr{\'e}}, Ph. and {Men'shchikov}, A. and {Schneider}, N. and {Arzoumanian}, D. and {Bontemps}, S. and {Attard}, M. and {Motte}, F. and {Didelon}, P. and {Maury}, A. and {Abergel}, A. and {Ali}, B. and {Baluteau}, J. -P. and {Bernard}, J. -Ph. and {Cambr{\'e}sy}, L. and {Cox}, P. and {di Francesco}, J. and {di Giorgio}, A.~M. and {Griffin}, M.~J. and {Hargrave}, P. and {Huang}, M. and {Kirk}, J. and {Li}, J.~Z. and {Martin}, P. and {Minier}, V. and {Molinari}, S. and {Olofsson}, G. and {Pezzuto}, S. and {Russeil}, D. and {Roussel}, H. and {Saraceno}, P. and {Sauvage}, M. and {Sibthorpe}, B. and {Spinoglio}, L. and {Testi}, L. and {Ward-Thompson}, D. and {White}, G. and {Wilson}, C.~D. and {Woodcraft}, A. and {Zavagno}, A.},
        title = "{The Aquila prestellar core population revealed by Herschel}",
      journal = {\aap},
         year = 2010,
        month = jul,
       volume = {518},
          eid = {L106},
        pages = {L106},
          doi = {10.1051/0004-6361/201014689},
archivePrefix = {arXiv},
       eprint = {1005.2981},
 primaryClass = {astro-ph.SR},
       adsurl = {https://ui.adsabs.harvard.edu/abs/2010A&A...518L.106K}
}

@ARTICLE{2013A&A...554A..55H,
       author = {{Hacar}, A. and {Tafalla}, M. and {Kauffmann}, J. and {Kov{\'a}cs}, A.},
        title = "{Cores, filaments, and bundles: hierarchical core formation in the L1495/B213 Taurus region}",
      journal = {\aap},
         year = 2013,
        month = jun,
       volume = {554},
          eid = {A55},
        pages = {A55},
          doi = {10.1051/0004-6361/201220090},
archivePrefix = {arXiv},
       eprint = {1303.2118},
 primaryClass = {astro-ph.GA},
       adsurl = {https://ui.adsabs.harvard.edu/abs/2013A&A...554A..55H}
}

@ARTICLE{2016A&A...587A..97H,
       author = {{Hacar}, A. and {Kainulainen}, J. and {Tafalla}, M. and {Beuther}, H. and {Alves}, J.},
        title = "{The Musca cloud: A 6 pc-long velocity-coherent, sonic filament}",
      journal = {\aap},
         year = 2016,
        month = mar,
       volume = {587},
          eid = {A97},
        pages = {A97},
          doi = {10.1051/0004-6361/201526015},
archivePrefix = {arXiv},
       eprint = {1511.06370},
 primaryClass = {astro-ph.GA},
       adsurl = {https://ui.adsabs.harvard.edu/abs/2016A&A...587A..97H}
}

@ARTICLE{2021RAA....21...24Y,
       author = {{Yue}, Nan-Nan and {Li}, Di and {Zhang}, Qi-Zhou and {Zhu}, Lei and {Henshaw}, Jonathan and {Mardones}, Diego and {Ren}, Zhi-Yuan},
        title = "{Resolution-dependent subsonic non-thermal line dispersion revealed by ALMA}",
      journal = {Research in Astronomy and Astrophysics},
         year = 2021,
        month = jan,
       volume = {21},
       number = {1},
          eid = {024},
        pages = {024},
          doi = {10.1088/1674-4527/21/1/24},
archivePrefix = {arXiv},
       eprint = {2006.04168},
 primaryClass = {astro-ph.GA},
       adsurl = {https://ui.adsabs.harvard.edu/abs/2021RAA....21...24Y}
}

@ARTICLE{2003ApJ...599..258R,
       author = {{Rosolowsky}, E. and {Engargiola}, G. and {Plambeck}, R. and {Blitz}, L.},
        title = "{Giant Molecular Clouds in M33. II. High-Resolution Observations}",
      journal = {\apj},
         year = 2003,
        month = dec,
       volume = {599},
       number = {1},
        pages = {258-274},
          doi = {10.1086/379166},
archivePrefix = {arXiv},
       eprint = {astro-ph/0307322},
 primaryClass = {astro-ph},
       adsurl = {https://ui.adsabs.harvard.edu/abs/2003ApJ...599..258R}
}

@ARTICLE{2011ApJ...732...78I,
       author = {{Imara}, Nia and {Blitz}, Leo},
        title = "{Angular Momentum in Giant Molecular Clouds. I. The Milky Way}",
      journal = {\apj},
         year = 2011,
        month = may,
       volume = {732},
       number = {2},
          eid = {78},
        pages = {78},
          doi = {10.1088/0004-637X/732/2/78},
archivePrefix = {arXiv},
       eprint = {1103.3741},
 primaryClass = {astro-ph.GA},
       adsurl = {https://ui.adsabs.harvard.edu/abs/2011ApJ...732...78I}
}

@ARTICLE{2014MNRAS.440.2860H,
       author = {{Henshaw}, J.~D. and {Caselli}, P. and {Fontani}, F. and {Jim{\'e}nez-Serra}, I. and {Tan}, J.~C.},
        title = "{The dynamical properties of dense filaments in the infrared dark cloud G035.39-00.33}",
      journal = {\mnras},
         year = 2014,
        month = may,
       volume = {440},
       number = {3},
        pages = {2860-2881},
          doi = {10.1093/mnras/stu446},
archivePrefix = {arXiv},
       eprint = {1403.1444},
 primaryClass = {astro-ph.SR},
       adsurl = {https://ui.adsabs.harvard.edu/abs/2014MNRAS.440.2860H}
}

@ARTICLE{2014A&A...561A..83P,
       author = {{Peretto}, N. and {Fuller}, G.~A. and {Andr{\'e}}, Ph. and {Arzoumanian}, D. and {Rivilla}, V.~M. and {Bardeau}, S. and {Duarte Puertas}, S. and {Guzman Fernandez}, J.~P. and {Lenfestey}, C. and {Li}, G. -X. and {Olguin}, F.~A. and {R{\"o}ck}, B.~R. and {de Villiers}, H. and {Williams}, J.},
        title = "{SDC13 infrared dark clouds: Longitudinally collapsing filaments?}",
      journal = {\aap},
         year = 2014,
        month = jan,
       volume = {561},
          eid = {A83},
        pages = {A83},
          doi = {10.1051/0004-6361/201322172},
archivePrefix = {arXiv},
       eprint = {1311.0203},
 primaryClass = {astro-ph.GA},
       adsurl = {https://ui.adsabs.harvard.edu/abs/2014A&A...561A..83P}
}

@ARTICLE{2014ApJ...790L..19F,
       author = {{Fern{\'a}ndez-L{\'o}pez}, M. and {Arce}, H.~G. and {Looney}, L. and {Mundy}, L.~G. and {Storm}, S. and {Teuben}, P.~J. and {Lee}, K. and {Segura-Cox}, D. and {Isella}, A. and {Tobin}, J.~J. and {Rosolowsky}, E. and {Plunkett}, A. and {Kwon}, W. and {Kauffmann}, J. and {Ostriker}, E. and {Tassis}, K. and {Shirley}, Y.~L. and {Pound}, M.},
        title = "{CARMA Large Area Star Formation Survey: Observational Analysis of Filaments in the Serpens South Molecular Cloud}",
      journal = {\apjl},
         year = 2014,
        month = aug,
       volume = {790},
       number = {2},
          eid = {L19},
        pages = {L19},
          doi = {10.1088/2041-8205/790/2/L19},
archivePrefix = {arXiv},
       eprint = {1407.0755},
 primaryClass = {astro-ph.GA},
       adsurl = {https://ui.adsabs.harvard.edu/abs/2014ApJ...790L..19F}
}

@ARTICLE{2008ApJS..177..341N,
       author = {{Narayanan}, Gopal and {Heyer}, Mark H. and {Brunt}, Christopher and {Goldsmith}, Paul F. and {Snell}, Ronald and {Li}, Di},
        title = "{The Five College Radio Astronomy Observatory CO Mapping Survey of the Taurus Molecular Cloud}",
      journal = {\apjs},
         year = 2008,
        month = jul,
       volume = {177},
       number = {1},
        pages = {341-361},
          doi = {10.1086/587786},
archivePrefix = {arXiv},
       eprint = {0802.2556},
 primaryClass = {astro-ph},
       adsurl = {https://ui.adsabs.harvard.edu/abs/2008ApJS..177..341N}
}

@ARTICLE{2006AJ....131.2921R,
       author = {{Ridge}, Naomi A. and {Di Francesco}, James and {Kirk}, Helen and {Li}, Di and {Goodman}, Alyssa A. and {Alves}, Jo{\~a}o F. and {Arce}, H{\'e}ctor G. and {Borkin}, Michelle A. and {Caselli}, Paola and {Foster}, Jonathan B. and {Heyer}, Mark H. and {Johnstone}, Doug and {Kosslyn}, David A. and {Lombardi}, Marco and {Pineda}, Jaime E. and {Schnee}, Scott L. and {Tafalla}, Mario},
        title = "{The COMPLETE Survey of Star-Forming Regions: Phase I Data}",
      journal = {\aj},
         year = 2006,
        month = jun,
       volume = {131},
       number = {6},
        pages = {2921-2933},
          doi = {10.1086/503704},
archivePrefix = {arXiv},
       eprint = {astro-ph/0602542},
 primaryClass = {astro-ph},
       adsurl = {https://ui.adsabs.harvard.edu/abs/2006AJ....131.2921R}
}

@ARTICLE{2020A&A...644A.151J,
       author = {{Juvela}, Mika},
        title = "{LOC program for line radiative transfer}",
      journal = {\aap},
         year = 2020,
        month = dec,
       volume = {644},
          eid = {A151},
        pages = {A151},
          doi = {10.1051/0004-6361/202039456},
archivePrefix = {arXiv},
       eprint = {2009.12609},
 primaryClass = {astro-ph.IM},
       adsurl = {https://ui.adsabs.harvard.edu/abs/2020A&A...644A.151J}
}

@ARTICLE{2025arXiv250319259L,
       author = {{Liu}, Xunchuan and {Mai}, Xiaofeng},
        title = "{Growing 3D clouds from 2D maps via full spherization}",
      journal = {arXiv e-prints},
         year = 2025,
        month = mar,
          eid = {arXiv:2503.19259},
        pages = {arXiv:2503.19259},
          doi = {10.48550/arXiv.2503.19259},
archivePrefix = {arXiv},
       eprint = {2503.19259},
 primaryClass = {astro-ph.GA},
       adsurl = {https://ui.adsabs.harvard.edu/abs/2025arXiv250319259L}
}

@ARTICLE{2020A&A...644A..27B,
       author = {{Bonne}, L. and {Bontemps}, S. and {Schneider}, N. and {Clarke}, S.~D. and {Arzoumanian}, D. and {Fukui}, Y. and {Tachihara}, K. and {Csengeri}, T. and {Guesten}, R. and {Ohama}, A. and {Okamoto}, R. and {Simon}, R. and {Yahia}, H. and {Yamamoto}, H.},
        title = "{Formation of the Musca filament: evidence for asymmetries in the accretion flow due to a cloud-cloud collision}",
      journal = {\aap},
         year = 2020,
        month = dec,
       volume = {644},
          eid = {A27},
        pages = {A27},
          doi = {10.1051/0004-6361/202038281},
archivePrefix = {arXiv},
       eprint = {2010.12479},
 primaryClass = {astro-ph.GA},
       adsurl = {https://ui.adsabs.harvard.edu/abs/2020A&A...644A..27B}
}

@ARTICLE{1979ApJ...231..720P,
       author = {{Phillips}, T.~G. and {Huggins}, P.~J. and {Wannier}, P.~G. and {Scoville}, N.~Z.},
        title = "{Observations of CO(J = 2-1) emission from molecular clouds.}",
      journal = {\apj},
         year = 1979,
        month = aug,
       volume = {231},
        pages = {720-731},
          doi = {10.1086/157237},
       adsurl = {https://ui.adsabs.harvard.edu/abs/1979ApJ...231..720P}
}

@ARTICLE{2015PASA...32...20B,
       author = {{Braiding}, Catherine and {Burton}, M.~G. and {Blackwell}, R. and {Gl{\"u}ck}, C. and {Hawkes}, J. and {Kulesa}, C. and {Maxted}, N. and {Rebolledo}, D. and {Rowell}, G. and {Stark}, A. and {Tothill}, N. and {Urquhart}, J.~S. and {Voisin}, F. and {Walsh}, A.~J. and {de Wilt}, P. and {Wong}, G.~F.},
        title = "{The Mopra Southern Galactic Plane CO Survey - Data Release 1}",
      journal = {\pasa},
         year = 2015,
        month = may,
       volume = {32},
          eid = {e020},
        pages = {e020},
          doi = {10.1017/pasa.2015.20},
archivePrefix = {arXiv},
       eprint = {1504.04068},
 primaryClass = {astro-ph.GA},
       adsurl = {https://ui.adsabs.harvard.edu/abs/2015PASA...32...20B}
}

@inproceedings{Lorensen87,
  author = {Lorensen, William E. and Cline, Harvey E.},
  booktitle = {SIGGRAPH},
  editor = {Stone, Maureen C.},
  ee = {https://doi.org/10.1145/37401.37422},
  isbn = {0-89791-227-6},
  pages = {163-169},
  publisher = {ACM},
  title = {Marching cubes: A high resolution 3D surface construction algorithm.},
  url = {http://dblp.uni-trier.de/db/conf/siggraph/siggraph1987.html#LorensenC87},
  year = 1987
}

@ARTICLE{1998A&A...336..697S,
       author = {{Stutzki}, J. and {Bensch}, F. and {Heithausen}, A. and {Ossenkopf}, V. and {Zielinsky}, M.},
        title = "{On the fractal structure of molecular clouds}",
      journal = {\aap},
         year = 1998,
        month = aug,
       volume = {336},
        pages = {697-720},
       adsurl = {https://ui.adsabs.harvard.edu/abs/1998A&A...336..697S}
}

@ARTICLE{2018PASJ...70S..46F,
       author = {{Fukui}, Yasuo and {Ohama}, Akio and {Kohno}, Mikito and {Torii}, Kazufumi and {Fujita}, Shinji and {Hattori}, Yusuke and {Nishimura}, Atsushi and {Yamamoto}, Hiroaki and {Tachihara}, Kengo},
        title = "{Molecular clouds toward three Spitzer bubbles S116, S117, and S118: Evidence for a cloud-cloud collision which formed the three H II regions and a 10 pc scale molecular cavity}",
      journal = {\pasj},
         year = 2018,
        month = may,
       volume = {70},
          eid = {S46},
        pages = {S46},
          doi = {10.1093/pasj/psy005},
archivePrefix = {arXiv},
       eprint = {1706.08720},
 primaryClass = {astro-ph.GA},
       adsurl = {https://ui.adsabs.harvard.edu/abs/2018PASJ...70S..46F}
}

@ARTICLE{2020A&A...644A...5B,
       author = {{Bracco}, A. and {Bresnahan}, D. and {Palmeirim}, P. and {Arzoumanian}, D. and {Andr{\'e}}, Ph. and {Ward-Thompson}, D. and {Marchal}, A.},
        title = "{Compressed magnetized shells of atomic gas and the formation of the Corona Australis molecular cloud}",
      journal = {\aap},
         year = 2020,
        month = dec,
       volume = {644},
          eid = {A5},
        pages = {A5},
          doi = {10.1051/0004-6361/202039282},
archivePrefix = {arXiv},
       eprint = {2010.10110},
 primaryClass = {astro-ph.GA},
       adsurl = {https://ui.adsabs.harvard.edu/abs/2020A&A...644A...5B}
}

@ARTICLE{2020MNRAS.499.4918A,
       author = {{Armijos-Abenda{\~n}o}, J. and {Banda-Barrag{\'a}n}, W.~E. and {Mart{\'\i}n-Pintado}, J. and {D{\'e}nes}, H. and {Federrath}, C. and {Requena-Torres}, M.~A.},
        title = "{Structure and kinematics of shocked gas in Sgr B2: further evidence of a cloud-cloud collision from SiO emission maps}",
      journal = {\mnras},
         year = 2020,
        month = dec,
       volume = {499},
       number = {4},
        pages = {4918-4939},
          doi = {10.1093/mnras/staa3119},
archivePrefix = {arXiv},
       eprint = {2010.02757},
 primaryClass = {astro-ph.GA},
       adsurl = {https://ui.adsabs.harvard.edu/abs/2020MNRAS.499.4918A}
}

@ARTICLE{2016ApJ...827L..27P,
       author = {{Park}, Geumsook and {Koo}, Bon-Chul and {Kang}, Ji-hyun and {Gibson}, Steven J. and {Peek}, J.~E.~G. and {Douglas}, Kevin A. and {Korpela}, Eric J. and {Heiles}, Carl E.},
        title = "{A High-velocity Cloud Impact Forming a Supershell in the Milky Way}",
      journal = {\apjl},
         year = 2016,
        month = aug,
       volume = {827},
       number = {2},
          eid = {L27},
        pages = {L27},
          doi = {10.3847/2041-8205/827/2/L27},
archivePrefix = {arXiv},
       eprint = {1607.07699},
 primaryClass = {astro-ph.GA},
       adsurl = {https://ui.adsabs.harvard.edu/abs/2016ApJ...827L..27P}
}

@ARTICLE{2016MNRAS.458.3479M,
       author = {{Marton}, G. and {T{\'o}th}, L.~V. and {Paladini}, R. and {Kun}, M. and {Zahorecz}, S. and {McGehee}, P. and {Kiss}, Cs.},
        title = "{An all-sky support vector machine selection of WISE YSO candidates}",
      journal = {\mnras},
         year = 2016,
        month = jun,
       volume = {458},
       number = {4},
        pages = {3479-3488},
          doi = {10.1093/mnras/stw398},
archivePrefix = {arXiv},
       eprint = {1602.05777},
 primaryClass = {astro-ph.IM},
       adsurl = {https://ui.adsabs.harvard.edu/abs/2016MNRAS.458.3479M}
}

@ARTICLE{2014ApJ...791..131K,
       author = {{Koenig}, X.~P. and {Leisawitz}, D.~T.},
        title = "{A Classification Scheme for Young Stellar Objects Using the Wide-field Infrared Survey Explorer AllWISE Catalog: Revealing Low-density Star Formation in the Outer Galaxy}",
      journal = {\apj},
         year = 2014,
        month = aug,
       volume = {791},
       number = {2},
          eid = {131},
        pages = {131},
          doi = {10.1088/0004-637X/791/2/131},
archivePrefix = {arXiv},
       eprint = {1407.2262},
 primaryClass = {astro-ph.GA},
       adsurl = {https://ui.adsabs.harvard.edu/abs/2014ApJ...791..131K}
}

@ARTICLE{2023A&A...674A...1G,
       author = {{Gaia Collaboration} and {Vallenari}, A. and {Brown}, A.~G.~A. and {Prusti}, T. and {de Bruijne}, J.~H.~J. and {Arenou}, F. and {Babusiaux}, C. and {Biermann}, M. and {Creevey}, O.~L. and {Ducourant}, C. and {Evans}, D.~W. and {Eyer}, L. and {Guerra}, R. and {Hutton}, A. and {Jordi}, C. and {Klioner}, S.~A. and {Lammers}, U.~L. and {Lindegren}, L. and {Luri}, X. and {Mignard}, F. and {Panem}, C. and {Pourbaix}, D. and {Randich}, S. and {Sartoretti}, P. and {Soubiran}, C. and {Tanga}, P. and {Walton}, N.~A. and {Bailer-Jones}, C.~A.~L. and {Bastian}, U. and {Drimmel}, R. and {Jansen}, F. and {Katz}, D. and {Lattanzi}, M.~G. and {van Leeuwen}, F. and {Bakker}, J. and {Cacciari}, C. and {Casta{\~n}eda}, J. and {De Angeli}, F. and {Fabricius}, C. and {Fouesneau}, M. and {Fr{\'e}mat}, Y. and {Galluccio}, L. and {Guerrier}, A. and {Heiter}, U. and {Masana}, E. and {Messineo}, R. and {Mowlavi}, N. and {Nicolas}, C. and {Nienartowicz}, K. and {Pailler}, F. and {Panuzzo}, P. and {Riclet}, F. and {Roux}, W. and {Seabroke}, G.~M. and {Sordo}, R. and {Th{\'e}venin}, F. and {Gracia-Abril}, G. and {Portell}, J. and {Teyssier}, D. and {Altmann}, M. and {Andrae}, R. and {Audard}, M. and {Bellas-Velidis}, I. and {Benson}, K. and {Berthier}, J. and {Blomme}, R. and {Burgess}, P.~W. and {Busonero}, D. and {Busso}, G. and {C{\'a}novas}, H. and {Carry}, B. and {Cellino}, A. and {Cheek}, N. and {Clementini}, G. and {Damerdji}, Y. and {Davidson}, M. and {de Teodoro}, P. and {Nu{\~n}ez Campos}, M. and {Delchambre}, L. and {Dell'Oro}, A. and {Esquej}, P. and {Fern{\'a}ndez-Hern{\'a}ndez}, J. and {Fraile}, E. and {Garabato}, D. and {Garc{\'\i}a-Lario}, P. and {Gosset}, E. and {Haigron}, R. and {Halbwachs}, J. -L. and {Hambly}, N.~C. and {Harrison}, D.~L. and {Hern{\'a}ndez}, J. and {Hestroffer}, D. and {Hodgkin}, S.~T. and {Holl}, B. and {Jan{\ss}en}, K. and {Jevardat de Fombelle}, G. and {Jordan}, S. and {Krone-Martins}, A. and {Lanzafame}, A.~C. and {L{\"o}ffler}, W. and {Marchal}, O. and {Marrese}, P.~M. and {Moitinho}, A. and {Muinonen}, K. and {Osborne}, P. and {Pancino}, E. and {Pauwels}, T. and {Recio-Blanco}, A. and {Reyl{\'e}}, C. and {Riello}, M. and {Rimoldini}, L. and {Roegiers}, T. and {Rybizki}, J. and {Sarro}, L.~M. and {Siopis}, C. and {Smith}, M. and {Sozzetti}, A. and {Utrilla}, E. and {van Leeuwen}, M. and {Abbas}, U. and {{\'A}brah{\'a}m}, P. and {Abreu Aramburu}, A. and {Aerts}, C. and {Aguado}, J.~J. and {Ajaj}, M. and {Aldea-Montero}, F. and {Altavilla}, G. and {{\'A}lvarez}, M.~A. and {Alves}, J. and {Anders}, F. and {Anderson}, R.~I. and {Anglada Varela}, E. and {Antoja}, T. and {Baines}, D. and {Baker}, S.~G. and {Balaguer-N{\'u}{\~n}ez}, L. and {Balbinot}, E. and {Balog}, Z. and {Barache}, C. and {Barbato}, D. and {Barros}, M. and {Barstow}, M.~A. and {Bartolom{\'e}}, S. and {Bassilana}, J. -L. and {Bauchet}, N. and {Becciani}, U. and {Bellazzini}, M. and {Berihuete}, A. and {Bernet}, M. and {Bertone}, S. and {Bianchi}, L. and {Binnenfeld}, A. and {Blanco-Cuaresma}, S. and {Blazere}, A. and {Boch}, T. and {Bombrun}, A. and {Bossini}, D. and {Bouquillon}, S. and {Bragaglia}, A. and {Bramante}, L. and {Breedt}, E. and {Bressan}, A. and {Brouillet}, N. and {Brugaletta}, E. and {Bucciarelli}, B. and {Burlacu}, A. and {Butkevich}, A.~G. and {Buzzi}, R. and {Caffau}, E. and {Cancelliere}, R. and {Cantat-Gaudin}, T. and {Carballo}, R. and {Carlucci}, T. and {Carnerero}, M.~I. and {Carrasco}, J.~M. and {Casamiquela}, L. and {Castellani}, M. and {Castro-Ginard}, A. and {Chaoul}, L. and {Charlot}, P. and {Chemin}, L. and {Chiaramida}, V. and {Chiavassa}, A. and {Chornay}, N. and {Comoretto}, G. and {Contursi}, G. and {Cooper}, W.~J. and {Cornez}, T. and {Cowell}, S. and {Crifo}, F. and {Cropper}, M. and {Crosta}, M. and {Crowley}, C. and {Dafonte}, C. and {Dapergolas}, A. and {David}, M. and {David}, P. and {de Laverny}, P. and {De Luise}, F. and {De March}, R.},
        title = "{Gaia Data Release 3. Summary of the content and survey properties}",
      journal = {\aap},
         year = 2023,
        month = jun,
       volume = {674},
          eid = {A1},
        pages = {A1},
          doi = {10.1051/0004-6361/202243940},
archivePrefix = {arXiv},
       eprint = {2208.00211},
 primaryClass = {astro-ph.GA},
       adsurl = {https://ui.adsabs.harvard.edu/abs/2023A&A...674A...1G}
}

@ARTICLE{2016A&A...590A.104O,
       author = {{Ossenkopf-Okada}, V. and {Csengeri}, T. and {Schneider}, N. and {Federrath}, C. and {Klessen}, R.~S.},
        title = "{The reliability of observational measurements of column density probability distribution functions}",
      journal = {\aap},
         year = 2016,
        month = may,
       volume = {590},
          eid = {A104},
        pages = {A104},
          doi = {10.1051/0004-6361/201628095},
archivePrefix = {arXiv},
       eprint = {1603.03344},
 primaryClass = {astro-ph.IM},
       adsurl = {https://ui.adsabs.harvard.edu/abs/2016A&A...590A.104O}
}

@ARTICLE{2004EPJB...41..345B,
       author = {{Barndorff-Nielsen}, O.~E. and {Bl{\ae}sild}, P. and {Schmiegel}, J.},
        title = "{A parsimonious and universal description of turbulent velocity increments}",
      journal = {European Physical Journal B},
         year = 2004,
        month = oct,
       volume = {41},
       number = {3},
        pages = {345-363},
          doi = {10.1140/epjb/e2004-00328-1},
       adsurl = {https://ui.adsabs.harvard.edu/abs/2004EPJB...41..345B}
}

@ARTICLE{2022A&A...668L...9M,
       author = {{Monaci}, Marco and {Magnani}, Loris and {Shore}, Steven N.},
        title = "{The mixing of dust and gas in the high latitude translucent cloud MBM 40}",
      journal = {\aap},
         year = 2022,
        month = dec,
       volume = {668},
          eid = {L9},
        pages = {L9},
          doi = {10.1051/0004-6361/202245021},
archivePrefix = {arXiv},
       eprint = {2301.05388},
 primaryClass = {astro-ph.GA},
       adsurl = {https://ui.adsabs.harvard.edu/abs/2022A&A...668L...9M}
}

@ARTICLE{2000ApJ...535..211I,
       author = {{Ingalls}, James G. and {Bania}, T.~M. and {Lane}, Adair P. and {Rumitz}, Matthias and {Stark}, Antony A.},
        title = "{Physical State of Molecular Gas in High Galactic Latitude Translucent Clouds}",
      journal = {\apj},
         year = 2000,
        month = may,
       volume = {535},
       number = {1},
        pages = {211-226},
          doi = {10.1086/308831},
archivePrefix = {arXiv},
       eprint = {astro-ph/9912079},
 primaryClass = {astro-ph},
       adsurl = {https://ui.adsabs.harvard.edu/abs/2000ApJ...535..211I}
}

@ARTICLE{2021PASJ...73S...1F,
       author = {{Fukui}, Yasuo and {Habe}, Asao and {Inoue}, Tsuyoshi and {Enokiya}, Rei and {Tachihara}, Kengo},
        title = "{Cloud-cloud collisions and triggered star formation}",
      journal = {\pasj},
         year = 2021,
        month = jan,
       volume = {73},
        pages = {S1-S34},
          doi = {10.1093/pasj/psaa103},
archivePrefix = {arXiv},
       eprint = {2009.05077},
 primaryClass = {astro-ph.GA},
       adsurl = {https://ui.adsabs.harvard.edu/abs/2021PASJ...73S...1F}
}

@article{10.1093/mnras/sty2813,
    author = {Bennett, Morgan and Bovy, Jo},
    title = {Vertical waves in the solar neighbourhood in Gaia DR2},
    journal = {Monthly Notices of the Royal Astronomical Society},
    volume = {482},
    number = {1},
    pages = {1417-1425},
    year = {2018},
    month = {10},
    issn = {0035-8711},
    doi = {10.1093/mnras/sty2813},
    url = {https://doi.org/10.1093/mnras/sty2813},
    eprint = {https://academic.oup.com/mnras/article-pdf/482/1/1417/26492855/sty2813.pdf},
}

@ARTICLE{2019FrASS...6....5H,
       author = {{Hennebelle}, Patrick and {Inutsuka}, Shu-ichiro},
        title = "{The role of magnetic field in molecular cloud formation and evolution}",
      journal = {Frontiers in Astronomy and Space Sciences},
         year = 2019,
        month = mar,
       volume = {6},
          eid = {5},
        pages = {5},
          doi = {10.3389/fspas.2019.00005},
archivePrefix = {arXiv},
       eprint = {1902.00798},
 primaryClass = {astro-ph.GA},
       adsurl = {https://ui.adsabs.harvard.edu/abs/2019FrASS...6....5H}
}

@ARTICLE{2016MNRAS.462.3602T,
       author = {{Tritsis}, Aris and {Tassis}, Konstantinos},
        title = "{Striations in molecular clouds: streamers or MHD waves?}",
      journal = {\mnras},
         year = 2016,
        month = nov,
       volume = {462},
       number = {4},
        pages = {3602-3615},
          doi = {10.1093/mnras/stw1881},
archivePrefix = {arXiv},
       eprint = {1607.08615},
 primaryClass = {astro-ph.SR},
       adsurl = {https://ui.adsabs.harvard.edu/abs/2016MNRAS.462.3602T}
}

@ARTICLE{2019MNRAS.485.4509L,
       author = {{Li}, Pak Shing and {Klein}, Richard I.},
        title = "{Magnetized interstellar molecular clouds - II. The large-scale structure and dynamics of filamentary molecular clouds}",
      journal = {\mnras},
         year = 2019,
        month = jun,
       volume = {485},
       number = {4},
        pages = {4509-4528},
          doi = {10.1093/mnras/stz653},
archivePrefix = {arXiv},
       eprint = {1901.04593},
 primaryClass = {astro-ph.GA},
       adsurl = {https://ui.adsabs.harvard.edu/abs/2019MNRAS.485.4509L}
}

@ARTICLE{2021ApJS..256...32S,
       author = {{Sun}, Yan and {Yang}, Ji and {Yan}, Qing-Zeng and {Lin}, Zehao and {Zhang}, Shaobo and {Su}, Yang and {Xu}, Ye and {Chen}, Xuepeng and {Wang}, Hongchi and {Zhou}, Xin},
        title = "{Examinations of CO Completeness Based on Three Independent CO Surveys}",
      journal = {\apjs},
         year = 2021,
        month = oct,
       volume = {256},
       number = {2},
          eid = {32},
        pages = {32},
          doi = {10.3847/1538-4365/ac11fe},
archivePrefix = {arXiv},
       eprint = {2107.05288},
 primaryClass = {astro-ph.GA},
       adsurl = {https://ui.adsabs.harvard.edu/abs/2021ApJS..256...32S}
}

@ARTICLE{2016MNRAS.461.3918H,
       author = {{Heyer}, M. and {Goldsmith}, P.~F. and {Y{\i}ld{\i}z}, U.~A. and {Snell}, R.~L. and {Falgarone}, E. and {Pineda}, J.~L.},
        title = "{Striations in the Taurus molecular cloud: Kelvin-Helmholtz instability or MHD waves?}",
      journal = {\mnras},
         year = 2016,
        month = oct,
       volume = {461},
       number = {4},
        pages = {3918-3926},
          doi = {10.1093/mnras/stw1567},
archivePrefix = {arXiv},
       eprint = {1606.08858},
 primaryClass = {astro-ph.GA},
       adsurl = {https://ui.adsabs.harvard.edu/abs/2016MNRAS.461.3918H}
}

@ARTICLE{2023A&A...673A..76S,
       author = {{Skalidis}, R. and {Gkimisi}, K. and {Tassis}, K. and {Panopoulou}, G.~V. and {Pelgrims}, V. and {Tritsis}, A. and {Goldsmith}, P.~F.},
        title = "{CO enhancement by magnetohydrodynamic waves. Striations in the Polaris Flare}",
      journal = {\aap},
         year = 2023,
        month = may,
       volume = {673},
          eid = {A76},
        pages = {A76},
          doi = {10.1051/0004-6361/202345880},
archivePrefix = {arXiv},
       eprint = {2303.04172},
 primaryClass = {astro-ph.GA},
       adsurl = {https://ui.adsabs.harvard.edu/abs/2023A&A...673A..76S}
}

@ARTICLE{2019A&A...622A..32L,
       author = {{Liu}, X. -C. and {Wu}, Y. and {Zhang}, C. and {Liu}, T. and {Yuan}, J. and {Qin}, S. -L. and {Ju}, B. -G. and {Li}, L. -X.},
        title = "{C$_{2}$H N = 1 - 0 and N$_{2}$H$^{+}$ J = 1 - 0 observations of Planck Galactic cold clumps}",
      journal = {\aap},
         year = 2019,
        month = feb,
       volume = {622},
          eid = {A32},
        pages = {A32},
          doi = {10.1051/0004-6361/201834411},
archivePrefix = {arXiv},
       eprint = {1901.01124},
 primaryClass = {astro-ph.SR},
       adsurl = {https://ui.adsabs.harvard.edu/abs/2019A&A...622A..32L}
}

@ARTICLE{2003ApJ...583..280B,
       author = {{Brunt}, Christopher M.},
        title = "{Large-Scale Turbulence in Molecular Clouds}",
      journal = {\apj},
         year = 2003,
        month = jan,
       volume = {583},
       number = {1},
        pages = {280-295},
          doi = {10.1086/345294},
       adsurl = {https://ui.adsabs.harvard.edu/abs/2003ApJ...583..280B}
}

@ARTICLE{2022ApJ...932...47W,
       author = {{Wong}, Tony and {Oudshoorn}, Luuk and {Sofovich}, Eliyahu and {Green}, Alex and {Shah}, Charmi and {Indebetouw}, R{\'e}my and {Meixner}, Margaret and {Hacar}, Alvaro and {Nayak}, Omnarayani and {Tokuda}, Kazuki and {Bolatto}, Alberto D. and {Chevance}, M{\'e}lanie and {De Marchi}, Guido and {Fukui}, Yasuo and {Hirschauer}, Alec S. and {Jameson}, K.~E. and {Kalari}, Venu and {Lebouteiller}, Vianney and {Looney}, Leslie W. and {Madden}, Suzanne C. and {Onishi}, Toshikazu and {Roman-Duval}, Julia and {Rubio}, M{\'o}nica and {Tielens}, A.~G.~G.~M.},
        title = "{The 30 Doradus Molecular Cloud at 0.4 pc Resolution with the Atacama Large Millimeter/submillimeter Array: Physical Properties and the Boundedness of CO-emitting Structures}",
      journal = {\apj},
         year = 2022,
        month = jun,
       volume = {932},
       number = {1},
          eid = {47},
        pages = {47},
          doi = {10.3847/1538-4357/ac723a},
archivePrefix = {arXiv},
       eprint = {2206.06528},
 primaryClass = {astro-ph.GA},
       adsurl = {https://ui.adsabs.harvard.edu/abs/2022ApJ...932...47W}
}

@ARTICLE{2025ApJ...989...25Y,
       author = {{Yuan}, Lixia and {Yang}, Ji},
        title = "{Cloud-to-cloud Velocity Dispersions Across a Local Arm Segment}",
      journal = {\apj},
         year = 2025,
        month = aug,
       volume = {989},
       number = {1},
          eid = {25},
        pages = {25},
          doi = {10.3847/1538-4357/ade98f},
archivePrefix = {arXiv},
       eprint = {2506.13424},
 primaryClass = {astro-ph.GA},
       adsurl = {https://ui.adsabs.harvard.edu/abs/2025ApJ...989...25Y}
}

@ARTICLE{2009A&A...503..323V,
       author = {{Visser}, R. and {van Dishoeck}, E.~F. and {Black}, J.~H.},
        title = "{The photodissociation and chemistry of CO isotopologues: applications to interstellar clouds and circumstellar disks}",
      journal = {\aap},
         year = 2009,
        month = aug,
       volume = {503},
       number = {2},
        pages = {323-343},
          doi = {10.1051/0004-6361/200912129},
archivePrefix = {arXiv},
       eprint = {0906.3699},
 primaryClass = {astro-ph.GA},
       adsurl = {https://ui.adsabs.harvard.edu/abs/2009A&A...503..323V}
}

@ARTICLE{1980ApJ...235L..39L,
       author = {{Langer}, W.~D. and {Goldsmith}, P.~F. and {Carlson}, E.~R. and {Wilson}, R.~W.},
        title = "{Evidence for isotopic fractionation of carbon monoxide in dark clouds}",
      journal = {\apjl},
         year = 1980,
        month = jan,
       volume = {235},
        pages = {L39-L44},
          doi = {10.1086/183153},
       adsurl = {https://ui.adsabs.harvard.edu/abs/1980ApJ...235L..39L}
}

@ARTICLE{1992ApJ...395..140B,
       author = {{Bertoldi}, Frank and {McKee}, Christopher F.},
        title = "{Pressure-confined Clumps in Magnetized Molecular Clouds}",
      journal = {\apj},
         year = 1992,
        month = aug,
       volume = {395},
        pages = {140},
          doi = {10.1086/171638},
       adsurl = {https://ui.adsabs.harvard.edu/abs/1992ApJ...395..140B}
}

@ARTICLE{1996ApJ...465..815O,
       author = {{Onishi}, Toshikazu and {Mizuno}, Akira and {Kawamura}, Akiko and {Ogawa}, Hideo and {Fukui}, Yasuo},
        title = "{A C 18O Survey of Dense Cloud Cores in Taurus: Core Properties}",
      journal = {\apj},
         year = 1996,
        month = jul,
       volume = {465},
        pages = {815},
          doi = {10.1086/177465},
       adsurl = {https://ui.adsabs.harvard.edu/abs/1996ApJ...465..815O}
}
\bibliographystyle{aa} 



\end{document}